\documentclass[twocolumn,superscriptaddress,showpacs,prd,aps,amsmath,amssymb,nofootinbib]{revtex4-1}
\usepackage{graphicx}
\usepackage{amssymb}
\usepackage{bm}
\usepackage{color}

\newcommand{\beq}{\begin{equation}} 
\newcommand{\eeq}{\end{equation}} 
\newcommand{\beqn}{\begin{eqnarray}} 
\newcommand{\eeqn}{\end{eqnarray}} 
\newcommand{\pa}{\partial}
\newcommand{\na}{\nabla}

\newcommand{\gabu}{g^{\alpha\beta}}

\newcommand{\gmabu}{\gamma^{ab}}

\newcommand{\Gabd}{G_{\alpha\beta}}

\newcommand{\zD}{{\raise1.0ex\hbox{${}^{\ \circ}$}}\!\!\!\!\!D}
\newcommand{\alone}{{\raise0.5ex\hbox{${}^{\ 1}$}}\!\!\!\!\alpha}
\newcommand{\Od}{{O}}

\newcommand{\dl}{\delta}

\newcommand{\Dl}{\Delta}

\newcommand{\Lie}{\mbox{\pounds}}

\newcommand{\nalam}{\mathrel{\raise0.9ex\hbox{$^\lambda$}\mkern-14mu
\lower0.0ex\hbox{$\nabla$}}}

\newcommand{\Phiint}{{\Phi_{\rm INT}}}
\newcommand{\Nrf}{{N_r^{\rm f}}}
\newcommand{\Nrm}{{N_r^{\rm m}}}

\newcommand{\Kabd}{K_{ab}}

\newcommand{\zeroD}{{\raise1.0ex\hbox{${}^{\ \circ}$}}\!\!\!\!\!D}
\newcommand{\Lap}{\Delta}
\newcommand{\zLap}{{\raise1.0ex\hbox{${}^{\ \circ}$}}\!\!\!\!\Delta}
\newcommand{\zna}{{\raise1.0ex\hbox{${}^{\ \circ}$}}\!\!\!\!\!\nabla}
\newcommand{\zS}{{\raise1.0ex\hbox{${}^{\ \circ}$}}\!\!\!\!\!S}
\newcommand{\GA}{\alpha}
\newcommand{\GB}{\beta}
\newcommand{\GG}{\gamma}
\newcommand{\GD}{\delta}
\newcommand{\GE}{\epsilon}

\newcommand{\GC}{\psi}
\newcommand{\GX}{\chi}

\newcommand{\GP}{\phi}

\newcommand{\GJ}{\theta}
\newcommand{\pd}{\partial}

\newcommand{\be}{\begin{equation}}
\newcommand{\ee}{\end{equation}}

\begin{document}

\title{
%Compact Object CALculator -- {\sc COCAL} :
%A new code for equilibriums and quasiequilibrium initial data 
New code for equilibriums and quasiequilibrium initial data 
of compact objects}

\author{K\=oji Ury\=u}
%\email{}
\affiliation{Department of Physics, University of the Ryukyus, Senbaru, Nishihara, 
Okinawa 903-0213, Japan}

\author{Antonios Tsokaros}
%\email{}
\affiliation{Department of I.C.S.E., University of Aegean, Karlovassi 83200, Samos, Greece} 

\date{\today}

\begin{abstract}
We present a new code, named {\sc COCAL} - Compact Object CALculator, for 
the computation of equilibriums and quasiequilibrium initial data sets 
of single or binary compact objects of all kinds.  
In the {\sc cocal} code, those solutions are calculated 
on one or multiple spherical coordinate patches 
covering the initial hypersurface up to the asymptotic region.  
The numerical method used to solve field equations written in elliptic 
form is an adaptation of self-consistent field iterations 
in which Green's integral formula is computed using multipole 
expansions and standard finite difference schemes.  
We extended the method so that it can be used on a computational domain 
with excised regions for a black hole and a binary companion.  
Green's functions are constructed for various types of boundary 
conditions imposed at
the surface of the excised regions for black holes.  
The numerical methods used in {\sc cocal} are chosen 
to make the code simpler than any other recent initial data codes, accepting 
the second order accuracy for the finite difference schemes.  
We perform convergence tests for time symmetric single
black hole data on a single coordinate patch, and 
binary black hole data on multiple patches.  
Then, we apply the code to obtain spatially conformally flat binary black hole 
initial data using boundary conditions including the one based on the existence 
of equilibrium apparent horizons. 
\end{abstract}

\maketitle

\section{Introduction}
\label{sec:int}

In the last decades, simulation codes for 
compact objects have been successfully developed 
in the field of numerical relativity, and various 
dynamical simulations have been performed 
including binary neutron stars and black holes 
inspirals to merger \cite{NRreview}, 
a massive core-collapse 
to a proto neutron star or a black hole (BH) formation 
\cite{SN}, and black hole dynamics \cite{HigherDimBH}.  
Recent efforts on these subjects are, for example, 
to study more realistic situations by including microphysics 
of the neutron stars (NS) \cite{RecentNS}, 
wider range of parameter space 
such as mass ratio and spins of binary black hole (BBH) 
mergers\cite{RecentBH} or BBH mergers in an ambient disk 
\cite{RecentBBH}.  
Accordingly, more realistic and accurate construction 
of initial data for such compact objects is required.

Several researchers have developed methods for 
computing various types of initial data sets for those 
simulations \cite{Cook:2000vr,BBH_QE,BNSCF,BHNS_QE,UryuCF,
UryuWL,TU2007,Huang:2008vp}, and equilibriums 
of rotating compact objects (see e.g.~\cite{RNSreview}).  
Many of such initial data codes are specialized to a certain 
problem such as a single stationary and axisymmetric neutron 
star or BBH data on a conformally flat initial hypersurface.  
An exception is {\sc lorene} \cite{lorene}, which is 
one of most used code for computing initial data for 
the merger simulations of binary neutron stars and black holes.  
The {\sc lorene} code was originally developed for computing 
rapidly rotating neutron stars, but has been extended to be 
capable of computing various kinds of equilibriums and 
quasiequilibrium initial data sets.

In this paper, we introduce our project for developing 
new codes for computing initial data of astrophysical 
compact objects, a single as well as binary compact 
objects of all kinds, and present several tests for 
the new codes.  
Our aim is to develop a set of codes for computing, 
on an initial hypersurface, a single neutron star 
(or a compact star such as a quark star), binary neutron 
stars and black holes, a central neutron star or black hole 
surrounded by a toroidal disk, and all these systems 
with magnetic fields.  We call our new codes ``{\sc cocal}'' 
as the abbreviation for Compact Object CALculator
\footnote{``Coc\`al'' means ``seagull'' in Trieste dialect of Italian.}.  
A noteworthy idea of the ``{\sc cocal}'' project 
is to develop a code using less technical numerical methods 
than the recent initial data solvers with spectral methods 
\cite{BBH_QE,BNSCF,BHNS_QE,lorene}.  Also the modules and subroutines of the 
fortran90 code are structured simply so that the code may be 
accessible by those who mastered introductory courses for 
programing.  
Such feature will help future developments to incorporate 
more complex physics in the code such as radiation, neutrino 
radiation transfers, or realistic equations of state for 
the high density nuclear matter.

The numerical method used in {\sc cocal} is based on 
Komatsu-Eriguchi-Hachisu (KEH) method for computing 
equilibrium of a rotating neutron star \cite{KEH89}.  
In our previous works \cite{UryuCF,UryuWL}, we have extended KEH method 
for computing initial data for binary 
compact objects in quasiequilibriums.  
Among all, in the paper \cite{TU2007}, we have 
introduced multiple spherical coordinate 
patches for computing binary compact objects.  
We improve the idea of the multiple patches in all aspects 
in the new {\sc cocal} code.  
In the paper \cite{Huang:2008vp}, we have presented convergence tests 
and solution sequences for rotating neutron star initial data, 
which was calculated by the first version of {\sc cocal}.  
In this paper for introducing the {\sc cocal} code, 
we focus on the basic setup of the multiple spherical coordinate 
patches and the coordinate grids, 
the method of elliptic equation solver on the multiple 
patches, and convergence tests for binary black hole initial data.  
The paper is organized as follows: in Sec.\ref{sec:nm}, we introduce 
an overview of the {\sc cocal} project, then coordinate setups, elliptic solver 
and other materials on numerical computing.  In Sec.\ref{sec:Codetests} 
the results of convergence tests are presented.  In Sec.\ref{sec:IWMid}
solutions of BBH data on a conformally flat initial 
hypersurface are presented.  We use geometric units with $G=c=1$ 
throughout the paper.

\section{{\sc COCAL} code}
\label{sec:nm}

\subsection{Overview}

In the {\sc cocal} project, we aim to develop numerical codes for 
computing a single compact object as well as 
binary compact objects in (quasi)equilibrium using 
common numerical method as much as possible.  
A plan for such codes also depends on 
how to formulate the problem to solve such compact objects.  
Usually, a system of equations to describe equilibrium systems 
of compact objects involves a set of elliptic equations for the 
gravitational fields, and relativistic hydrodynamical equations 
including Euler equation and the rest mass conservation 
equation, to which a stationary condition, either a time or 
a helical symmetry \cite{Friedman:2001pf}, is imposed.  
When the magnetic field is present, elliptic equations 
for the electromagnetic fields are added, and the equations 
for the fluid is replaced by magnetohydrodynamical (MHD) Euler 
equation.  Because the stationary Euler, or MHD-Euler, equations 
are difficult to integrate numerically, 
a set of the first integrals in the form of algebraic 
equations, a sufficient condition for the stationary 
(MHD) Euler equations, is derived and solved simultaneously with 
the field equations (see e.g. \cite{{Uryu:2010su},Gourgoulhon:2011gz}).  

A choice of a numerical method is therefore made according to 
what kind of solver is used for solving the system of elliptic equations.  
The numerical method used in {\sc cocal} is based on KEH method 
for computing equilibriums of rotating neutron stars \cite{KEH89}.  
In this method, the elliptic equations are solved on spherical 
coordinates using the Green's formula iteratively.  
This is done by separating the flat Laplacian or Helmholtz operator 
on the variable to be solved for, then moving remaining (possibly non-linear) 
terms to the source, and rewriting it in the integral form using Green's formula.  
Expanding the Green's function using spherical harmonics, 
the formula is integrated on the spherical coordinate grids 
numerically (see, e.g. \cite{UryuCF,UryuWL,YBRUF06}).  
The method is extended for computations of binary compact objects 
as discussed in this section.  

We choose simple finite difference formulas which are mostly 
second order accurate, and in some cases choose third or fourth 
order formulas only if they are necessary (see Sec.~\ref{sec:singleBH}).  
No symmetry, such as an equatorial plane symmetry, is assumed a priori on 
the 3D spherical computational domain.  
The {\sc cocal} code is written in Fortran 90 language, 
and runs with a few GB of memory for a model with a moderate resolution.  
%%% For example, those servers with Xeon 5500 with a few tenths of GB 
%%% memory would be so far enough for computing typical quasiequilibrium 
%%% solution.  The codes are well structured, so optimizing them 
%%% for a particular platform would be straightforward.  

We have developed basic subroutines for the {\sc cocal} code, including 
the coordinate grid setups for a single and multiple spherical coordinates, 
as well as 
the elliptic solvers for a single or binary compact objects, which we discuss 
in detail below.  
Mainly, two types of initial value formulation for Einstein's equation have been 
coded so far; one is assuming spatial conformal flatness 
(Isenberg-Wilson-Mathews formulation \cite{ISEN78,WM89}), and the other 
non-conformal flatness (waveless formulation \cite{SUF04,UryuWL}).  

Also, the quadrupole formula to compute the gravitational wave amplitude 
and luminosity, and a Helmholtz solver have been developed.  
We are in the phase to test all basic subroutines 
by computing simple test problems as well as known problems such as 
BBH initial data, or rotating neutron star solutions.  In the next step, 
we will combine these developments and start computing new equilibriums 
and initial data sets such as helically symmetric binary compact objects, 
or magnetized compact objects.

\subsection{Coordinate patches for binary systems}
\label{sec:patch}

\begin{figure*}
\includegraphics[width=14cm,clip]
{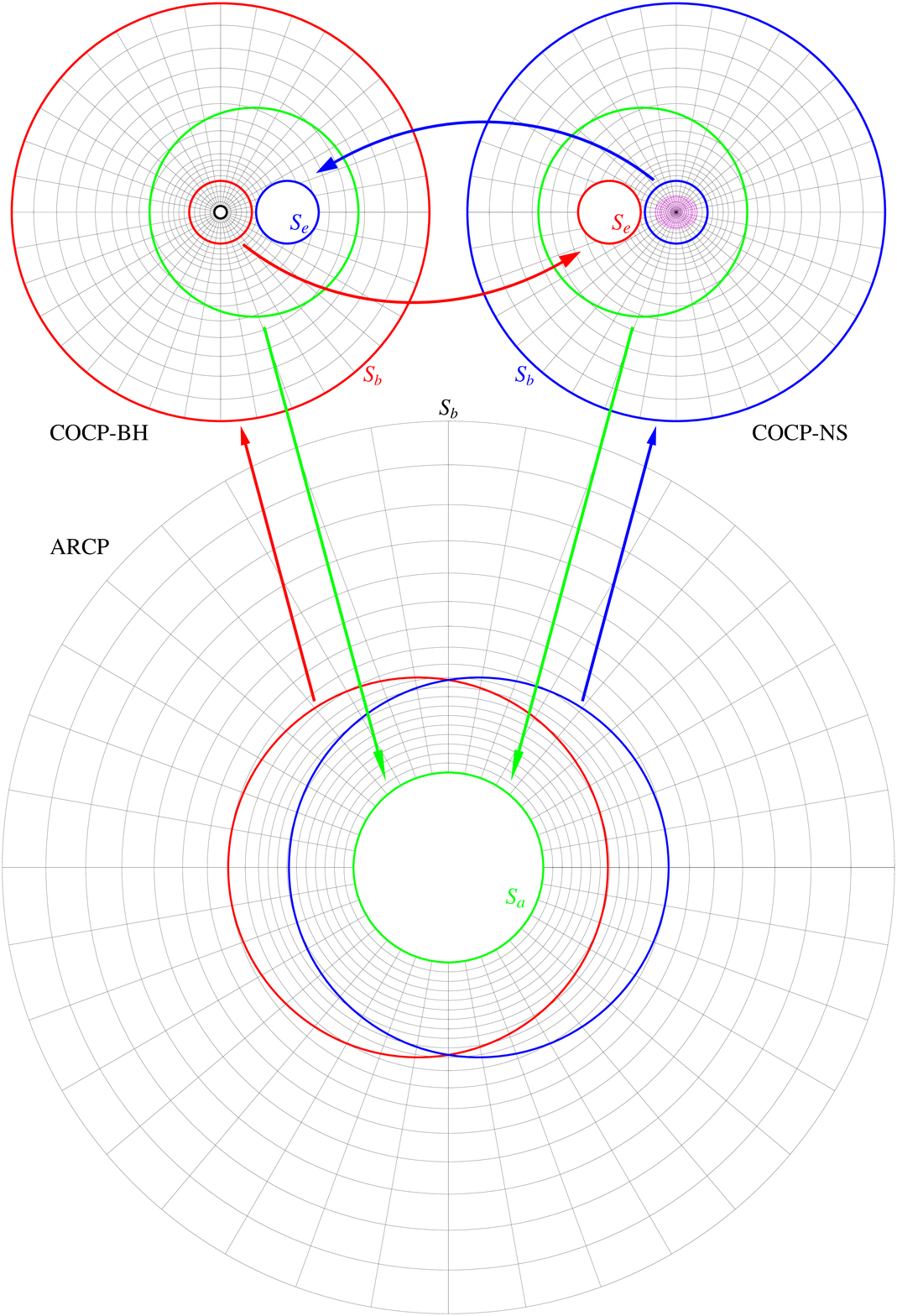}
\caption{A typical setup for multiple coordinate grid patches in the {\sc cocal} code 
for a BH-NS system.  Left and right top patches are those 
for compact object coordinate patches (COCP) centered at each compact object 
(BH for the left, and NS for the right).  
The smallest circle with thick curve in COCP-BH is the sphere $S_a$ where 
the interior region is excised and certain BH boundary conditions are imposed.  
The ovals drawn in COCP-NS denote NS.  
Bottom patch is that for asymptotic region coordinate patch (ARCP), centered 
at the mass center of the system, and extends to asymptotics.  
The arrows represent maps of potentials between the multiple patches.  
Alternatively, the radius of COCP may be extended to asymptotics, 
instead of using ARCP.  
Note that the spheres $S_a$, $S_b$, and $S_e$ of these coordinate patches are 
distinct ones on a spacelike hypersurface $\Sigma_t$.  
The radius of each coordinate patch doesn't reflect the 
size used in actual computations.}
\label{fig:multipatches}
\end{figure*}

We assume that the spacetime is $\cal M$ is foliated by a family of 
spacelike hypersurface $(\Sigma_t)_{t\in {\mathbb R}}$, 
${\cal M} = {\mathbb R} \times \Sigma$ parametrized by 
$t\in {\mathbb R}$.  
In the {\sc cocal} code, we solve fields on a initial 
hypersurface $\Sigma_t$ which may be stationary (in equilibrium), 
or quasi-stationary (in quasi-equilibrium).  
The initial hypersurface $\Sigma_t$ is covered by overlapping 
multiple spherical coordinate patches whose coordinates 
are denoted by $(r,\theta,\phi)$.  
Angular coordinates cover all directions 
$(\theta,\phi) \in [0,\pi]\times [0,2\pi]$
without any symmetry imposed.  
We also introduce Cartesian coordinates as 
a convenient reference frame in a standard manner, 
that is, to have the positive side of x-axis coincide with 
a $(\theta,\phi)=(\pi/2,0)$ line, that of y-axis 
with a $(\theta,\phi)=(\pi/2,\pi/2)$ line, and that of 
z-axis with a $\theta=0$ line.  

In Fig.~\ref{fig:multipatches}, a schematic figure of three 
spherical coordinate patches whose coordinates are discretized 
in grid points is shown for the case of computing BH-NS 
binary systems by {\sc cocal}.  Shown is the 2D section of 
the 3D hypersurface, that may agree with the equatorial or 
meridional plane of the compact objects.  Even though this 
may be the most complex setup for coordinate grids 
in {\sc cocal}, it is not technical at all compared to those 
of existing codes in which adaptive coordinates are used.

For the computation of binary systems, 
two compact objects are placed at the centers of the two patches.  
We call these two patches the compact object coordinate patch (COCP),  
and the third patch the asymptotic region coordinate patch (ARCP).  
A domain of COCP is defined between two concentric spheres 
$S_a$ and $S_b$ from which an interior of an another 
sphere $S_e$ is excised.  Writing radii of $S_a$, $S_b$, 
and $S_e$ as $r_a$, $r_b$, and $r_e$ respectively, 
we define spherical coordinates of COCP as 
$(r,\theta,\phi) \in [r_a,r_b]\times[0,\pi]\times[0,2\pi]$, 
and locate the center of the excised sphere at 
$(r,\theta,\phi)=(d_s,\pi/2,0)$, that is on the positive side of the x-axis
\footnote{The positive side of x-axis of COCP-NS in Fig.~\ref{fig:multipatches} 
is pointing from the center of the coordinate grids toward left.}.  
We introduce the excision of a domain interior of the sphere $S_e$ 
for computing binary systems, and elucidate its role in the following 
section.  When a single and/or axisymmetric object is computed, the 
excision inside $S_e$ is not used.  
When BH is computed on COCP-BH with certain BH boundary conditions 
on $S_a$ $(r_a>0)$, the region inside of $S_a$ is excised.  
When a NS or a puncture BH is calculated, the sphere $S_a$ of COCP is 
removed by setting $r_a=0$, so the radial coordinate covers up to $r=0$.  
A domain of ARCP is defined between two concentric spheres 
$S_a$ and $S_b$, and its spherical coordinates are defined as 
$(r,\theta,\phi) \in [r_a,r_b]\times[0,\pi]\times[0,2\pi]$.  
When values of field potentials or other variables are communicated 
from one patch to the other, those values on a certain sphere are mapped 
to a corresponding boundary sphere as indicated by arrows 
in Fig.~\ref{fig:multipatches}.

Values of radii $r_a$, $r_b$, and $r_e$ for each of the coordinate 
patches used in actual computations will be summarized in 
Sec.~\ref{sec:Codetests}.  Typically, they are set as follows.  
For the case of using three patches as in Fig.~\ref{fig:multipatches}, 
the radius $r_a$ of the inner boundary $S_a$ of ARCP is taken 
large enough to be placed outside of the excised spheres 
$S_e$ for compact objects on COCP, 
but small compare to the size of the domain $r_b$ of the COCP.  
The outer boundary of ARCP, the radius $r_b$ of the sphere $S_b$, 
is extended to the asymptotic region when a field 
to be calculated behaves as a Coulomb type fall off, 
while it is truncated at the near zone when a radiation field 
is calculated.  The center of ARCP is located at the 
center of mass of binary compact objects.  
Therefore typically, for compact objects with a mass $M$, 
$r_a =\Od(M)$ (0 for NS), $r_b =\Od(100 M)$, and $r_e = \Od(M)-\Od(10M)$ 
for COCP, and $r_a =\Od(10M)$, and $r_b =\Od(10^6 M)$ or larger for ARCP.

As an another option for the choice of coordinate patch, 
the outer radius of each COCP $r_b$ may be extended to asymptotics,
say $r_b =\Od(10^6 M)$, and ARCP is removed.  This option 
simplifies the code, but it can not be used when a radiation field 
is computed by solving Helmholtz equation.  We present tests for 
the Helmholtz solver in a separate paper.

\subsection{Elliptic equation solver}
\label{sec:solver}

As mentioned earlier, the formulation for computing  
(quasi-)equilibrium configurations of compact objects 
results in a coupled system of elliptic equations, either 
Poisson or Helmholtz equations with non-linear source terms, 
coupled with algebraic equations.  The numerical method used 
in {\sc cocal} to solve such a system of equations is an extension of 
KEH method which is an application of self-consistent field method 
for computing equilibriums of self-gravitating fluids to 
general relativistic stars \cite{KEH89}.  A distinctive feature of
these methods is the use of Green's formula for an elliptic equation solver.  
We have introduced in previous papers our implementation of KEH method 
to compute binary neutron stars and black holes 
\cite{TU2007,Huang:2008vp,UryuCF,UryuWL}.

In the {\sc cocal} code, we have made a major change in the choice of 
coordinate patch, and accordingly in the elliptic equation solver.  
Our new implementation is better in all aspects for 
computing binary compact objects than our previous ones.  
We come back to discuss this point after we introduce 
the elliptic equation solver in {\sc cocal}.

In solving each field equation, we separate out a flat Laplacian or Helmholtz 
operator $\cal L$, and write it with a 
non-linear source S, 
\be
{\cal L} \Phi = S,
\label{eq:Poisson}
\ee
on an initial slice $\Sigma_t$, where $\Phi$ represents metric potentials.  
For the case of Laplacian ${\cal L} = \Lap$, 
using the Green's function without boundary 
$G(x,x')=1/|x-x'|$ that satisfies 
\be
\Lap G(x,x') = -4\pi \dl(x-x'), 
\label{eq:Green}
\ee
Green's identity is obtained by 
\beqn
\Phi(x)= -\frac1{4\pi}\int_{V} G(x,x')S(x') d^{3}x' 
\qquad\qquad\qquad\nonumber \\
+ \frac{1}{4\pi} \int_{\pd V} \left[G(x,x')\na'^{a} \Phi(x')
- \Phi(x')\na'^a G(x,x') \right]dS'_a. 
\ \ \ 
\label{eq:GreenIde}
\eeqn
where $V$ is the domain of integration, $x,x' \in V\subseteq\Sigma_0$, and 
$\pa V$ is its boundary.  
For the case of BH-NS system shown in Fig.~\ref{fig:multipatches}, 
the boundary of COCP-BH becomes $\pa V = S_a \cup S_b \cup S_e$, 
that of COCP-NS $\pa V = S_b \cup S_e$, and that of 
ARCP $\pa V = S_a \cup S_b$.  
For the evaluation of the integrals in Eq.(\ref{eq:GreenIde}), 
a multipole expansion of $G(x,x')$ in associated Legendre functions
on the spherical coordinate is used; 
\beqn
G(x,x')&=&
\frac{1}{\left|{x}-{x'}\right|}\,=\, 
\sum_{\ell=0}^\infty g_\ell(r,r') \sum_{m=0}^\ell \epsilon_m \,
\frac{(\ell-m)!}{(\ell+m)!}
\nonumber\\
&&\!\!\!\!\!\!\!
\times
P_\ell^{~m}(\cos\theta)\,P_\ell^{~m}(\cos\theta')
\cos m(\varphi-\varphi'), 
\eeqn
where the radial Green's function $g_\ell(r,r')$ is defined by 
\beq
g_\ell(r,r')=\frac{r_<^\ell}{r_>^{\ell+1}}, 
\eeq
with 
$
r_> := \sup\{r,r'\}, \ r_< := \inf\{r,r'\}, 
$
and the coefficients $\epsilon_m$ are equal to $\epsilon_0 = 1$, 
and $\epsilon_m = 2$ for $m\ge 1$.

Eq.(\ref{eq:GreenIde}) is an integral identity but is not a solution of 
Eq.(\ref{eq:Poisson}) in a sense that both of  
$\Phi$ and its derivative $n^a \na_a \Phi$ can not be specified freely.  
Eq.~(\ref{eq:GreenIde}) can be used to compute a potential over $V$, 
only if correct values of $\Phi$ and $n^a \na_a \Phi$ are 
known at the boundary $\pa V$.  Here, $n^a$ is an outward normal to $\pa V$.  
Therefore, as it is commonly found in 
standard text books for electromagnetism \cite{Jackson}, 
a homogeneous function $F(x,x')$ for the Laplacian is added to evaluate the % appropriate
Green's function that satisfies the boundary condition at $\pa V$.  For example,  
a Green's function $G(x,x')+F(x,x')=0$ at $\pa V$ is used to impose Dirichlet 
boundary condition.  In our previous paper \cite{TU2007}, we have developed 
an elliptic equation solver on multiple coordinate patches that uses such Green's 
functions and solve Eq.(\ref{eq:GreenIde}) by iteration.

A construction of such a Green's function that satisfies a boundary condition 
is, however, possible only when a certain specific geometry of the domain of 
computation is adapted to the coordinate systems.  
In the present case for COCP of the {\sc cocal} code in Fig.~\ref{fig:multipatches}, 
a Green's function that satisfy boundary conditions 
at $S_a$ and $S_b$ may not be derived in a practical form of equation, 
because we excised the region inside of the sphere $S_e$.  
To impose boundary conditions at $S_a$ and $S_b$, we introduce a homogeneous 
solution $\chi(x)$, and write a formal solution as
\beq
\Phi(x) \,=\, \chi(x) \,+\, \Phiint(x), 
\label{eq:solver}
\eeq
where $\Phiint$ is equal to the right hand side of Eq.(\ref{eq:GreenIde}); 
\beqn
\Phiint(x)= -\frac1{4\pi}\int_{V} G(x,x')S(x') d^{3}x' 
\qquad\qquad\qquad\nonumber \\
+ \frac{1}{4\pi} \int_{\pd V} \left[G(x,x')\na'^{a} \Phi(x')
- \Phi(x')\na'^a G(x,x') \right]dS'_a.  
\ \ \ 
\label{eq:Green_int}
\eeqn

The homogeneous solution is computed so that the potential $\Phi$ satisfies 
the boundary conditions, which are either one of Dirichlet, Neumann, or 
Robin boundary conditions at the boundary spheres $S_a$ or $S_b$; 
\beqn
&{\rm Dirichlet :}\quad & \Phi_{\rm BC} \,=\, f_{\rm D}
\label{eq:bcon_dir}
\\
&{\rm Neumann :}\quad & n^a \na_a\Phi_{\rm BC} \,=\, f_{\rm N}
\label{eq:bcon_neu}
\\
&{\rm Robin :}\quad & n^a \na_a\Phi_{\rm BC} + \frac14 \na_a n^a \Phi_{\rm BC} \,=\, f_{\rm R}
\label{eq:bcon_robin}
\eeqn
where $f_{\rm D}$, $f_{\rm N}$, and $f_{\rm R}$ are given functions on 
the spheres $S_a$ or $S_b$.  
Formulas for $\chi(x)$ are derived by using a Legendre expansion 
in an usual manner as shown in Appendix.  Noticing 
$\chi(x) = \Phi(x) \,-\, \Phiint(x)$ 
the formulas for $\chi(x)$ can be written analogously 
to the surface integral terms 
of Green's formula, but with a different kernel function $G^{\rm BC}$ 
\beqn
\chi(x)
&=& 
%\frac{1}{4\pi} \int_{\pd V\setminus S_e} 
\frac{1}{4\pi} \int_{S_a \cup S_b} 
\left[G^{\rm BC}(x,x')\na'^{a} (\Phi_{\rm BC} - \Phiint)(x')
\right.
\nonumber\\
&& \left. 
- (\Phi_{\rm BC} - \Phiint)(x')\na'^a G^{\rm BC}(x,x') \right]dS'_a.  
\label{eq:Green_chi}
\eeqn
The function $G^{\rm BC}(x,x')$ is 
expanded in terms of the associated Legendre functions
\beqn
G^{\rm BC}(x,x') &=& 
\sum_{\ell=0}^\infty g^{\rm BC}_\ell(r,r') \sum_{m=0}^\ell \epsilon_m \,
\frac{(\ell-m)!}{(\ell+m)!}
\nonumber\\
&&\!\!\!\!\!\!\!\!\!\!\!\!\!\!\!\!\!
\times
P_\ell^{~m}(\cos\theta)\,P_\ell^{~m}(\cos\theta')
\cos m(\varphi-\varphi'), 
\label{eq:general_G}
\eeqn
where the radial function $g^{\rm BC}_\ell(r,r')$ is chosen 
according to the type of boundary conditions used.
We derive such radial functions used in the corresponding surface integrals
for various cases of boundary conditions as listed in Table \ref{tab:Greenfn}.  
Concrete forms of these functions are presented in Appendix \ref{sec:SurfInt}.  
\begin{table}
\begin{tabular}{lll}
\hline
$G^{\rm BC}(x,x')$ & Boundary $S_a$ & Boundary $S_b$ \\
\hline
$G^{\rm NB}(x,x')$ & None & None \\
$G^{\rm DD}(x,x')$ & Dirichlet & Dirichlet \\
$G^{\rm ND}(x,x')$ & Neumann   & Dirichlet \\
$G^{\rm DN}(x,x')$ & Dirichlet & Neumann \\
$G^{\rm NN}(x,x')$ & Neumann   & Neumann \\
$G^{\rm RD}(x,x')$ & Robin     & Dirichlet \\
$G^{\rm DR}(x,x')$ & Dirichlet & Robin     \\
\hline
\end{tabular}
\caption{List of Green's function available in the {\sc cocal} code.  
The second and third columns correspond to the types of boundary conditions 
imposed on the boundary spheres $S_a$ and $S_b$, respectively.  
The case with no boundary condition is denoted by None.} 
\label{tab:Greenfn}
\end{table}

\subsection{Iteration procedure}
\label{sec:iteration}

The final solution will be obtained from the iteration of Eq.~(\ref{eq:solver}), 
with Eqs.~(\ref{eq:Green_int}) and (\ref{eq:Green_chi}), 
where explicit form of Eq.~(\ref{eq:Green_chi}) for $\chi(x)$ 
depends on the boundary condition, 
for example, Eq.~(\ref{eq:chiioND}) or (\ref{eq:chiioDD}). 

We summarize the $n^{\rm th}$ step of 
the Poisson solver in the {\sc cocal} code as follows:
\begin{itemize}
\item[1)] Compute the volume source term $S(\Phi^{(n-1)})$ as well as the surface source terms
on all possible surfaces $S_a$, $S_b$, $S_e$. 
\item[2)] Compute the volume integral and the surface integral at $S_e$ for obtaining
$\widehat{\Phi}_{\rm INT}(x)$ from Eq.~(\ref{eq:Green_int}).
\item[3)] Compute the effective source for the integral on $S_a$ and $S_b$. 
For Dirichlet boundary condition
it will be $\Phi_{\rm BC}-\widehat{\Phi}_{\rm INT}$, while for Neumann
$\frac{\pd\Phi_{\rm BC}}{\pd r}-\frac{\pd\widehat{\Phi}_{\rm INT}}{\pd r}$.
\item[4)] Compute the surface integrals at $S_a$ and $S_b$ for obtaining 
$\widehat{\chi}(x)$ according to Eq.~(\ref{eq:Green_chi}) 
using the appropriate function $G^{\rm BC}$ for the boundary
conditions of the problem.  
\item[5)] Add the results from steps 2) and 4) to obtain $\widehat{\Phi}(x)$ from 
Eq.~(\ref{eq:solver}).
\item[6)] Update $\Phi^{(n)}$ according to 
\[ \Phi^{(n)}(x)\,:=\,c\widehat{\Phi}(x)+(1-c)\Phi^{(n-1)}(x) \]
where $0.1\leq c\leq 0.4$.
\item[7)] Check if 
\[2\frac{|\Phi^{(n)}-\Phi^{(n-1)}|}{|\Phi^{(n)}|+|\Phi^{(n-1)}|}< \epsilon_c  \]
for all points of the grids, 
where $\epsilon_c = 10^{-6}-10^{-8}$ is taken in typical computations, 
and $\epsilon_c = 10^{-7}$ in this paper.  
If yes exit. If no go back to step 1).
\end{itemize} 
Here, intermediate variables during an iteration step are 
denoted with a hat as $\widehat{\Phi}_{\rm INT}$, $\widehat{\chi}$, 
and $\widehat{\Phi}$.  
The above iteration procedure is applied to each coordinate patch 
one after other.  
In step 1), the sources of the surface terms are computed 
either from boundary conditions to be imposed on the surface, 
or from data of corresponding surface on the other patch
(see, Fig.~\ref{fig:multipatches} how the potentials are transferred 
from a boundary surface to the other).  
Several different iteration schemes are possible for solving 
a set of elliptic equations for more than one variable.  As long as 
we experimented, a convergence of the iteration does not depend on 
the order of computing those variables at each iteration step.  

We will see this elliptic equation solver produce accurate solutions 
for test problems of binary black hole data.  
Two comments on the elliptic equation solver are made here.  
Although, $\Phiint$ in Eq.~(\ref{eq:Green_int}) involves 
surface integrals on all $S_a$, $S_b$, and $S_e$, those on $S_a$ and $S_b$ 
are not included in $\Phiint$ in an actual computation.  
Those computations are redundant because the homogeneous solution 
$\chi(x)$ is determined again from the surface integrals on $S_a$ and $S_b$
as in Eq.~(\ref{eq:Green_chi}).  
So far, we do not plan to develop elliptic solvers for vector (tensor) 
fields in which Green's functions are expanded in vector (tensor) spherical 
harmonics.  Instead, we write the Cartesian components of vector or tensor 
equations, and solve each components as scalar equations on spherical grids 
for simplicity.  
We will see an example in Sec.~\ref{sec:IWMid} (see also, \cite{UryuWL}).

\begin{figure*}
  \includegraphics[height=100mm,clip]{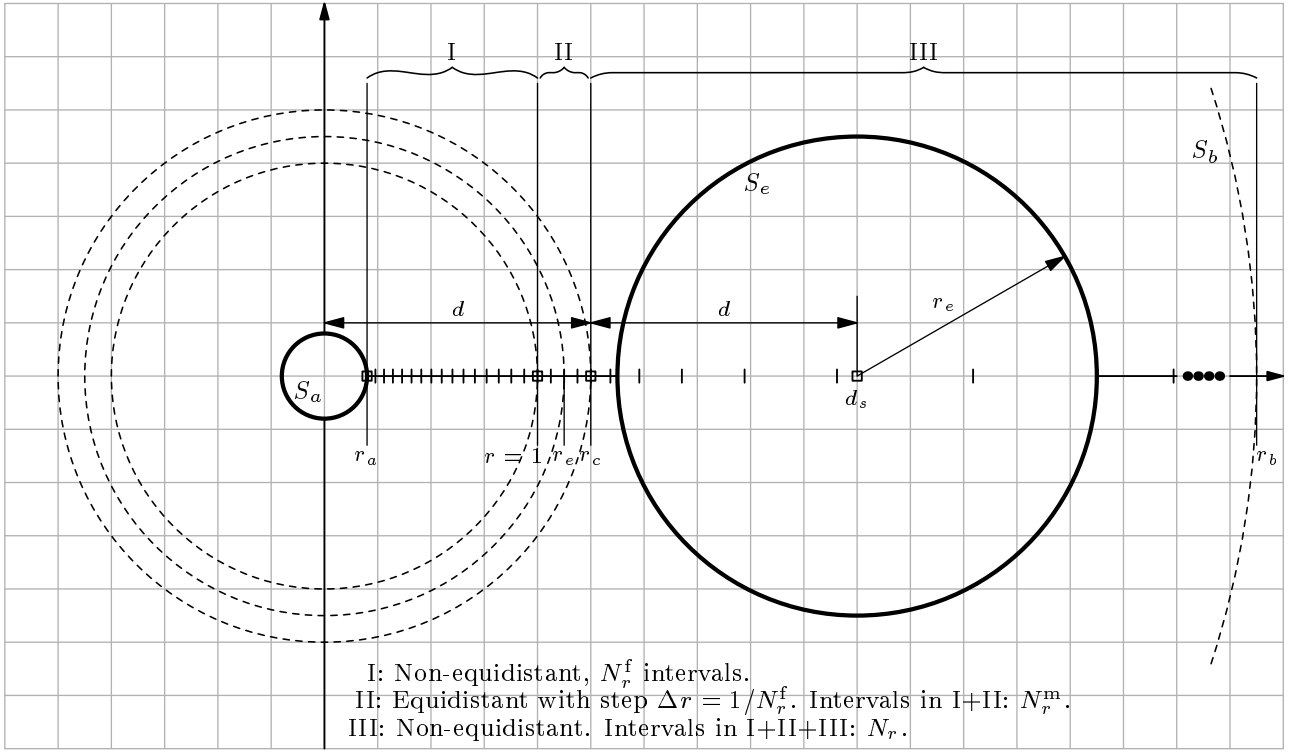}
  \caption{Radial coordinate grids for COCP for the case of 
  regions for BH and binary companion being excised.  The radial 
  coordinate grids corresponds to those of lowest resolutions
  A1, B1, D1 and F1 in Tables \ref{tab:1BHtest_grids} and \ref{tab:BBHtest_grids}.}
  \label{fig:radial}
\end{figure*}
\subsection{Grid spacing}
\label{sec:GridSp}

We apply finite difference scheme to solve the system of 
equations for compact objects on the spherical domain 
introduced in Sec~\ref{sec:patch} (see, Fig.~\ref{fig:multipatches}).  
Spherical coordinates $(r,\GJ,\GP)$ for COCP and ARCP are 
bounded by two concentric spheres $S_a$ and $S_b$ of radius $r_a$ and 
$r_b$, respectively, with the possible excision of a sphere 
$S_e$ of radius $r_e$ inside COCP.  
The origin of the radial coordinate $r$ is placed at the common 
center of $S_a$ and $S_b$, where the compact object is placed 
for the case of COCP.  Excised sphere for a binary companion $S_e$ 
is always positioned at a positive value on the x-axis at a distance $d_s$ 
from the origin.  Clearly $r_a < d_s-r_e$.  For neutron star calculations 
the sphere $S_a$ is absent and the coordinate system extends from $r=0$ to $r_b$.

In the {\sc cocal} code, the spacing of all coordinate grid points 
$(r_i,\theta_j,\phi_k)$ with $i = 0, \cdots, N_r$, $j = 0, \cdots, N_\theta$, 
and $k = 0, \cdots, N_\phi$, are freely specifiable.  However, 
in ($\GJ,\GP$) directions, uniform grids are recommended to resolve the 
trigonometric and associated Legendre functions 
used in the elliptic equation solver, as well as a structure of 
compact objects evenly.  That is, we set the grid interval in 
these directions as 
\beqn
\Delta\GJ_j &=& \theta_j - \theta_{j-1} \,=\, \Dl \theta \,=\, \frac{\pi}{N_\theta}, 
\\
\Delta\GP_k &=& \phi_k - \phi_{k-1} \,=\, \Dl \phi \,=\, \frac{2\pi}{N_\phi}. 
\eeqn

The grid spacing in the radial direction $r$ 
is usually constructed on one hand to resolve 
the vicinity of the compact object with finer grid spacings, 
and on the other hand to extend to asypmtotics using 
increasingly sparse spacings\footnote{For the case of 
solving Helmholtz equation the region extends to 
a near zone (a size of a several wavelengths of a dominant mode).}.
The setup for radial grids of COCP in the present computation is 
illustrated in Fig.~\ref{fig:radial}.  
The grid is composed from three regions I, II, and III.  
For the case with $r_a < 1$, the region I is set by $r\in[r_a,1]$ 
the region II by $r\in[1,r_c]$, and region III $r\in[r_c,r_b]$.  
We introduce grid numbers $\Nrf,\ \Nrm$ which
correspond to the numbers of intervals in regions I and I+II, respectively. 
We introduce a standard grid spacing $\Delta r$ as $\Delta r = 1/\Nrf$.  
For the case with $r_a < 1$, the grid intervals, $\Dl r_i := r_i - r_{i-1}$,
are defined by 
\beqn
\Dl r_{i+1} &=& h \Dl r_i,\ \ \mbox{for}\ \  i = 1, \cdots, \Nrf-1
\\
\Dl r_{i\phantom{+1}} &=& \phantom{h} \Dl r,\ \ \ \mbox{for}\ \  i = \Nrf, \cdots, \Nrm
\\
\Dl r_{i+1} &=& k \Dl r_i,\ \ \mbox{for}\ \  i = \Nrm, \cdots, N_r-1 \ \ 
\eeqn
which correspond to regions I, II, and III in Fig.~\ref{fig:radial}, respectively, 
where ratios $h(\leq 1)$ and $k(>1)$ are respectively determined from relations
\beqn
1-r_a &=& \Dl r \frac{1-h^\Nrf}{1-h}, 
\\
r_b - r_c &=& \Dl r \frac{k(k^{N_r-\Nrm}-1)}{k-1}.  
\label{eq:coord_r_ratio_k}
\eeqn
For the case with $r_a > 1$, which is mostly for ARCP, 
the grid intervals, $\Dl r_i$, are defined by
\beqn
\Dl r_{i} &=& \phantom{h} \Dl r\ \ \ \mbox{for}\ \  i = 1, \cdots, \Nrm, 
\\
\Dl r_{i+1} &=& k \Dl r_i\ \ \mbox{for}\ \  i = \Nrm, \cdots, N_r-1, \ \ 
\eeqn
where the ratio $k$ is determined from Eq.~(\ref{eq:coord_r_ratio_k}).  
Parameters for the grid setup are listed in Table \ref{tab:grid_param}.

\begin{table}
\begin{tabular}{lll}
\hline
$r_a$  &:& Radial coordinate where the radial grids start.       \\
$r_b$ &:& Radial coordinate where the radial grids end.     \\
$r_c$ &:& Radial coordinate between $r_a$ and $r_b$ where   \\
&\phantom{:}& the radial grid spacing changes from   \\
&\phantom{:}& equidistant to non-equidistant.  \\
$r_e$  &:& Radius of the excised sphere.       \\
$N_{r}$ &:& Number of intervals $\Dl r_i$ in $r \in[r_a,r_{b}]$. \\
$\Nrf$ &:& Number of intervals $\Dl r_i$ in $r \in[r_a,1]$ for $r_a < 1$. \\
&\phantom{:}& or $r \in[r_a,r_a+1]$ for $r_a \geq 1$.  \\
$\Nrm$ &:& Number of intervals $\Dl r_i$ in $r \in[r_a,r_{c}]$. \\
$N_{\theta}$ &:& Number of intervals $\Dl \theta_j$ in $\theta\in[0,\pi]$. \\
$N_{\phi}$ &:& Number of intervals $\Dl \phi_k$ in $\phi\in[0,2\pi]$. \\
%%% $\Nrex$ &:& Number of intervals $\Dl r_i$ in $r \in[r_a,r_{e}]$ (COCP). \\
$d$ &:& Coordinate distance between the center of $S_a$ ($r=0$) \\
&\phantom{:}& and the center of mass. \\
$d_s$ &:& Coordinate distance between the center of $S_a$ ($r=0$) \\
&\phantom{:}& and the center of $S_e$. \\
$L$ &:& Order of included multipoles. \\
\hline
\end{tabular}  
\caption{Summary of grid parameters. 
The radius $r_e$ is defined only for COCP.
%%%$\Nrex$ is defined only for COCP.
}
\label{tab:grid_param}
\end{table}

\subsection{Finite differences and multipole expansion}

Approximations made in our numerical method are 
a truncation of the series of Legendre expansion at a finite 
order of multipole, and an evaluation of a solution on 
discretized grids -- the finite differencing.  
The accuracy of the code is, therefore, determined from 
finite difference formulas to be used, the number of 
grid points and their spacings, and the number of 
multipoles being included.

In the {\sc cocal} code, we use second order mid-point rule for 
numerical integrations and differentiations, 
along with second order linear interpolation rule.  
In the elliptic equation solver (\ref{eq:solver}), 
the source terms are evaluated at the mid-points 
\[
%%%(r_{i-1/2},\theta_{j-1/2},\phi_{k-1/2})\,=\, 
(r_{i-\frac12},\theta_{j-\frac12},\phi_{k-\frac12}) = 
\left(\frac{r_i+r_{i-1}}{2}, \frac{\theta_j + \theta_{j-1}}{2}, 
\frac{\phi_k + \phi_{k-1}}{2}\right),   \] 
and integrated with the weights 
$\Dl r_i \Dl \theta_j \Dl \phi_k$ (other than a Jacobian).   
The mid-point rule has a few advantages.  The second order 
accuracy of the mid-point rule for a quadrature formula 
is maintained even with a discontinuity of the derivative of 
Green's function for a volume integral at $r = r'$.  
It may be possible to derive a higher order quadrature formula 
for numerically integrating such functions, but for instance, 
a Simpson rule doesn't guarantee fourth order accuracy at grid points 
$r_i$ with $i$ being odd integers.  
Also, an excision of a region inside of a sphere 
$S_e$ for a binary companion on COCP 
complicates a derivation of a higher order quadrature formula
which maintain the degree of precision near the sphere.  
Because of the simplicity of the mid-point rule, it is not difficult 
to modify the weights for an integration to maintain the accuracy.  
Another advantage of the mid-point rule is to avoid the coordinate 
singularities of the spherical coordinates.  

In some cases, we also use third or higher order finite difference 
formula for the numerical differentiations.  Especially, 
it is found that it is necessary to use the third order 
finite difference formula for the radial derivatives 
to maintain {\it second order} convergence of the field 
near the BH (see, Sec.\ref{sec:singleBH}).  
Interpolations of scalar functions 
from the grid points to the mid-points are done using 
a second order linear interpolation formula.  
We often need to interpolate 
a function from one coordinate patch to the other, such 
as to compute source term at $S_e$ of COCP.  In such case, 
the functions are interpolated using fourth order Lagrange formula.  
For example when the surface integral at the excised sphere is
computed we need the potential and its derivative at point $x'$ on $S_e$ 
as seen from the center of $S_e$. Theses values are taken by interpolating
the nearby 64 points of the other coordinate system.  Some examples of 
finite difference formulas frequently used in the {\sc cocal} 
code is summarized in Appendix \ref{sec:FDformulas}.

As discussed in \cite{TU2007}, the excised region is introduced 
to improve the resolution in angular directions, and accordingly 
to reduce the number of multipoles to resolve a companion object.  
Without the excised region, the size of the companion object
itself has to be resolved by angular grids, while in our setup, 
it is enough to resolve the size of the excised region, which is
usually taken as large as the half of separation $\sim d_s/2$.  
Then an angle to be resolved can be always about 
$\sim 2 \arcsin 1/2 = \pi/3$.  
Note that although it is, in principle, possible to excise $S_e$ 
with a different radius from each COCP, and it is allowed in {\sc cocal}, 
it is more practical to have the same size of excised region for 
the same reason above.  
To summarize, the angular resolution of a COCP is determined from 
the degree of accuracy to resolve the deformation of the compact objects 
centered at the patch, and to resolve the size of excised sphere, 
$\sim \pi/3$.  
The angular resolution of ARCP depends just on how many multipoles 
one wish to keep in the near zone to asymptotics.  For both COCP and 
ARCP, the number of Legendre expansions are in the range 
$\ell \sim 10-16$ for computing binary systems.  

Legendre $P_\ell^{~m}$ may have $\ell$ zero crossings in $\theta\in[0,\pi)$, and 
$\sin m\phi$ or $\cos m\phi$ have $2m$ zeros in $\phi\in[0,2\pi)$.  
The number of grid points along the angular coordinates has to be 
enough to resolve these multipoles with maximum $\ell$ or $m$, 
say 4 times more than the number of zeros.

\begin{table}
\begin{tabular}{ccccrrrrrr}
\hline
Type & $r_a$ & $r_b$ & $r_c$ & $\Nrf$ & $\Nrm$ & $N_r$ & $N_\theta$ & $N_\phi$ & $L$ \\
\hline
A1 & 0.2 & $10^4$ & 1.25 & 16  & 20    & 48    & 48  & 96 &  12 \\
A2 & 0.2 & $10^4$ & 1.25 & 32  & 40    & 96    & 48  & 96 &  12 \\
A3 & 0.2 & $10^4$ & 1.25 & 64  & 80    & 192   & 48  & 96 &  12 \\
A4 & 0.2 & $10^4$ & 1.25 & 128  & 160  & 384   & 48  & 96 &  12 \\
\hline
B1 & 0.2 & $10^4$ & 1.25 & 16  & 20    & 48    & 24  & 96 &  12 \\
B2 & 0.2 & $10^4$ & 1.25 & 16  & 20    & 48    & 48  & 96 &  12 \\
B3 & 0.2 & $10^4$ & 1.25 & 16  & 20    & 48    & 96  & 96 &  12 \\
B4 & 0.2 & $10^4$ & 1.25 & 16  & 20    & 48    & 192  & 96 &  12 \\
\hline
D1 & 0.2 & $10^4$ & 1.25 & 16  & 20    & 48    & 24  & 24 &  12 \\
D2 & 0.2 & $10^4$ & 1.25 & 32  & 40    & 96    & 48  & 48 &  12 \\
D3 & 0.2 & $10^4$ & 1.25 & 64  & 80    & 192   & 96  & 96 &  12 \\
D4 & 0.2 & $10^4$ & 1.25 & 128  & 160  & 384   & 192  & 192 &  12 \\
\hline
\end{tabular}
\caption{Grid parameters used in convergence tests for 
a single BH and equal mass BBH data solved on 
a single coordinate patch, COCP.}
\label{tab:1BHtest_grids}
\end{table}

\section{Code tests}
\label{sec:Codetests}

\subsection{A toy problem for black holes}

Convergence tests for a time symmetric BH and BBH data are 
performed to check the numerical method of {\sc cocal} presented 
in Sec.\ref{sec:nm}.  We assume the spacetime $\cal M$ is foliated 
by a family of spacelike hypersurfaces 
$(\Sigma_t)_{t\in {\mathbb R}}$, 
${\cal M} = {\mathbb R} \times \Sigma$ parametrized by 
$t\in {\mathbb R}$.  
To obtain simple black hole solutions on $\Sigma_t$, 
we assume time symmetric initial data, the extrinsic 
curvature $\Kabd$ on $\Sigma_t$ vanishes, or in other words, 
assume the line element at the neighborhood of $\Sigma_t$
\be
ds^{2}=-\GA^{2}dt^{2}+\GC^{4}f_{ij}dx^{i}dx^{j} \:,
\ee
where $f_{ij}$ is the flat spatial metric.  Decomposing Einstein's 
equation $\Gabd=0$ with respect to the foliation using 
hypersurface normal $n^\alpha$ to $\Sigma_t$, and the projection 
tensor $\gmabu = \gabu + n^\alpha n^\beta$ to it, 
we write the Hamiltonian constraint 
$\Gabd n^\alpha n^\beta=0$, and a combination of the spatial trace of 
Einstein's equation and the constraint 
$G_{\GA\GB}(\gamma^{\GA\GB} + \frac12n^\alpha n^\beta)=0$, as 
\be
\nabla^{2}\GC=0 \:\:\:\:\:\: \textrm{and} \:\:\:\:\:\: \nabla^{2}(\GA\GC)=0 \: .
\label{eq:Laplace}
\ee 
These equations have solutions, which correspond to the Schwarzschild metric 
in isotropic coordinates for a single BH.  For a two BH case, a BBH solution 
is given by Brill-Lindquist \cite{BL63}  
\beq
\GC \,=\, 1+\frac{M_1}{2r_1}+\frac{M_2}{2r_2}
%%% =1+\frac{a_1}{r_1}+\frac{a_2}{r_2} 
\:\:\:\: \textrm{and} \:\:\:\:
%%% \label{eq:Psi2bhem}  \\  
\alpha\psi \,=\, 1-\frac{M_1}{2r_1}-\frac{M_2}{2r_2}, 
%%% =1-\frac{a_1}{r_1}-\frac{a_2}{r_2},
%%% \label{eq:APsi2bhem}
\label{eq:psi_alps_2bhsol}
\eeq
where subscripts 1 and 2 corresponds to those of the first and second BH; 
$r_1$ and $r_2$ are distances from the first and second BH, respectively, 
and $M_1$ and $M_2$ are mass parameters.  
The coordinates $r_1$ and $r_2$ are written in terms of each other, for example
in the first coordinate system of COCP, 
the radial coordinate are 
\[ 
r_1 \,=\, r 
\:\:\:\: \textrm{and} \:\:\:\:
r_2 \,=\, \sqrt{r^2 + d_s^2 - 2 r d_s\, \sin\theta\cos\phi},
\]
where $\GJ,\GP$ the angular spherical coordinates of the first coordinate system, 
and $1\leftrightarrow 2$ for the second COCP.  
On the third coordinate system of ARCP, 
\beqn
r_1 &=& \sqrt{r^2 + d_1^2 - 2 r d_1\, \sin\theta\cos\phi}  ,
\nonumber\\
r_2 &=& \sqrt{r^2 + d_2^2 - 2 r d_2\, \sin\theta\cos\phi}  ,
\nonumber
\eeqn
where $d_1$ and $d_2$ are the distance between the center of 
ARCP and one of two COCP, and hence $d_s = d_1+d_2$.

Instead of solving two Laplace equations Eq.~(\ref{eq:Laplace}), we write 
a equation for $\alpha$ with a source on the whole domain of $\Sigma_t$
\be
\nabla^{2}\GC=0 \:\:\:\:\:\: \textrm{and} \:\:\:\:\:\: 
\nabla^{2}\GA=-\frac{2}{\GC}f^{ij}\pd_i\GC \pd_j\GA .
\label{eq:Laplace2}
\ee 
In an actual computation, the BH centered at the COCP 
is excised at the radii $r_a$ of $S_a$, and the binary companion 
is excised at the radii $r_e$ of $S_e$ which is centered at $x=d_s$.  
Boundary conditions for these elliptic equations at $S_a$ are 
taken from analytic solutions (\ref{eq:psi_alps_2bhsol})
%%% (\ref{eq:Psi2bhem}) and (\ref{eq:APsi2bhem})
when Dirichlet boundary conditions are imposed.  
Neumann boundary conditions can be imposed with the use of 
\beqn
\frac{\pd\psi}{\pd r_1} = -\frac{M_1}{2r^2_1} - \frac{M_2}{2r^2_2}\frac{\pd r_2}{\pd r_1} 
\qquad\qquad\qquad \mbox{at}\ \ r_1=r_a, \ \ 
\label{eq:2bhbcon_psi}   \\
\frac{\pd\alpha}{\pd r_1} =\frac1{\psi}
\left(\frac{M_1}{2r^2_1} + \frac{M_2}{2r^2_2}\frac{\pd r_2}{\pd r_1}
 - \GA\frac{\pd\GC}{\pd r_1}\right) \ \ \mbox{at}\ \ r_1=r_a. \ \ 
\label{eq:2bhbcon_alp}
\eeqn
For the outer boundary conditions, we choose Dirichlet boundary conditions 
whose data is taken from the analytic solution Eq.~(\ref{eq:psi_alps_2bhsol}) 
in all tests in this section.  When the third parch, ARCP, is not used, 
Dirichlet data is imposed on $S_b$ $(r=r_b)$ of COCP, while ARCP is used 
as in Fig.~\ref{fig:multipatches}, Dirichlet data is imposed only at $S_b$ of ARCP.

\begin{figure*}
\begin{tabular}{cc}
\begin{minipage}{.49\hsize}
\begin{center}
\includegraphics[height=60mm,clip]{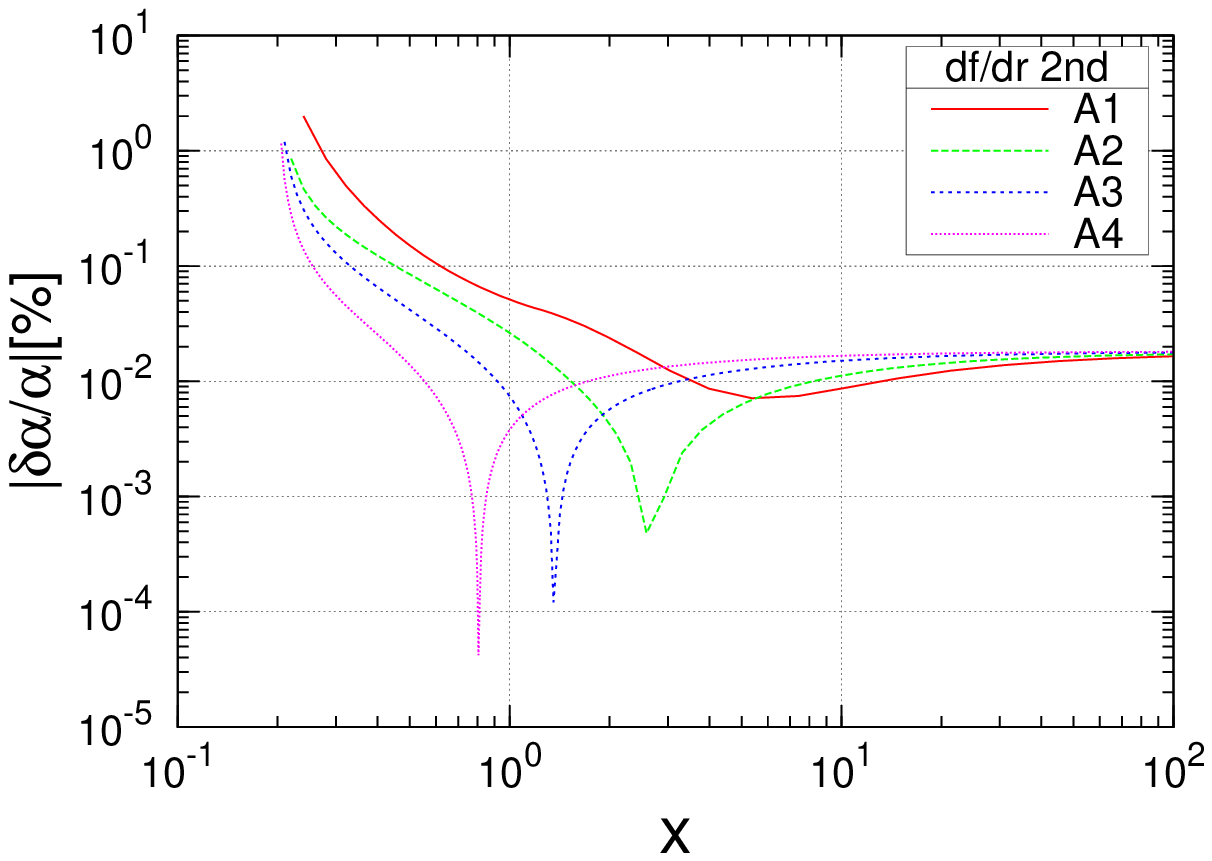}
\includegraphics[height=60mm,clip]{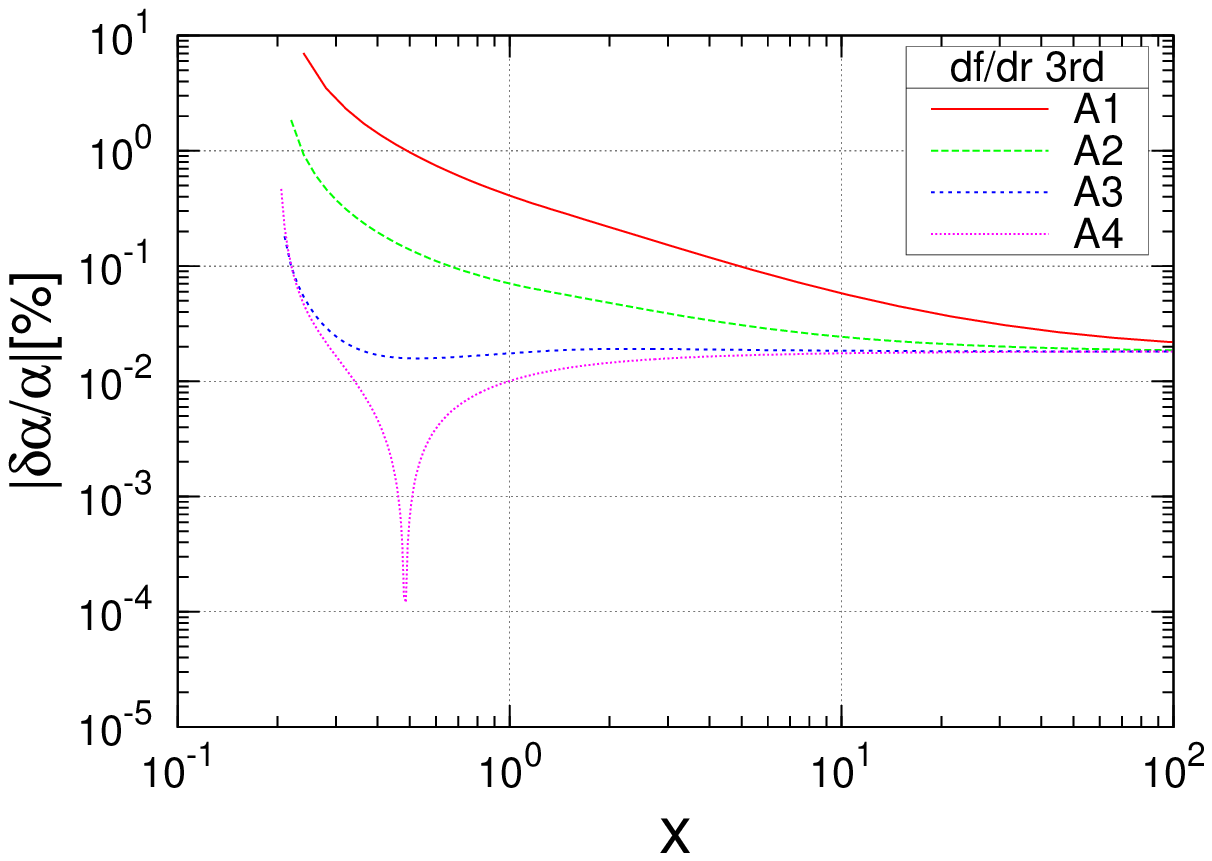}
\end{center}
\end{minipage} 
&
\begin{minipage}{.49\hsize}
\begin{center}
\includegraphics[height=60mm,clip]{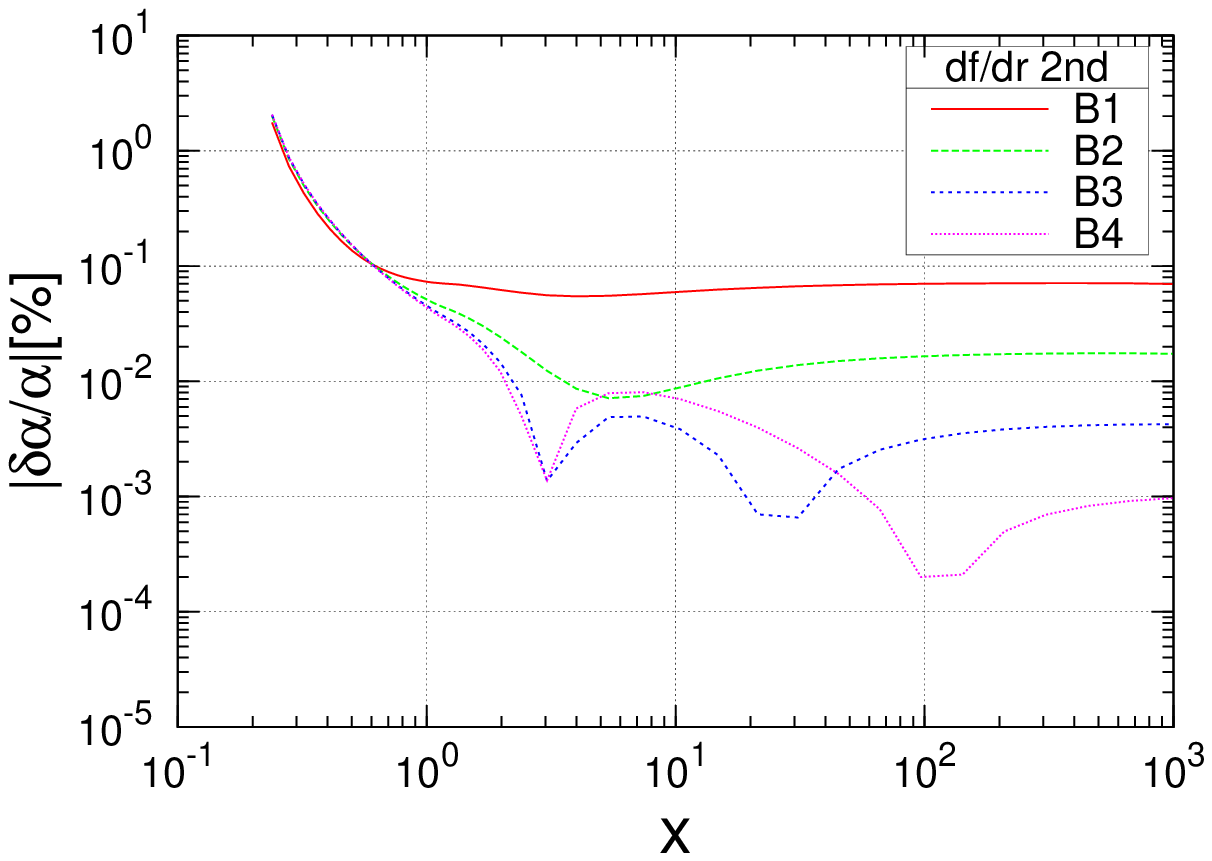}
\includegraphics[height=60mm,clip]{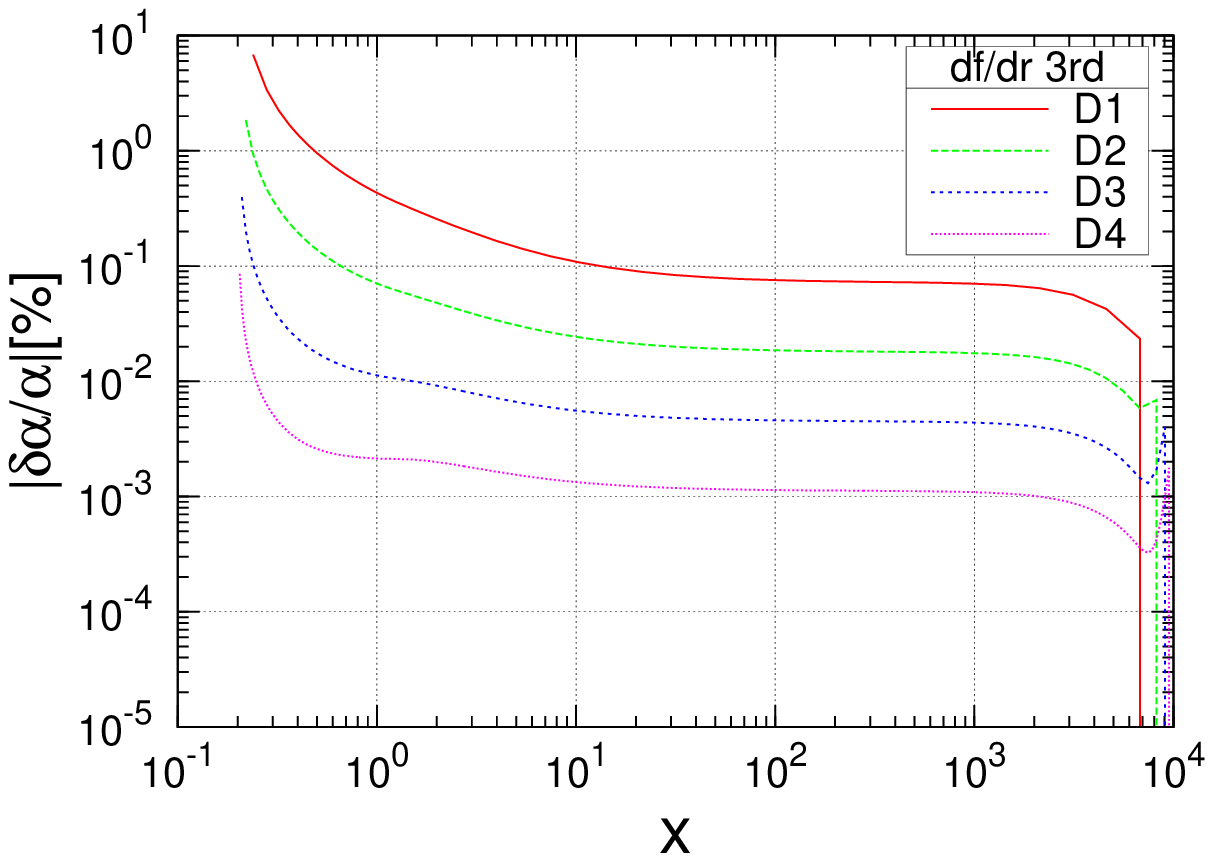}
\end{center}
\end{minipage} 
\end{tabular} 
\caption{Plots of fractional errors in the lapse $\dl \alpha/\alpha$ 
along the positive $x$-axis.  Plots show the errors with 
changing the number of radial grid points $r_i$ as A1-A4 (top left panel), 
zenith grid points $\theta_i$ as B1-B4 (top right panel), 
radial grid points $r_i$ as A1-A4 (bottom left panel), 
and all grid points $(r_i,\theta_j,\phi_k)$ as D1-D4 (bottom right panel).  
In each panel, solid (red), long dashed (green), dashed (blue), 
and dotted (magenta) lines are in order from lower to higher resolutions.  
In the top panels, second order 
finite difference formula is used for calculating radial derivatives,  
while the third order formula is used in the bottom panels.}
\label{fig:1BHtest}
\end{figure*}
\begin{figure}
\begin{center}
\includegraphics[height=60mm]{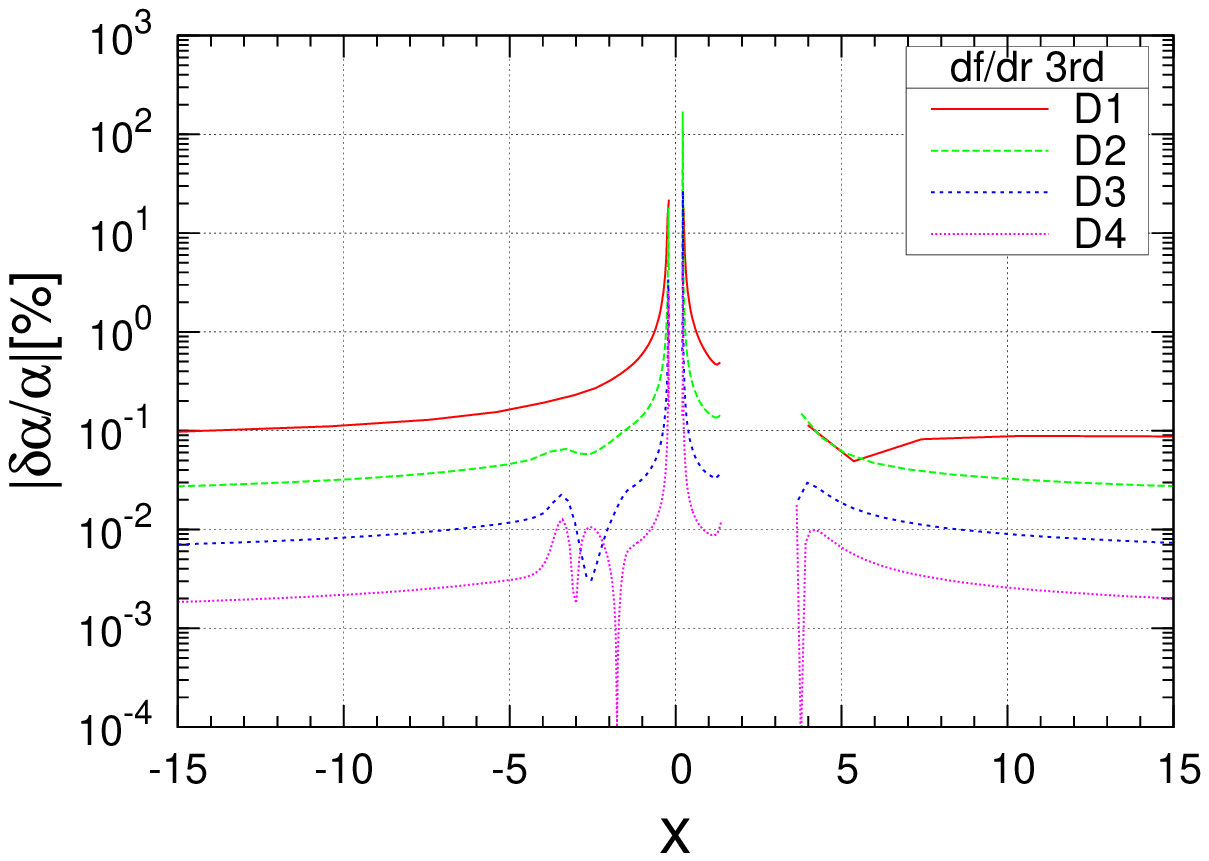}
\includegraphics[height=60mm]{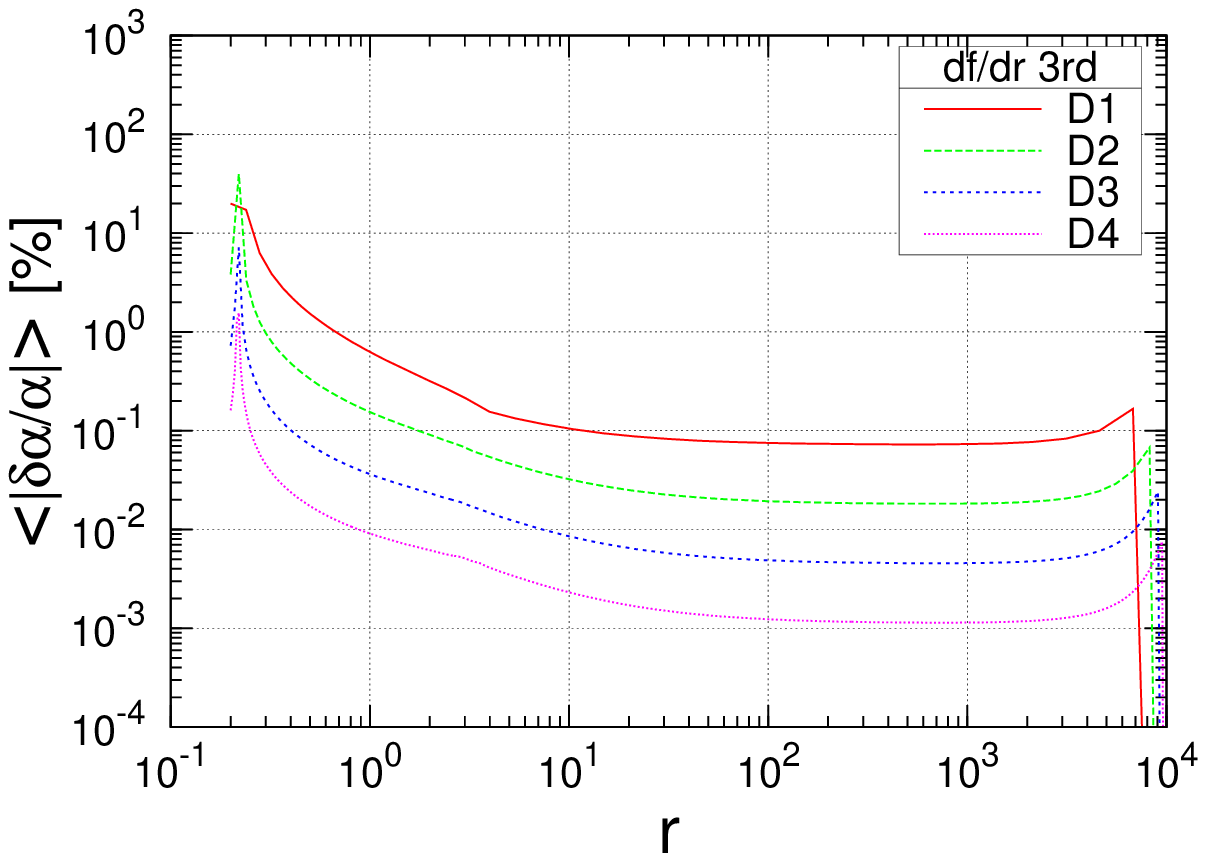}
\includegraphics[height=60mm]{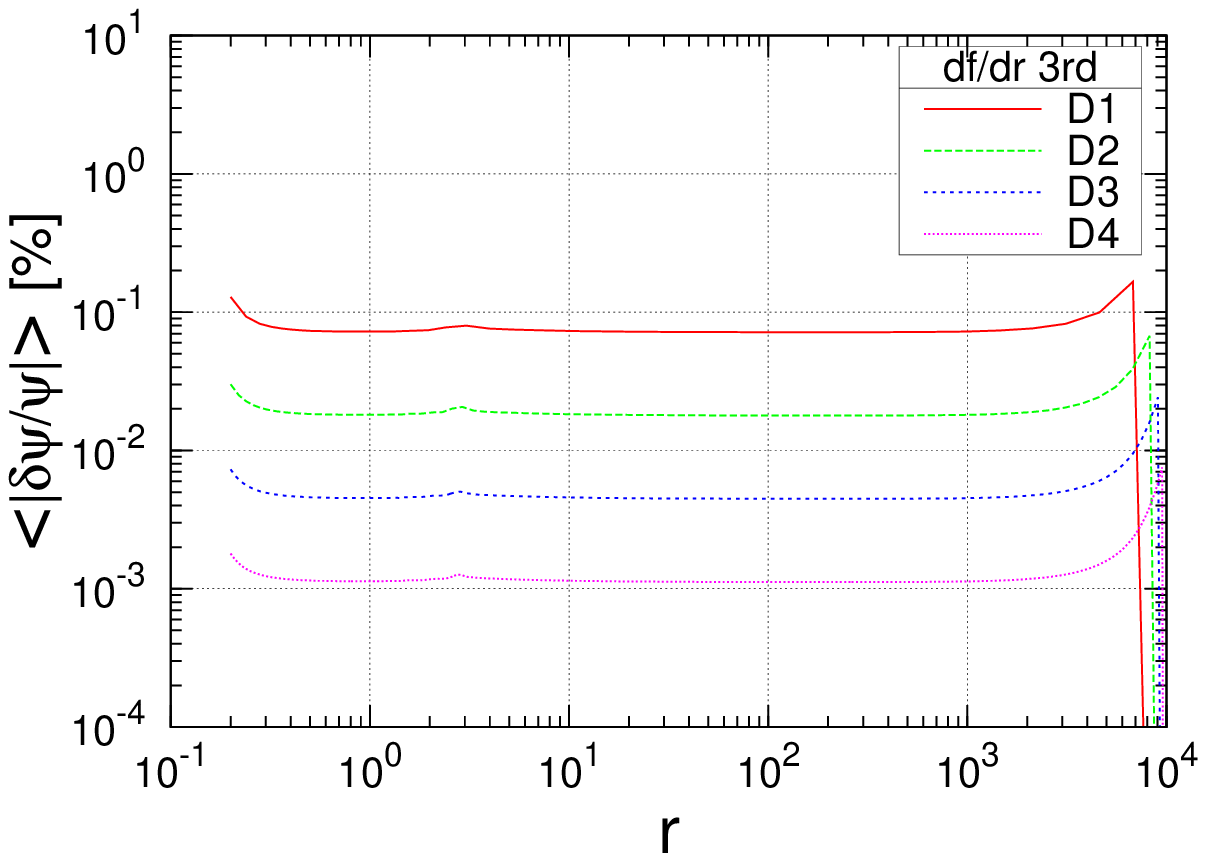}
\caption{Same as Fig.~\ref{fig:1BHtest} but for the equal mass BBH data 
calculated on a single patch Fig.~\ref{fig:singlepatch}.    
Top panel: Plots for the fractional errors in the lapse $\dl \alpha/\alpha$ 
for the equal mass BBH data along the positive $x$-axis.  
Middle panel: averaged fractional errors in the lapse 
$\left< \left| \dl \alpha/\alpha \right| \right>$.  
Bottom panel: averaged fractional errors in the conformal factor 
$\left< \left| \dl \psi/\psi \right| \right>$.  
Plots show the errors with changing the number of 
all grid points as D1-D4 in Table \ref{tab:1BHtest_grids}.}
\label{fig:eqmBBHtest}
\end{center}
\end{figure}
\begin{figure}
\begin{center}
\includegraphics[height=70mm]{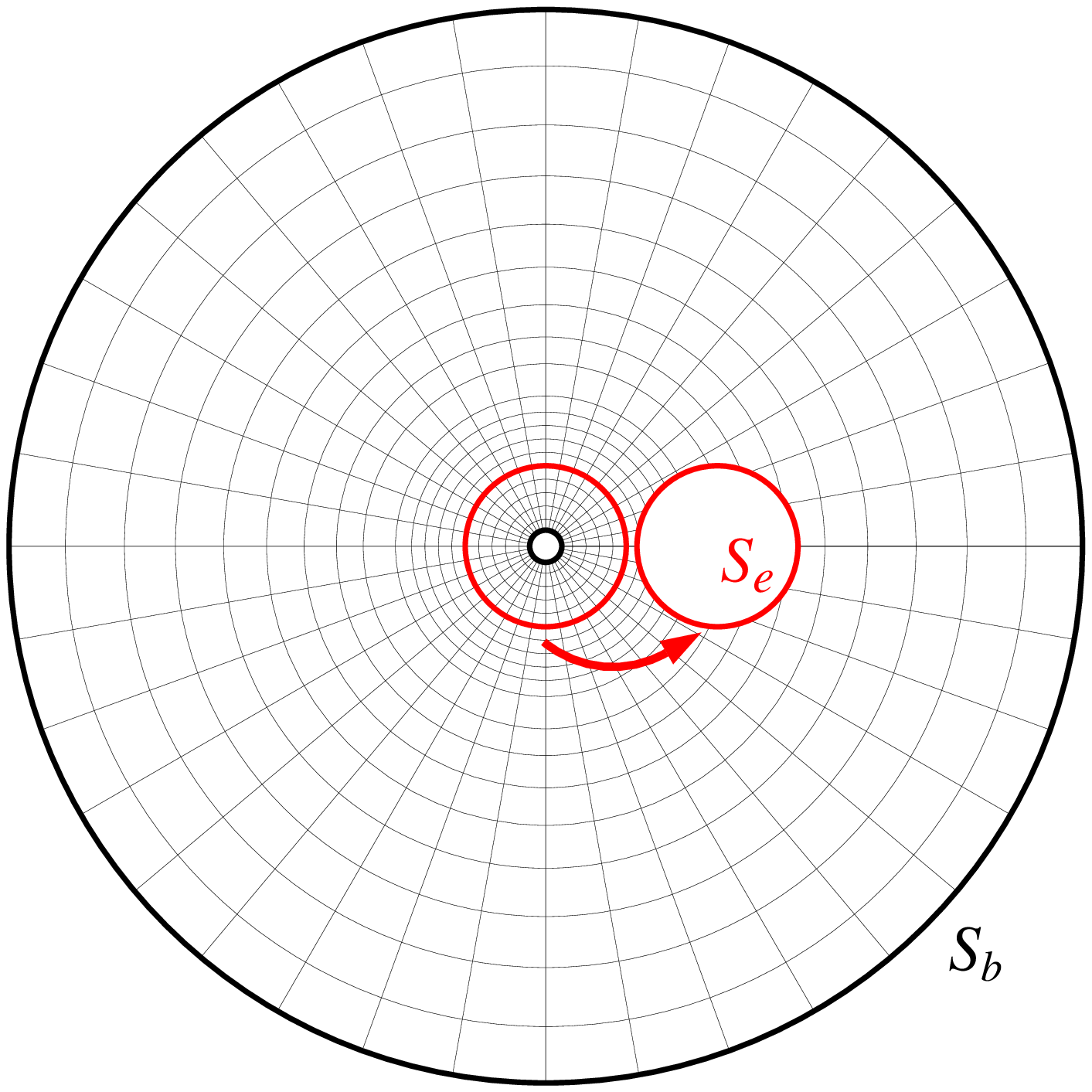}
\caption{A setup for a single coordinate grid patches for a calculation of equal mass BBH.  
The radius of coordinate patch doesn't reflect the actual size.}
\label{fig:singlepatch}  
\end{center}
\end{figure}

\subsection{Convergence tests}

Convergence tests are performed to examine that the code 
produces solutions with an expected order of finite difference errors, 
and to find experimentally an (almost) optimally balanced set 
of resolutions for each coordinate grid $(r_i,\theta_j,\phi_k)$ 
which is not over resolved in one coordinate direction so as not to 
waste the computational resources.  We find, from 
convergence tests for a single BH solution, it is 
necessary to use third order finite difference formula for 
a radial derivative in the volume source terms in 
Eq.~(\ref{eq:Green_int}).  We also find an optimally 
balanced resolutions between $r_i$ and $\theta_j$ grids.  
From convergence tests for BBH data, we find  
appropriate resolution for $\phi_i$ direction, 
and number of multipoles.  Results for the convergence 
tests are discussed in this section.

\subsubsection{Single BH}
\label{sec:singleBH}

For the first test, we compute a single BH solution with 
mass parameter $M_1 = 2r_a = 0.4$ (and $M_2 = 0$).  
Eqs.(\ref{eq:Laplace2}) are solved on a single patch with 
a single excision region interior of $S_a$ for the BH.  
In Fig.~\ref{fig:1BHtest}, 
a fractional error of the lapse $\alpha$ 
\beq
\left| \frac{\dl \alpha}{\alpha} \right|
:= \left|\frac{\alpha - \alpha_{\rm exact}}{ \alpha_{\rm exact}}\right|
\label{eq:frac_error}
\eeq
is plotted along the x-axis for different resolutions in 
the radial coordinate grids $r_i$ (top and bottom left panels), in 
the zenith angle grids $\theta_j$ (top right panel), 
and in all grids (bottom right panel). 
These resolutions are tabulated in Table \ref{tab:1BHtest_grids}, 
and are indicated by A1-A4, B1-B4, and D1-D4, respectively.  
In the set A1-A4, the radial resolution $\Dl r$ is doubled, 
in B1-B4, the zenith angle resolution $\Dl \theta$
is doubled, and in D1-D4, the resolutions in all directions 
are doubled at each level.  
Another difference in these results is 
the order of the finite difference 
formula used to compute a radial derivative in the volume 
source term in Eq.~(\ref{eq:Green_int}), where 
the second order (mid-point) formula is used in the top left and right panels, 
and third order (Lagrange) formula is used in the bottom left and right panels. 
It is noticed from the top left panel in Fig.\ref{fig:1BHtest} 
that the error does not decrease as $\Od(\Dl r^2)$ when 
the number of radial grid points is increased as the parameter 
sets A1-A4 in Table \ref{tab:1BHtest_grids} even for such a 
spherically symmetric solution.  

It appears that there are two reasons for that.  
In the top right panel, a convergence test is performed changing the 
number of grid points in zenith angle $\theta_j$ as the parameter sets B1-B4.  
It shows an improvement of the accuracy in $\Od(\Dl \theta^2)$ in the 
larger radius $r \agt 100$, that is the error in this region 
is dominated by the finite differencing in $\theta$ direction.  
However, the accuracy near the BH is not improved in both tests.  
In the bottom left panel of Fig.~\ref{fig:1BHtest}, the 
same convergence test as the top left panel is performed, but the 
finite difference formula for the radial derivatives is replaced 
by that of third order Lagrange formula $\Od(\Dl r^3)$.  It shows that 
it is necessary to set the order of finite difference formula for 
the radial derivative as $\Od(\Dl r^3)$ to see $\Od(\Dl r^2)$ 
accuracy near the BH.  This $\Od(\Dl r^2)$ error must be due to 
the mid-point rule used in the numerical integration.  
While, the error in the larger radius does not decrease in the 
outer region with the radius $r \agt 100$.  
Finally, as shown in the bottom right panel of Fig.\ref{fig:1BHtest} 
the error decreases in the second order for the set D1-D4, 
in which the third order finite difference formula is used for 
the radial derivatives.  
Convergence test are done also for 
increasing grid points in $\phi$ directions $\phi_k$ 
and also for changing the number of Legendre expansion as $L = 4 - 10$, 
but they didn't change the results for such spherically symmetric BH test.

\subsubsection{Equal mass BBH computed with a single patch}
\label{sec:eqmBBH}

In Fig.\ref{fig:eqmBBHtest}, results of convergence tests for 
equal mass BBH data are plotted.  In this test, we used only 
a single patch shown in Fig.\ref{fig:singlepatch}, where 
the potential at the radius $r = r_e$ rotated by $\pi$ in $\phi$ 
coordinate is mapped to the excision sphere $S_e$ on the same patch 
to compute the equal mass data when the elliptic equations are solved.  
This amounts to impose the $\pi$-rotation symmetry about the center 
of mass which is located at $(r,\theta,\phi) = (d_s/2,\pi/2,0)$.  
In this test, the number of 
grid points is chosen as D1-D4 in Table \ref{tab:1BHtest_grids}, 
separation between the coordinate centers of two BH 
(a distance between the centers of $S_a$ to $S_e$) is set 
as $d_s = 2.5$, the excision radius of the BH 
$r_a = 0.2$, the excision radius of the binary companion 
$r_e = 1.125$, and the mass parameters $M_1 = M_2 = 2 r_a$.  
Hereafter, the third order finite difference formula is always 
used for computing the radial derivatives as discussed in 
Sec.~\ref{sec:singleBH}.  

In the top panel of Fig.~\ref{fig:eqmBBHtest}, fractional errors of the lapse 
$| \dl\alpha/\alpha |$ defined in Eq.~(\ref{eq:frac_error})
are plotted along the x-axis near the BH.  Because of 
the excision of the interior of the sphere $S_e$ and 
of the use of Legendre expansion in the elliptic solver, 
a certain modulation is seen in the errors.
Therefore, hereafter we show fractional errors 
averaged over the number of $(\theta_j,\phi_k)$ grids points 
at a radius $r = r_i$ 
defined by 
\beq
\left< \left| \frac{\dl \alpha}{\alpha} \right| \right>
:= \frac1{\#({\cal G}_i)} \sum_{\theta_j,\phi_k \in {\cal G}_i}
\left|\frac{\alpha - \alpha_{\rm exact}}{ \alpha_{\rm exact}}\right|
\label{eq:frac_error_ave}
\eeq
where writing a grid point $(r_i,\theta_j,\phi_k)$ by $p$, 
we define a set ${\cal G}_i$ by 
${\cal G}_i := \left\{(r_i,\theta_j,\phi_k) \ | \  
p\in V\setminus S^{\rm in}_e \ \mbox{and}\  r_i = \mbox{const} \right\}$ 
where  $S^{\rm in}_e$ is an interior domain of $S_e$.  
Then, $\#({\cal G}_i)$ is the number of points included in ${\cal G}_i$.  

The averaged fractional errors for the lapse are plotted along 
the radial coordinate $r$ in the middle panel of Fig.~\ref{fig:1BHtest}.  
As expected, second order convergence is observed when the grid points 
are increased as D1-D4.  In the figure, it is seen that 
a couple of grid points at the vicinity of BH boundary $S_a$ have 
(averaged) errors as large as $\sim 1\%$ even for the 
highest resolution D4.  
This is due to our choice of the boundary $r_a = M/2$ as same as 
the single BH test in the previous section.  With this choice, 
the value of $\alpha$ becomes negative at the BH excision radius $r=r_a$.  
Hence, the fractional error diverges at radii r (depending on 
$\theta$, and $\phi$) where $\alpha$ crosses zero, 
even though the grid points are slightly off from the zeros.  
In this way, the worst possible error in computation for the 
metric potentials of BH is estimated.  Even near the radius for $\alpha=0$, 
the second order convergence is maintained as seen in the figure.  
We also show in the bottom panel of Fig.~\ref{fig:1BHtest}, 
averaged fractional errors for the conformal factor $\psi$.  
The value of $\psi$ is about 2 near $r=r_a$, and for such 
a potential the convergence of the solution is almost uniform 
in all radii as observed.

\begin{table}
\begin{tabular}{clccccrrrrrrr}
\hline
Type & Patch & $r_a$ & $r_b$ & $r_c$ & $r_e$ & $\Nrf$ & $\Nrm$ & $N_r$ & $N_\theta$ & $N_\phi$ & $L$ \\
\hline
E1 & COCP-1 & 0.2 & $10^4$ & 1.25 & 1.125 & 16  & 20    & 64    & 24  & 24 &  12 \\
   & COCP-2 & 0.4 & $10^4$ & 1.25 & 1.125 & 16  & 20    & 64    & 24  & 24 &  12 \\
E2 & COCP-1 & 0.2 & $10^4$ & 1.25 & 1.125 & 32  & 40    & 128   & 48  & 48 &  12 \\
   & COCP-2 & 0.4 & $10^4$ & 1.25 & 1.125 & 32  & 40    & 128   & 48  & 48 &  12 \\
E3 & COCP-1 & 0.2 & $10^4$ & 1.25 & 1.125 & 64  & 80    & 256   & 96  & 96 &  12 \\
   & COCP-2 & 0.4 & $10^4$ & 1.25 & 1.125 & 64  & 80    & 256   & 96  & 96 &  12 \\
E4 & COCP-1 & 0.2 & $10^4$ & 1.25 & 1.125 & 128  & 160  & 512   & 192  & 192 &  12 \\
   & COCP-2 & 0.4 & $10^4$ & 1.25 & 1.125 & 128  & 160  & 512   & 192  & 192 &  12 \\
\hline
F1 & COCP-1 & 0.2 & $10^2$ & 1.25 & 1.125 & 16  & 20   & 48   & 24  & 24 &  12 \\
   & COCP-2 & 0.4 & $10^2$ & 1.25 & 1.125 & 16  & 20   & 48   & 24  & 24 &  12 \\
   & ARCP   & 5.0 & $10^6$ & 6.25 &  ---  &  4  &  5   & 48   & 24  & 24 &  12 \\
F2 & COCP-1 & 0.2 & $10^2$ & 1.25 & 1.125 & 32  & 40   & 96   & 48  & 48 &  12 \\
   & COCP-2 & 0.4 & $10^2$ & 1.25 & 1.125 & 32  & 40   & 96   & 48  & 48 &  12 \\
   & ARCP   & 5.0 & $10^6$ & 6.25 &  ---  &  8  & 10   & 96   & 48  & 48 &  12 \\
F3 & COCP-1 & 0.2 & $10^2$ & 1.25 & 1.125 & 64  & 80   & 192  & 96  & 96 &  12 \\
   & COCP-2 & 0.4 & $10^2$ & 1.25 & 1.125 & 64  & 80   & 192  & 96  & 96 &  12 \\
   & ARCP   & 5.0 & $10^6$ & 6.25 &  ---  & 16  & 20   & 192  & 96  & 96 &  12 \\
F4 & COCP-1 & 0.2 & $10^2$ & 1.25 & 1.125 & 128 & 160  & 384  & 192  & 192 &  12 \\
   & COCP-2 & 0.4 & $10^2$ & 1.25 & 1.125 & 128 & 160  & 384  & 192  & 192 &  12 \\
   & ARCP   & 5.0 & $10^6$ & 6.25 &  ---  & 32  & 40   & 384  & 192  & 192 &  12 \\
\hline
\end{tabular}
\caption{Grid parameters used in convergence tests for 
non-equal mass BBH data solved on multiple coordinate patches.  
The separation of two BHs is set as $d_s=2.5$.}
\label{tab:BBHtest_grids}
\end{table}
\begin{figure}
\begin{center}
\includegraphics[height=60mm]{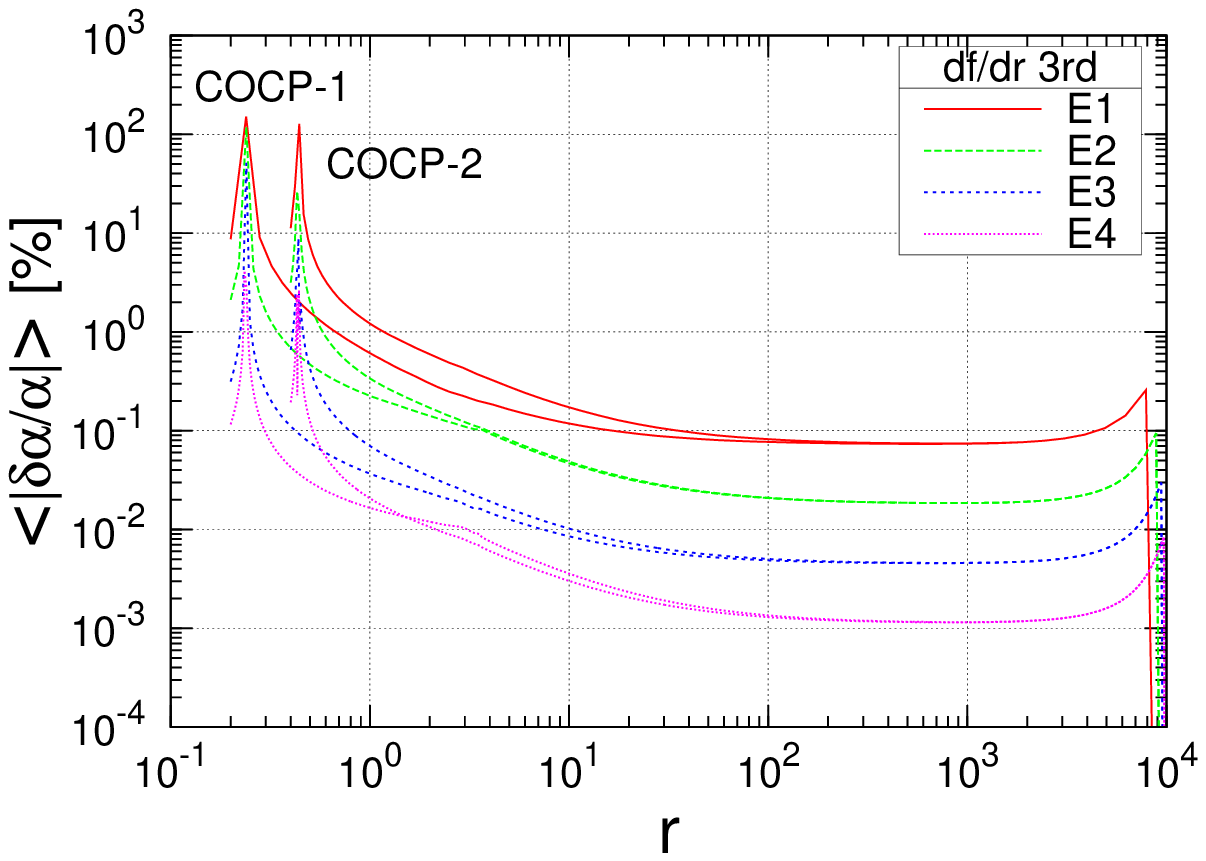}
\includegraphics[height=60mm]{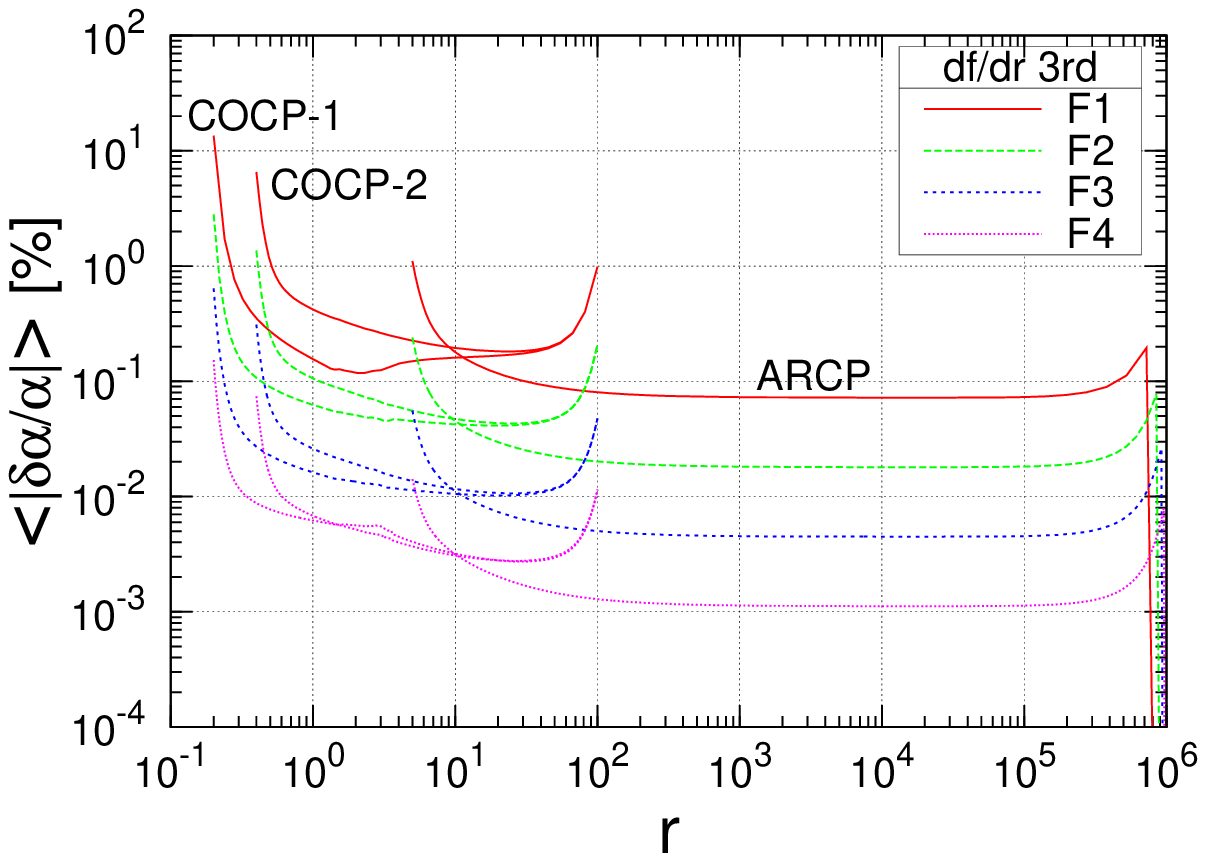}
\caption{Same as Fig.~\ref{fig:1BHtest} but 
averaged fractional errors in the lapse 
$\left< \left| \dl \alpha/\alpha \right| \right>$ are plotted 
along the radial coordinate $r$ 
for the non-equal mass BBH data 
calculated on multiple patches Fig.~\ref{fig:multipatches}.    
Top panel: data computed on two multiple patches 
with changing the number of 
grid points as E1-E4 in Table \ref{tab:1BHtest_grids}. 
Bottom panel: data computed on three multiple patches 
with changing the number of 
all grid points as F1-F4 in Table \ref{tab:1BHtest_grids}.}
\label{fig:noneqmBBHtest}
\end{center}
\end{figure}
\begin{figure}
\begin{center}
\includegraphics[height=60mm]{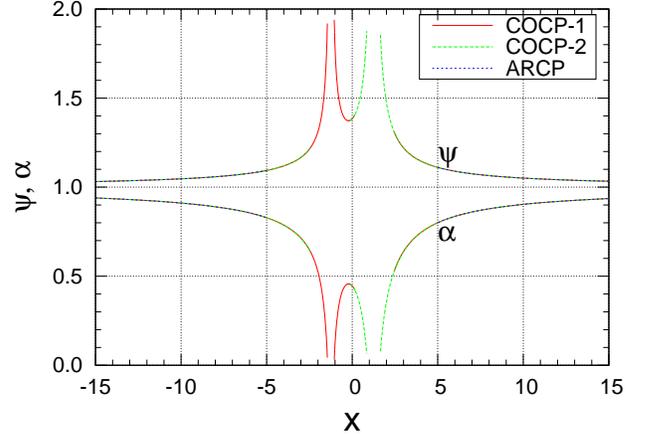}
\caption{Plots for the conformal factor $\psi$ and the lapse $\alpha$, 
computed on the three multi patches.  The model is the same as 
that of the bottom panel of Fig.~\ref{fig:noneqmBBHtest}.}
\label{fig:noneqmBBHsol}
\end{center}
\end{figure}

\subsubsection{Non equal mass BBH computed with multiple patches}

Convergence test for non-equal mass BBH data using multi coordinate patches 
discussed in Sec.\ref{sec:nm} is performed with grid parameters 
presented in Table \ref{tab:BBHtest_grids}.  In the first example, 
BBH data is computed on two COCP whose boundary radius $r=r_b$ is 
taken large enough to reach to asymptotics $r_b=10^4$.  
In this computation, the number of grid points 
is chosen as E1-E4 in Table \ref{tab:BBHtest_grids}, 
in which the resolution in radial grids $r_i$ in the region $r>r_c$ 
is increased by 44/24 times of the corresponding level of resolution 
for equal mass BBH case, D1-D4.  
Separation between the coordinate centers of two BH 
is set as $d_s = 2.5$, the excision radius and mass parameter 
of the first BH are $r_a = 0.2$ with $M_1 = 2 r_a = 0.4$, 
and those of second BH $r_a = 0.4$ with $M_2 = 2 r_a = 0.8$.  
The results of the averaged fractional error in 
the top panel of Fig.~\ref{fig:noneqmBBHtest} is similar to those 
of the equal mass BBH case in Fig.~\ref{fig:eqmBBHtest}, which 
proves our multiple patch methods works accurately as expected.  

Finally, the BBH data is computed using three multiple patches 
shown in Fig.~\ref{fig:multipatches}.  The number of grid points 
is chosen as F1-F4 in Table \ref{tab:BBHtest_grids},
and the separation is set as $d_s = 2.5$.  
In the bottom panel of Fig.~\ref{fig:noneqmBBHtest}, the 
results for the averaged fractional errors 
$\left<\left|\dl\alpha/\alpha\right|\right>$ are shown.  
In this computation we decreased the values of mass parameter 
as $M_1 = 0.8 \times 2 r_a = 0.32$ with $r_a = 0.2$ for the 
first BH and 
$M_2 = 0.8 \times 2 r_a = 0.64$ with $r_a = 0.4$ for the second BH, 
so that the lapse $\alpha$ is always positive even near the BH excision 
boundary at $r=r_a$.  
As seen in the figure, the error near the BH boundary is decreased 
about 1/10 of the previous case, although the resolutions near 
the BH are the same.  In the Fig.~\ref{fig:noneqmBBHsol}, 
we present the plots of the potentials $\alpha$ and $\psi$ 
for the same model to show a smooth transition of potentials 
from one to the other patch.

\begin{figure*}
  \includegraphics{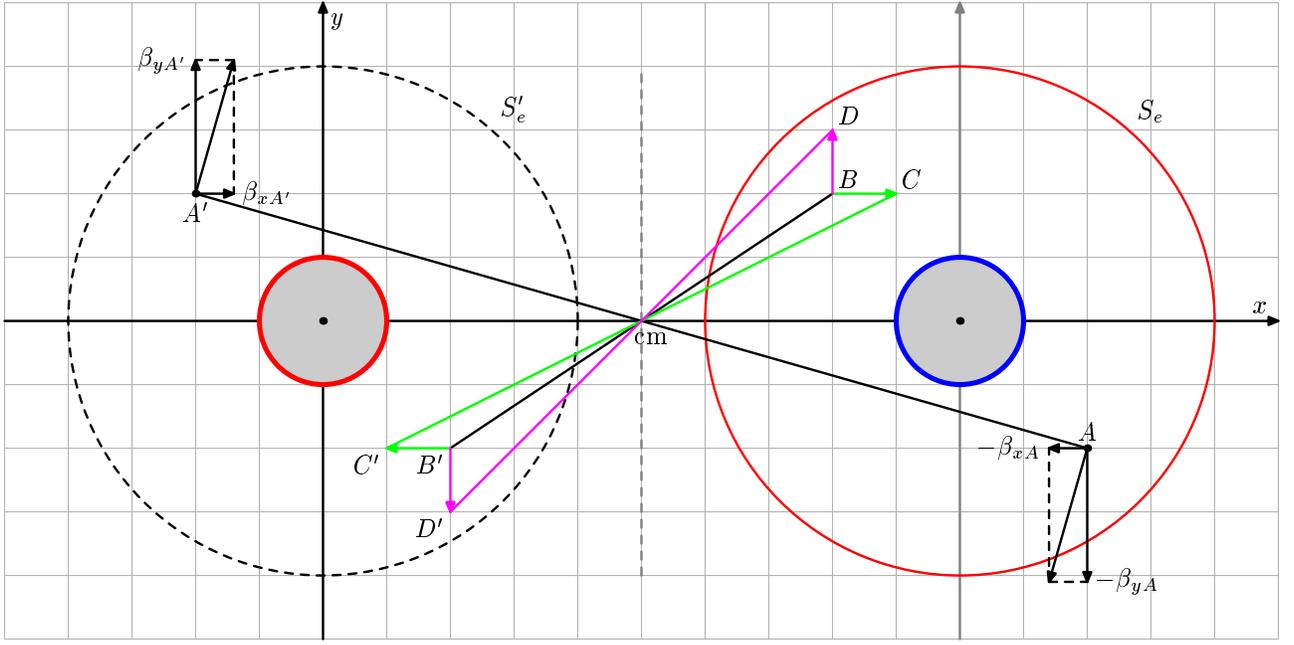}
  \caption{Schematic figure for the $\pi$ rotation symmetry of $x$ and $y$ 
  components of the shift from a region inside to $S'_e$ to the excised 
  region inside of the sphere $S_e$.}
  \label{fig:bvxy_symmetry}
\end{figure*}
\begin{table}
\begin{tabular}{lcllllc}
\hline
Type & $n_1$ & $n_0$ & $\Omega$ & $\Omega_{\rm s}$ & Spin axis & Figures  \\
\hline
TU & 3.0 & 1.0  & 0.3  & 0.0 & --- & Fig.~\ref{fig:bvyd_x_TU2007}              \\
TU & 2.2 & 0.005& 0.08 & 0.0 & --- & Fig.~\ref{fig:Psi_Alph_x}                 \\
AH & --- & 0.1  & 0.08 & 0.0 & --- & Figs.~\ref{fig:Bvxy}-\ref{fig:gspin_Bvxzd} \\
AH & --- & 0.005& 0.08 & 0.0 & --- & Figs.~\ref{fig:Bvxy}-\ref{fig:AH_Bvzd_z}   \\
AH & --- & 0.1  & 0.08 & 0.1 & $z$-axis & Fig.~\ref{fig:spin_Bvxyd}             \\
AH & --- & 0.1  & 0.08 & 0.1 & $y$-axis & Fig.~\ref{fig:gspin_Bvxzd}            \\
\hline
\end{tabular}
\caption{Boundary conditions and their parameters used in the computations 
for BBH initial data.  The first column, Type, denotes the types of 
boundary conditions used, TU corresponds to Eq.~(\ref{eq:Initial_bc}), and 
AH corresponds to apparent horizon boundary conditions 
Eqs.~(\ref{eq:AHpsiBC})-(\ref{eq:AHalphBC}).
The model of Fig.\ref{fig:bvyd_x_TU2007} 
is computed with a binary separation $d_s=2.8$, a radius of BH excision $r_a=0.1$, 
and a radius of binary excision $r_e=1.3$.  All the other models are 
computed with the parameter set D3 in Table \ref{tab:1BHtest_grids}.}
\label{tab:BBHinitial}
\end{table}
\begin{figure}
  \includegraphics[height=60mm,clip]{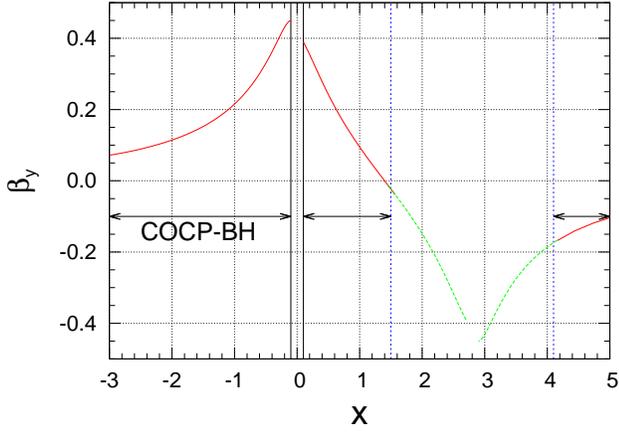}
  \caption{Plots for the $y$-component of shift vector along $x$-axis 
   of BBH initial data computed from the boundary conditions 
   (\ref{eq:Initial_bc}).  Parameters in the conditions are chosen 
   the same as a solution presented in \cite{TU2007} as 
   $n_1=3$, $n_0 = 0.1$, and $\Omega = 0.3$.  
  Two BH are located at $x=0$ and $x=2.8$.  The region inside of the 
  excised sphere $S_e$ is interpolated using $\pi$-rotation symmetry.  
  Thin solid (black) vertical lines are the boundaries at $S_a$ 
  with the radius $r_a = 0.1$, and 
  dotted (blue) vertical lines are the boundaries at $S_e$ 
  with the radius $r_e = 1.3$.}  
  \label{fig:bvyd_x_TU2007}
\end{figure}
\begin{figure}
  \includegraphics[height=60mm,clip]{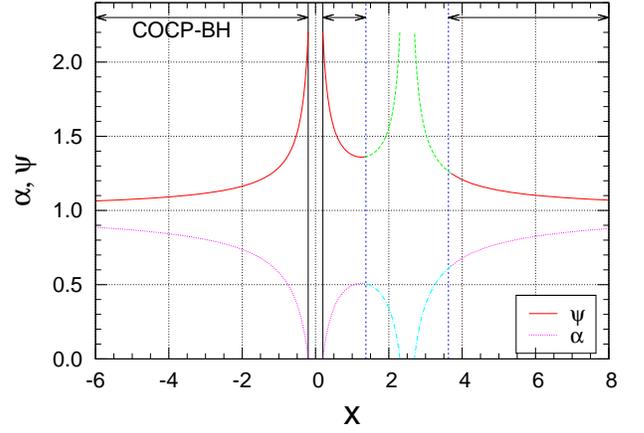}
  \caption{Plots for the conformal factor $\psi$ and lapse function $\alpha$ 
  along the $x$-axis of BBH initial data computed for the boundary condition 
  $(\ref{eq:Initial_bc})$.  Parameters in the conditions are chosen as 
  $n_1 = 0.1$, $n_0=0.005$ and $\Omega=0.08$.  
  The two BH are located at $x=0$ and $x=2.5$.
  The region inside of the 
  excised sphere $S_e$ is interpolated using $\pi$-rotation symmetry.  
  Thin solid (black) vertical lines are the boundaries at $S_a$ 
  with the radius $r_a = 0.2$, and 
  dotted (blue) vertical lines are the boundaries at $S_e$ 
  with the radius $r_e = 1.125$.}
  \label{fig:Psi_Alph_x}
\end{figure}
\begin{figure}
  \includegraphics[height=60mm,clip]{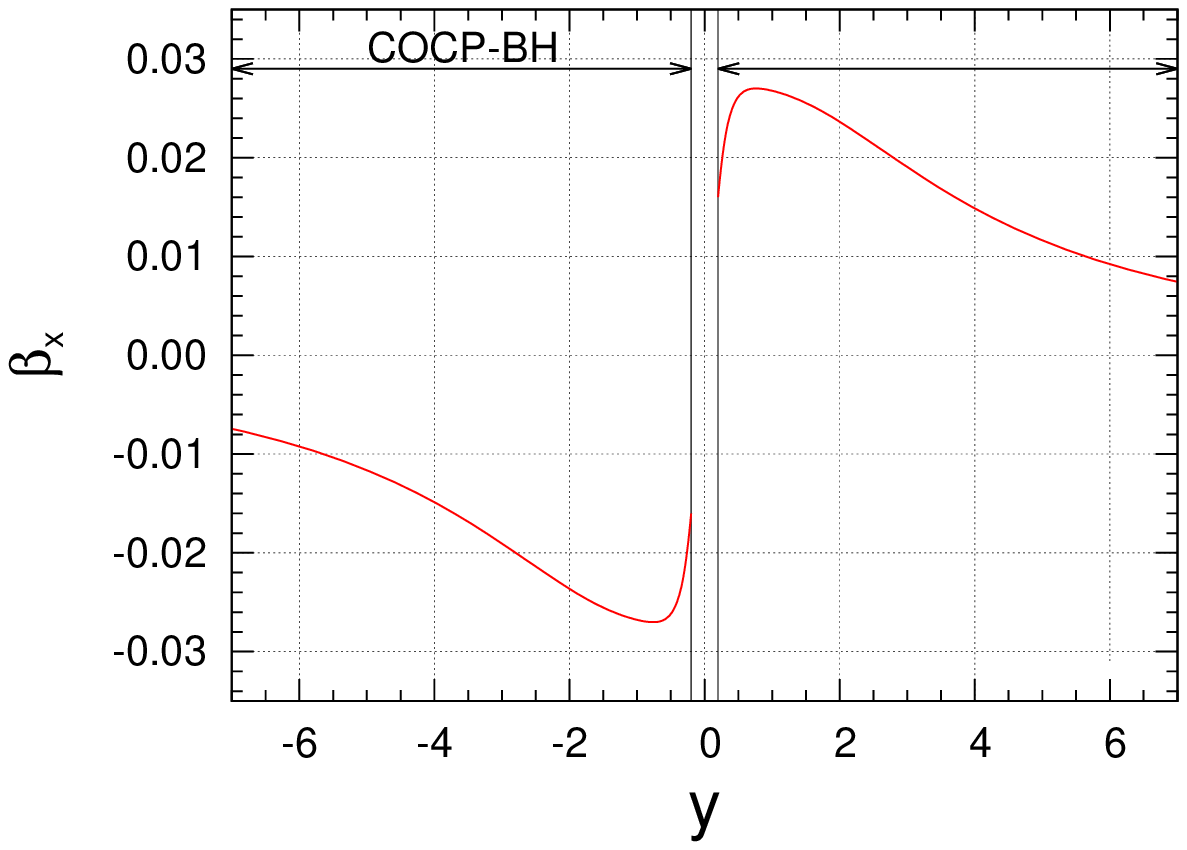}
  \includegraphics[height=60mm,clip]{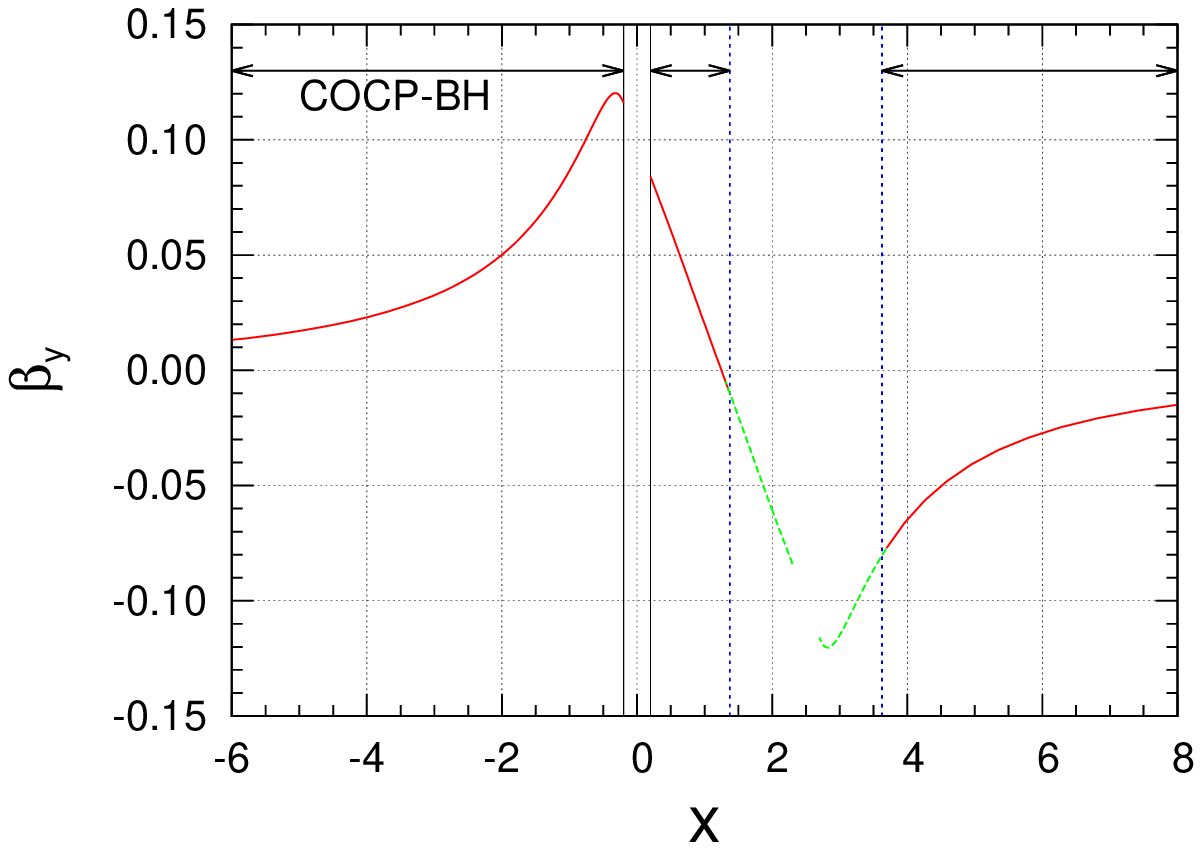}
  \includegraphics[height=60mm,clip]{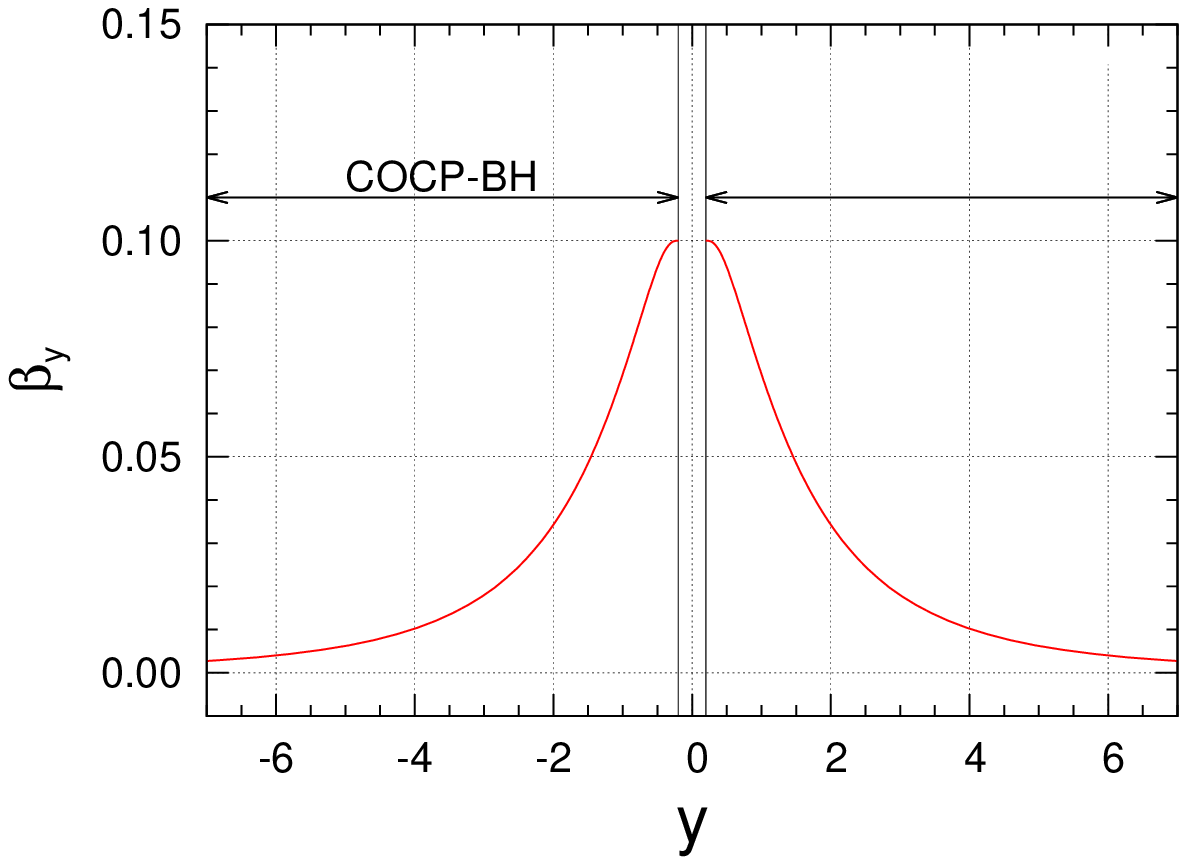}
  \caption{Plots for the same model as Fig.~\ref{fig:Psi_Alph_x} 
  but for the components of the shift $\beta_i$.  
  Top panel: $\GB_x$ along the $y$-axis.  
  Middle panel: $\GB_y$ across the $x$-axis. 
  Bottom panel: $\GB_y$ across the $y$-axis.}
  \label{fig:TU_bvd}
\end{figure}
\begin{figure}
  \includegraphics[height=80mm,clip]{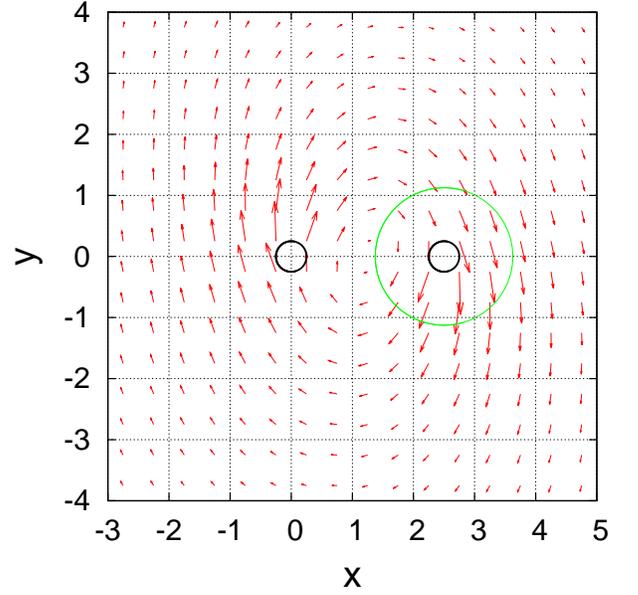}
  \caption{Shift vector on the $xy$-plane of the BBH initial data 
  for the case with AH boundary conditions 
  (\ref{eq:AHpsiBC})-(\ref{eq:AHalphBC}).  Parameters in the 
  conditions are chosen as $n_0=0.1$, $\Omega = 0.08$, 
  and $\Omega_{\rm s}=0$.  
  The center of mass is located at $x=1.25$ on the $x$-axis.  
  In the computation, the region inside of the left thick circle centered 
  at the origin with a radius $r_a=0.2$, and of the thin circle (green) with 
  a radius $r_e=1.125$ centered at $x=2.5$ on the $x$-axis are excised.  The 
  data inside of the thin circle is interpolated by a symmetry.  
  Note that the center of mass does not coincide with the center of 
  the x-axis.}
  \label{fig:Bvxy}
\end{figure}
\begin{figure}
  \includegraphics[height=80mm,clip]{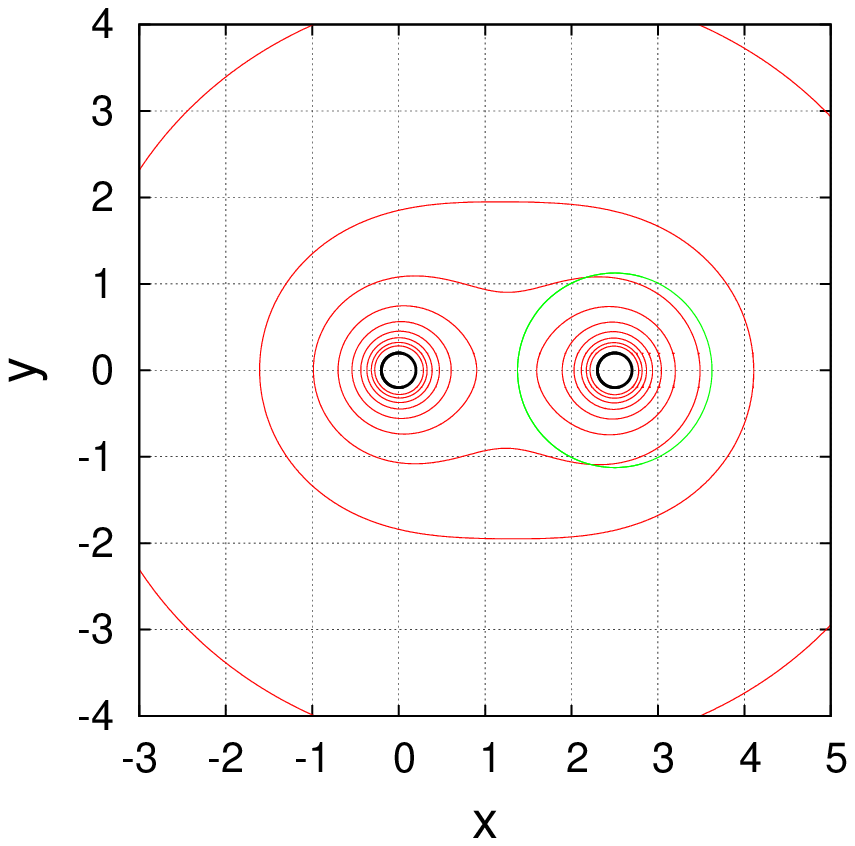}
  \includegraphics[height=80mm,clip]{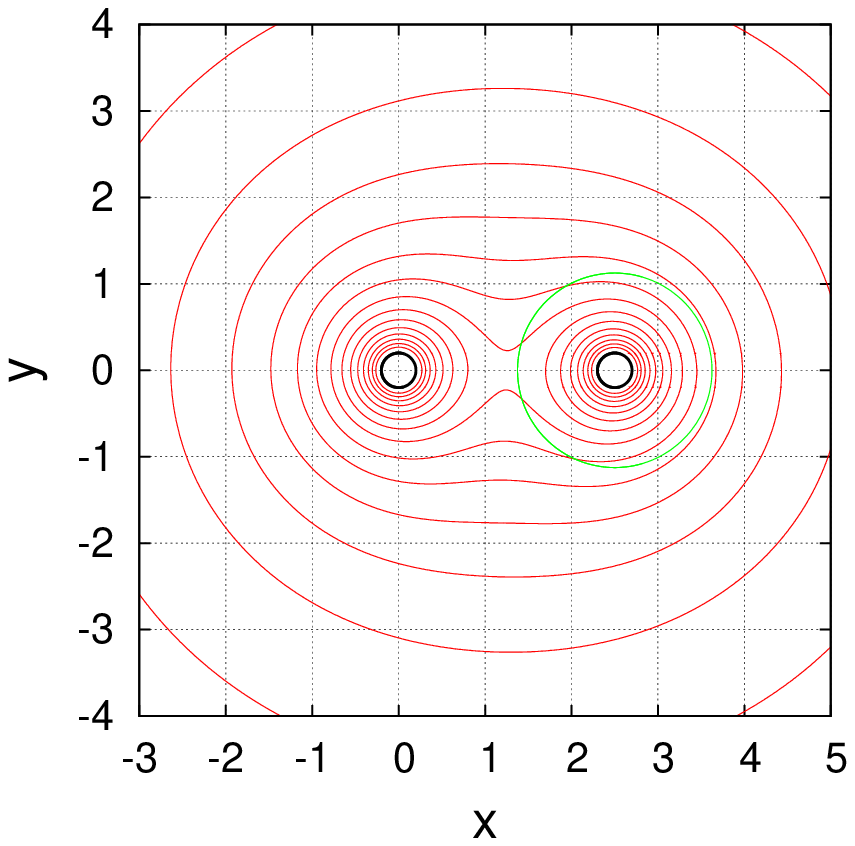}
  \caption{Contour plots on the $xy$-plane for the conformal factor $\psi$ (top panel)
  and the lapse function $\alpha$ (bottom panel) for the same model as Fig.~\ref{fig:Bvxy}.  
  For $\psi$, we draw isolines from $\GC=1.1$ to $\GC=2.0$ with step $0.1$. For $\alpha$, 
  we draw isolines from $\GA=0.2$ to $\GA=0.9$ with step $0.05$.}
  \label{fig:PsiAlph_contour}
\end{figure}
\begin{figure}[ht]
  \includegraphics[height=60mm,clip]{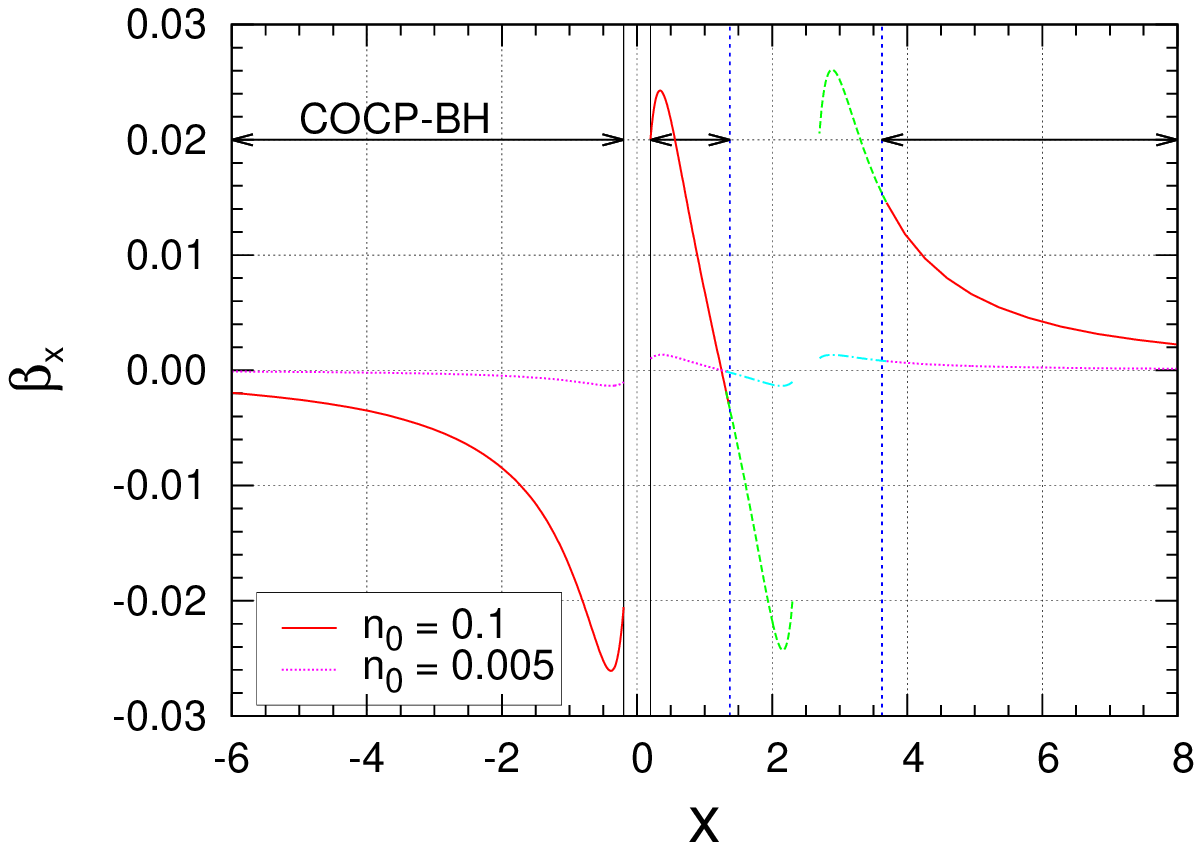}
  \includegraphics[height=60mm,clip]{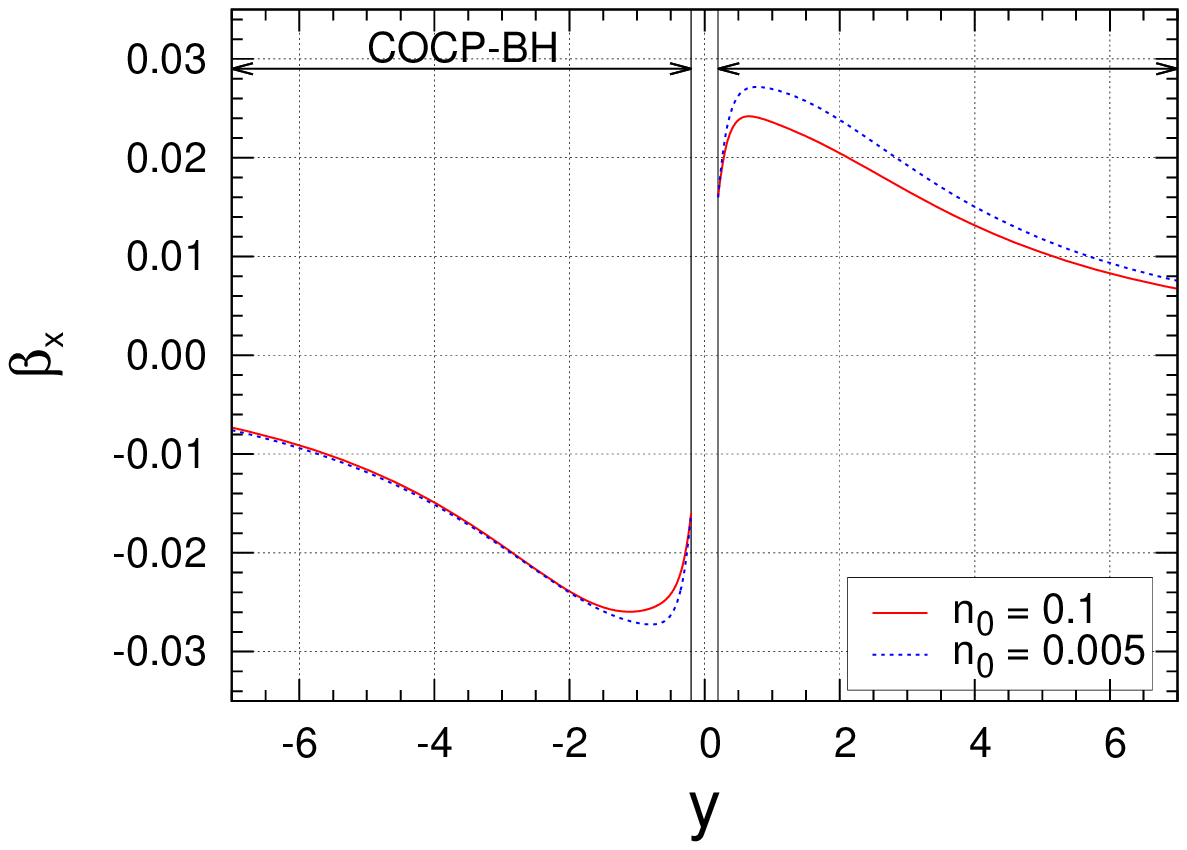}
  \includegraphics[height=60mm,clip]{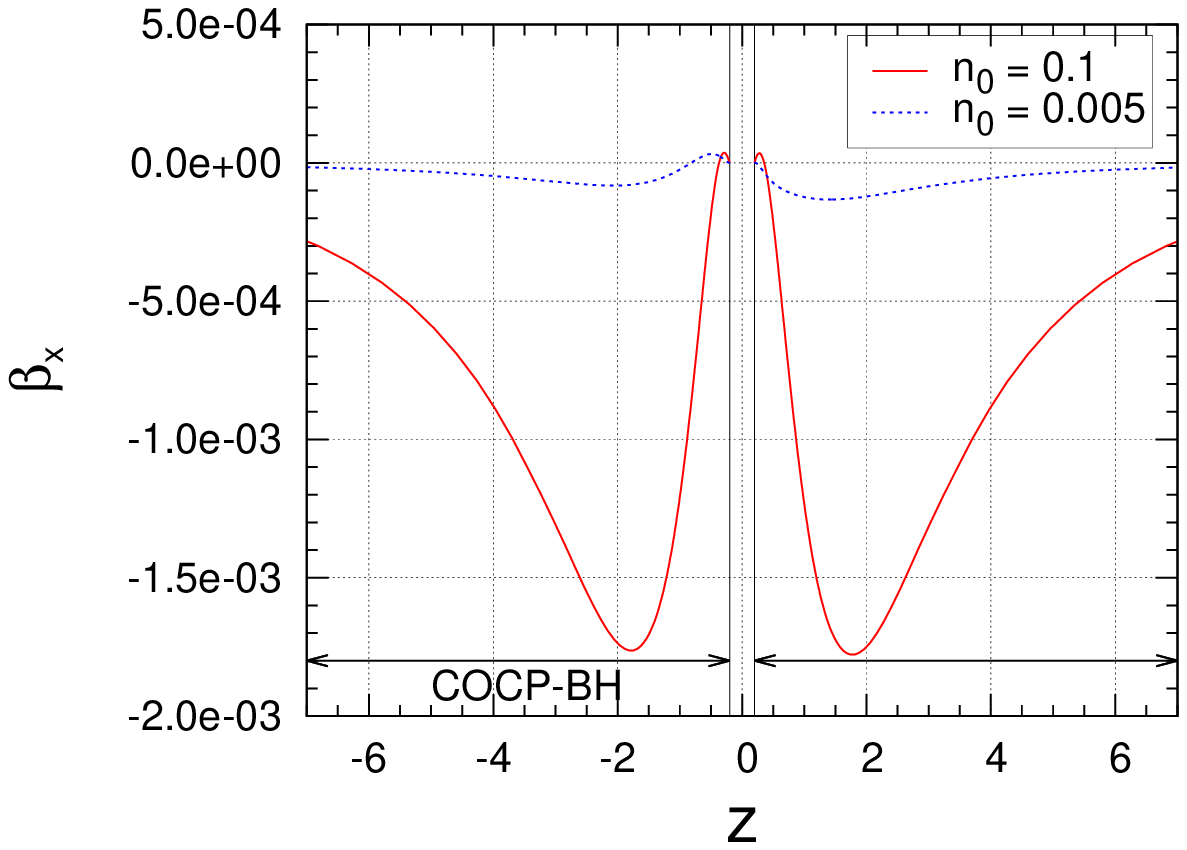}
  \caption{Plots for the $x$-component of the shift $\beta_x$
  of BBH initial data for the case with AH boundary conditions 
  (\ref{eq:AHpsiBC})-(\ref{eq:AHalphBC}).  Parameters in the 
  conditions are chosen as $n_0=0.1$ (solid red lines), and 
  $n_0=0.005$ (dashed green lines), with $\Omega = 0.08$ and 
  $\Omega_{\rm s}=0$.  
    Top panel: along the $x$-axis.  
    Middle panel: along the $y$-axis. 
    Bottom panel: along the $z$-axis.}
  \label{fig:AH_Bvxd}
\end{figure}

\begin{figure}
  \includegraphics[height=60mm,clip]{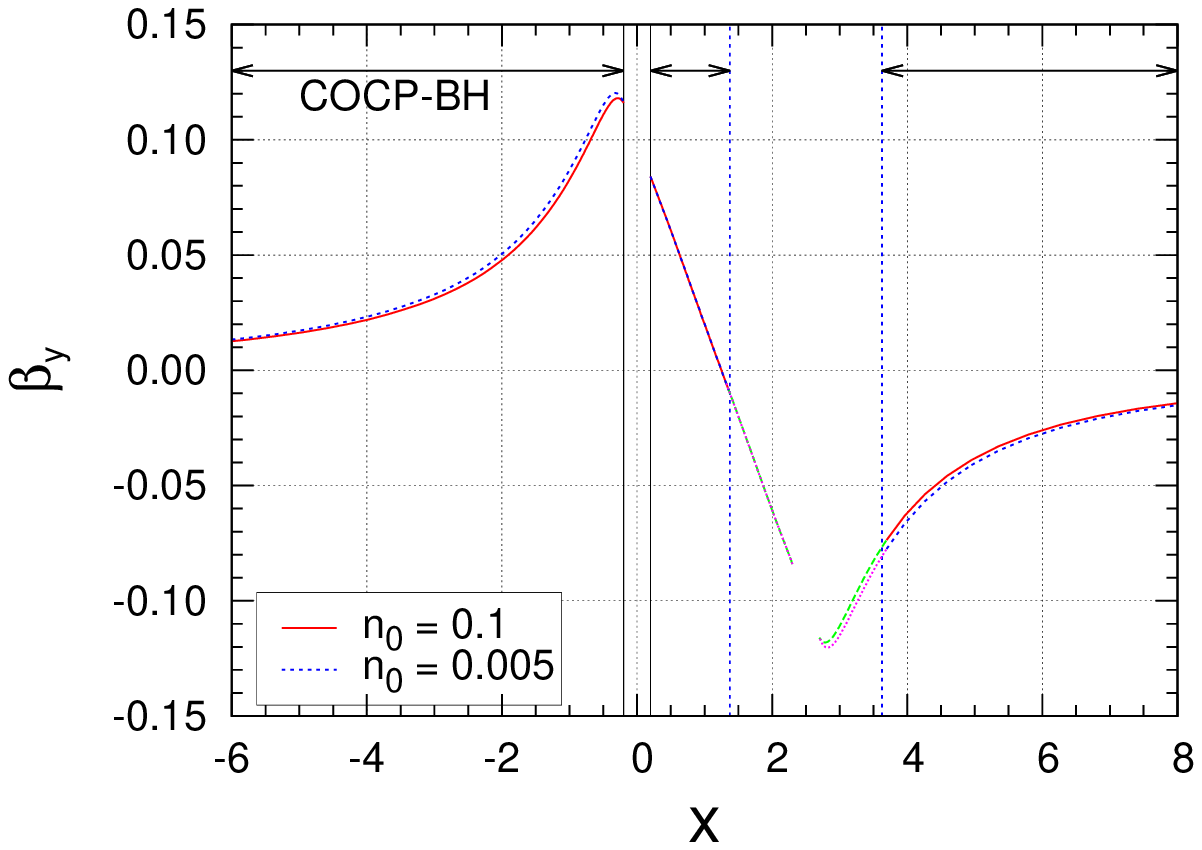}
  \includegraphics[height=60mm,clip]{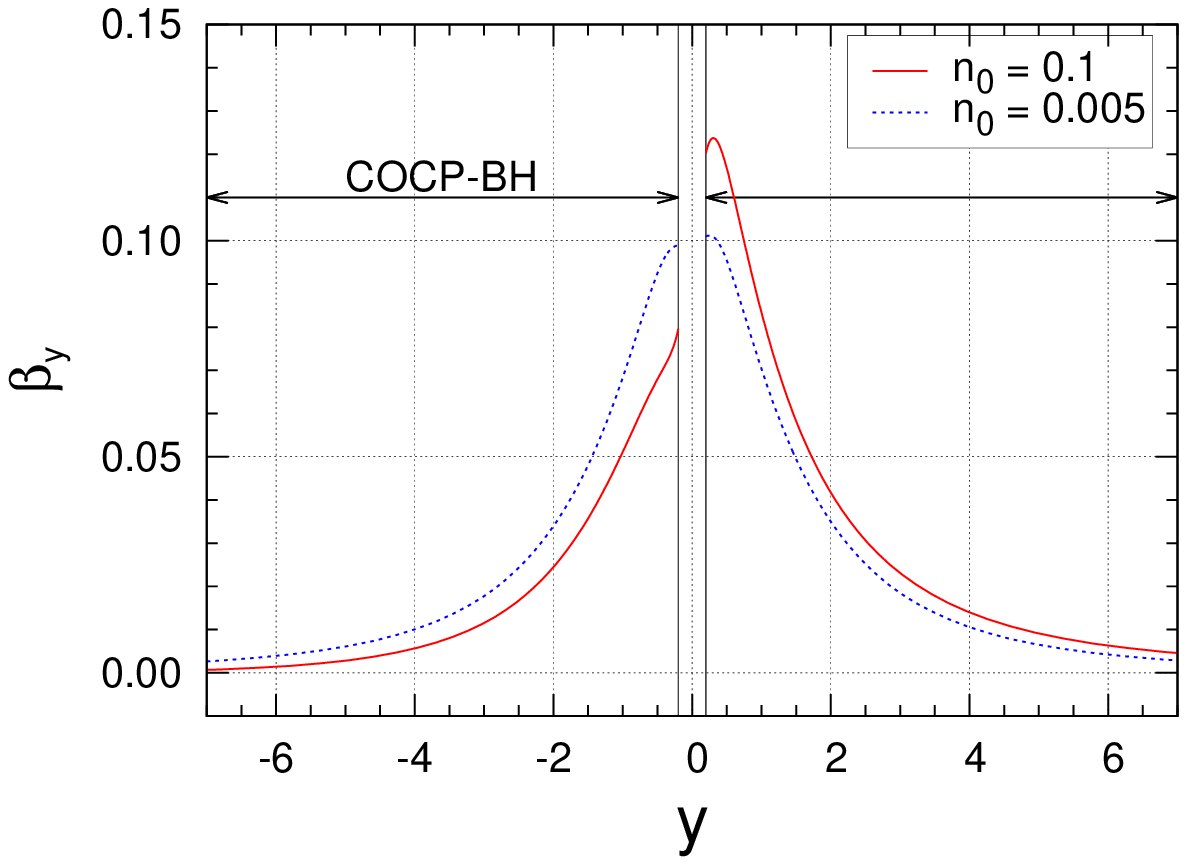}
  \includegraphics[height=60mm,clip]{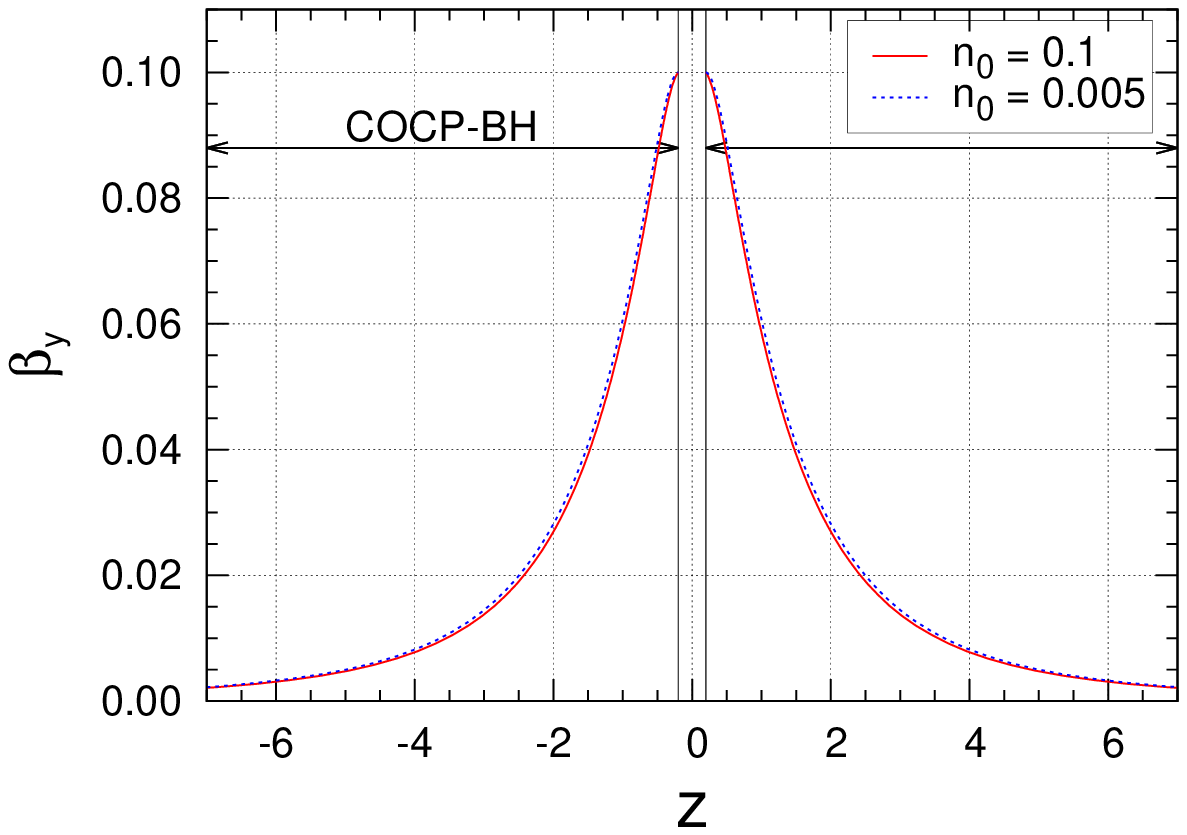}
  \caption{Plots for the same model as Fig.~\ref{fig:AH_Bvxd} but 
  for the $y$-component of the shift $\GB_y$.}
  \label{fig:AH_Bvyd}
\end{figure}

\begin{figure}
  \includegraphics[height=60mm,clip]{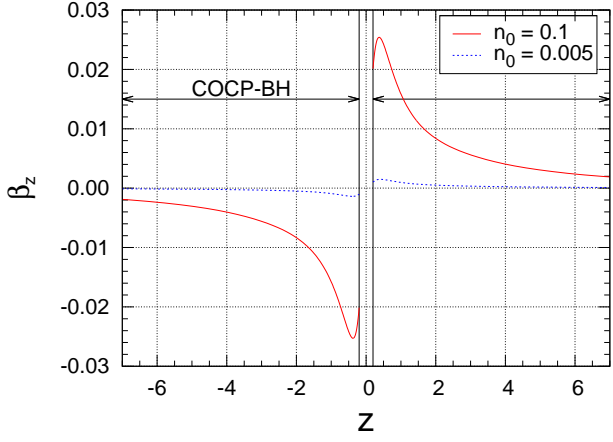}
  \caption{Plots for the same model as Fig.~\ref{fig:AH_Bvxd} but 
  for the $z$-component of the shift $\GB_z$ along the $z$-axis.}
  \label{fig:AH_Bvzd_z}
\end{figure}

\begin{figure}
  \includegraphics[height=60mm,clip]{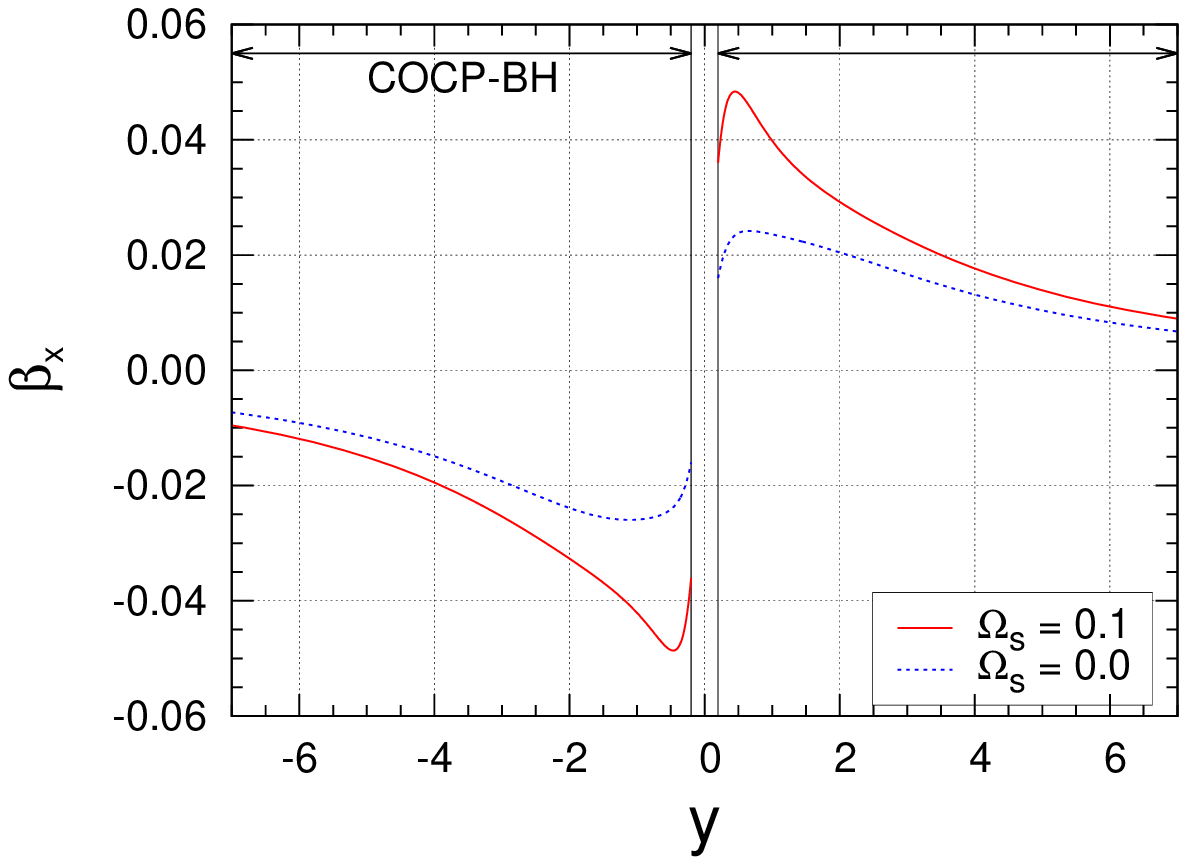}
  \includegraphics[height=60mm,clip]{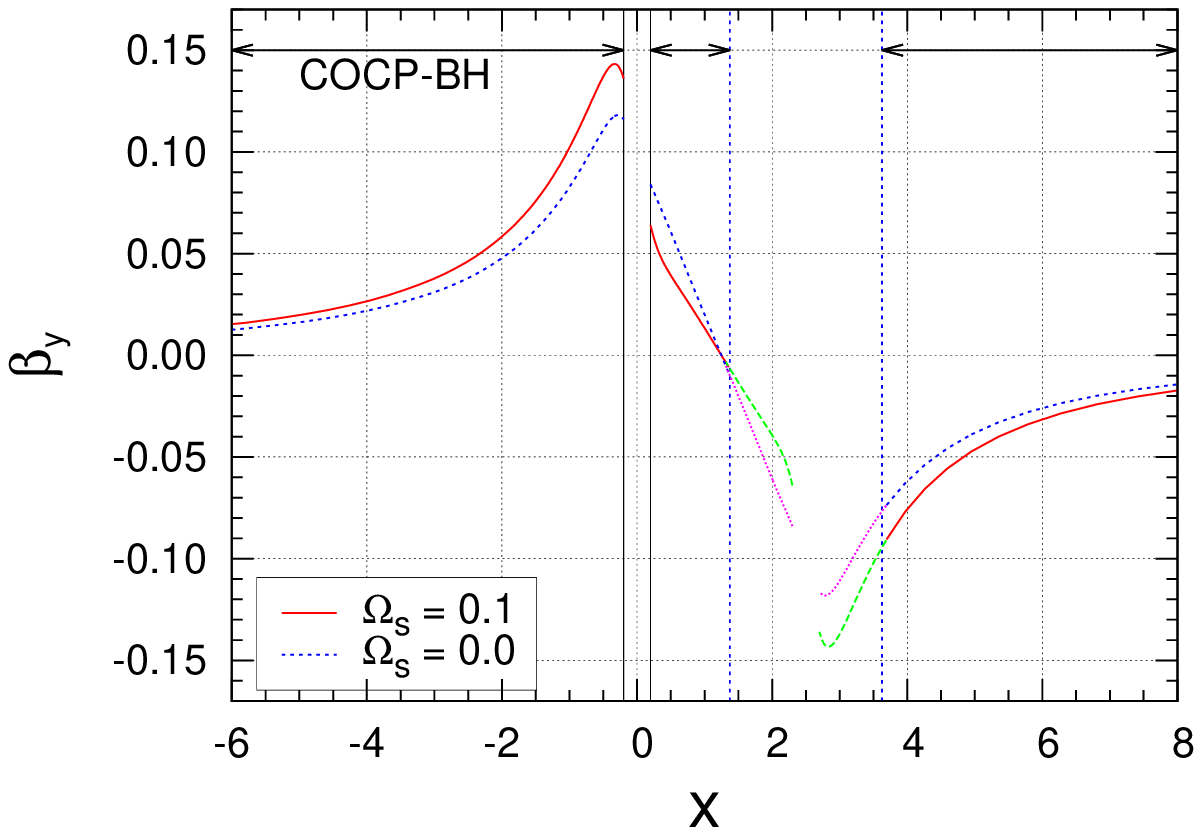}
  \caption{Plots for the components of the shift $\beta_i$
    of BBH initial data for the case with AH boundary conditions 
    (\ref{eq:AHpsiBC})-(\ref{eq:AHalphBC}).  Parameters in the 
    conditions are chosen as $n_0=0.1$ with a spin parameter 
    $\Omega_{\rm s}=0$ (solid red lines), and 
    $\Omega_{\rm s}=0.1$ (dashed green lines).  
    The spins are aligned to the orbital angular momentum 
   (i.e. parallel to $z$-axis).  
    Top panel: $\beta_x$ component along the $y$-axis.  
    Bottom panel: $\beta_y$ component along the $x$-axis. 
    Solid red curves correspond to those in Figs.~\ref{fig:AH_Bvxd}--\ref{fig:AH_Bvzd_z}.}
  \label{fig:spin_Bvxyd}
\end{figure}

\begin{figure}
\includegraphics[height=60mm,clip]{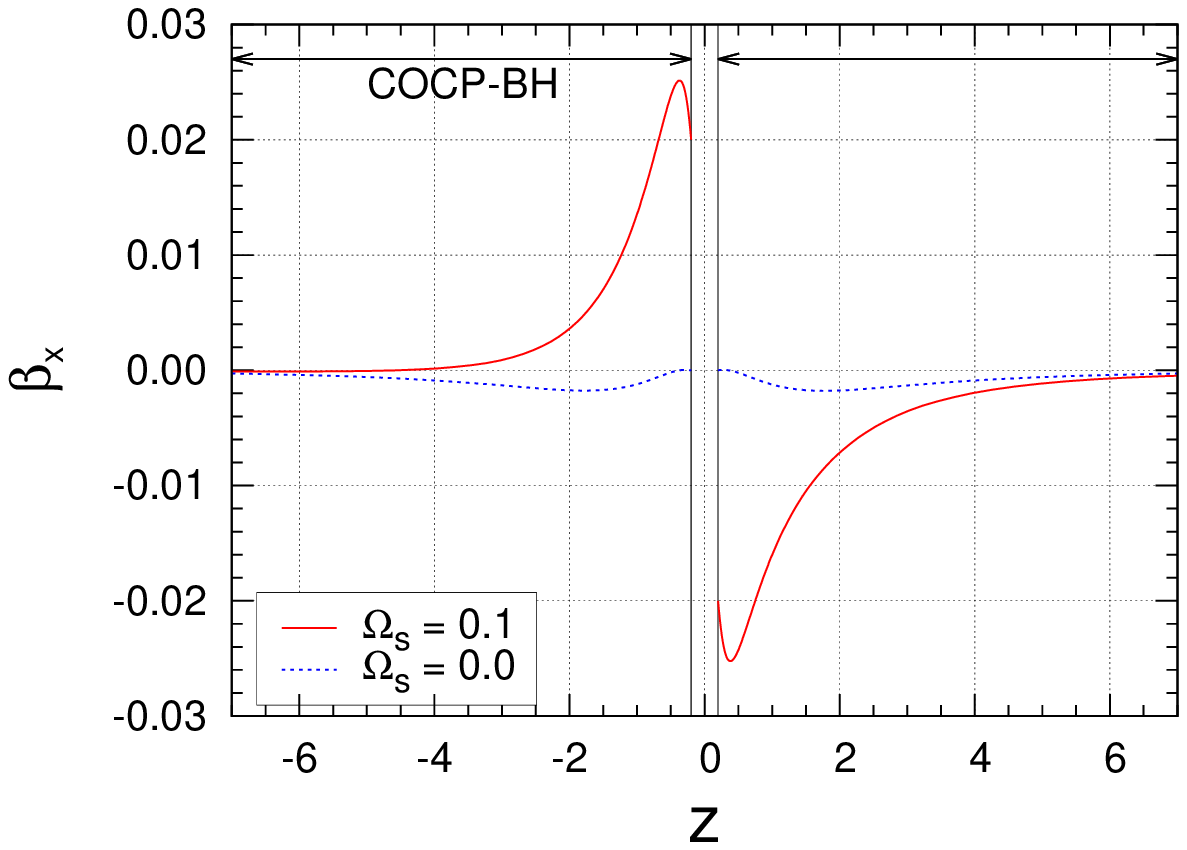}
\includegraphics[height=60mm,clip]{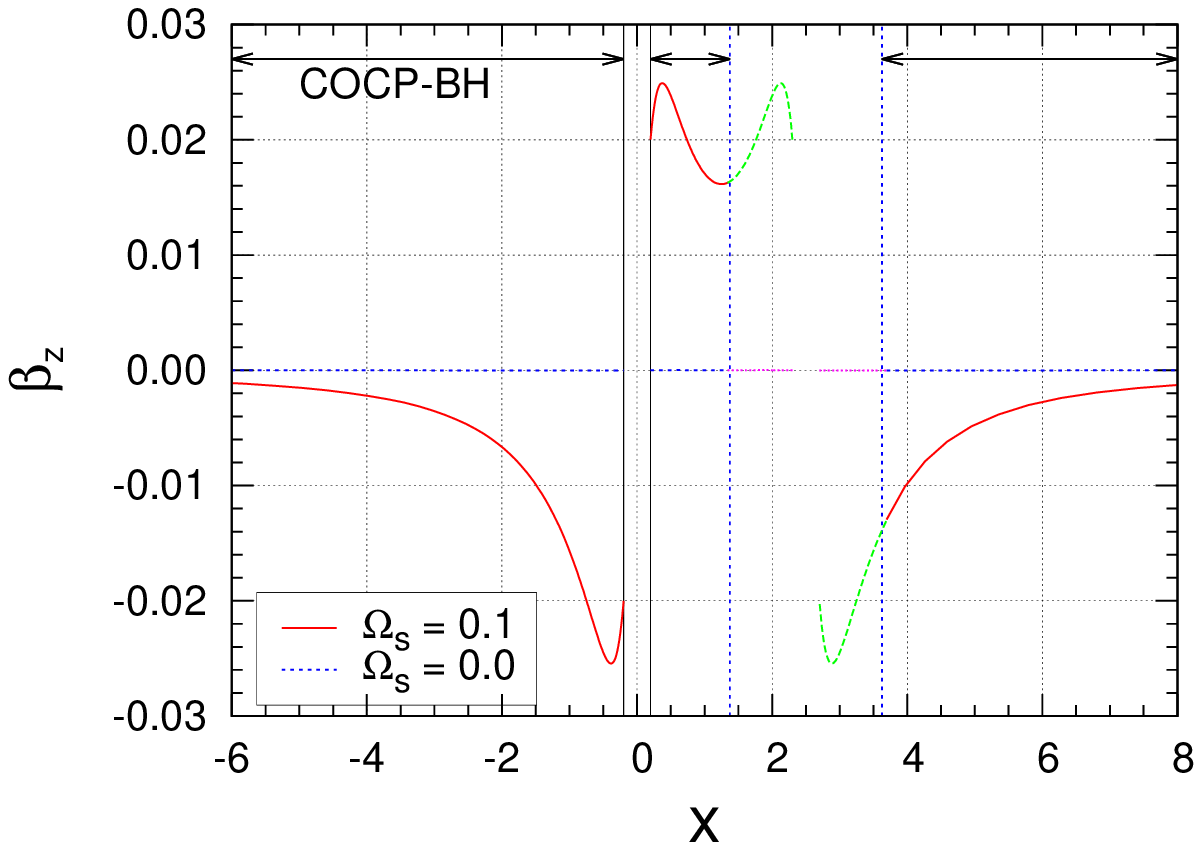}
\includegraphics[height=60mm,clip]{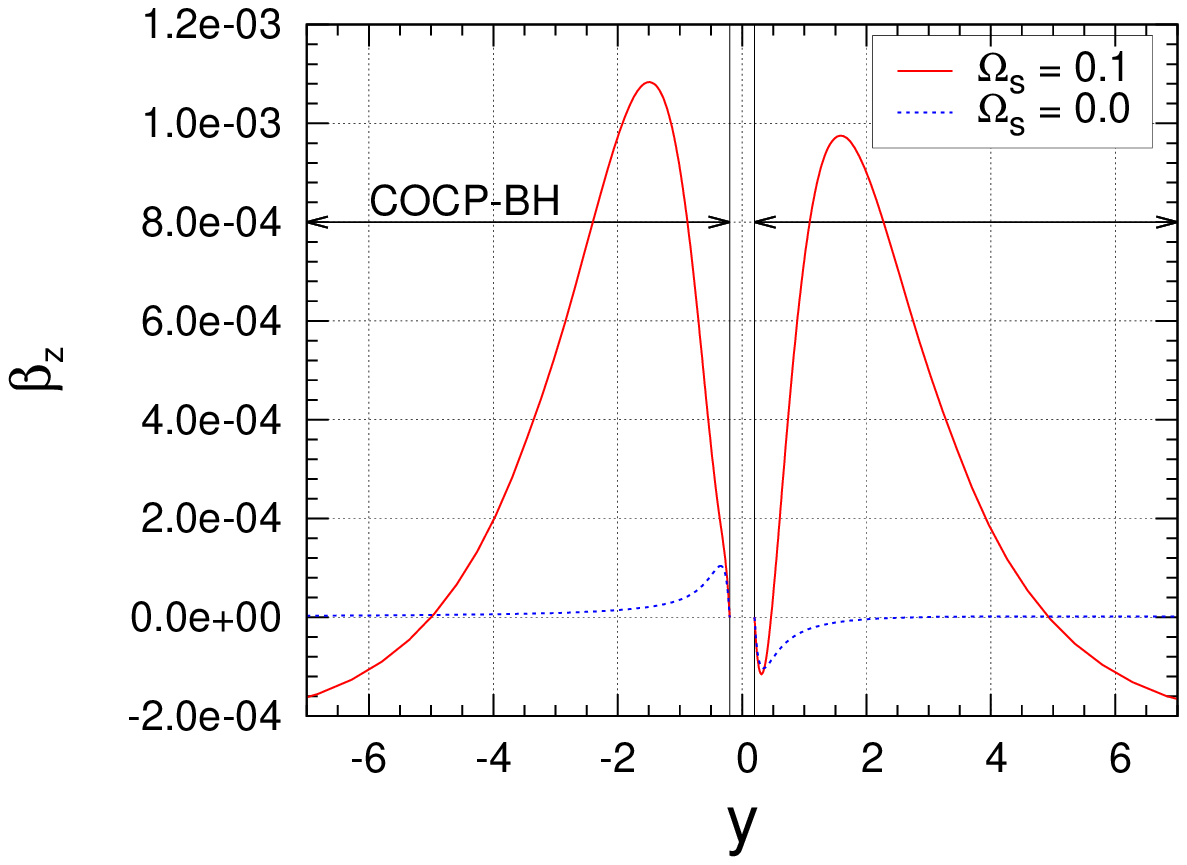}
  \caption{Same as Fig.\ref{fig:spin_Bvxyd}, but the 
  direction of the spin is aligned parallel to the $y$-axis.  
    Top panel: $\beta_x$ component along the $z$-axis.  
    Middle panel: $\beta_z$ component along the $x$-axis. 
    Bottom panel: $\beta_z$ component along the $y$-axis.} 
  \label{fig:gspin_Bvxzd}
\end{figure}
%%% 
%%% 

\section{Initial data for binary black holes}
\label{sec:IWMid}

Finally we present examples for initial data sets for 
equal mass BBH, which have been widely used for BBH merger 
simulations in the literature.  Among several formulations for 
computing such data sets (see e.g.~\cite{NRreview,Cook:2000vr} and 
references therein), we adopt Isenberg-Wilson-Mathews (IWM) formulation.  
In this section, we show the solutions computed from two different 
types of boundary conditions.  The first are simple boundary conditions 
used in our previous paper \cite{TU2007}.  The second are the apparent 
horizon boundary conditions, which have been used to compute quasi-circular 
initial data for BBH (see e.g. 
\cite{Gourgoulhon:2005ng,Booth:2005qc,Cook:2000vr,BBH_QE}).

\subsection{IWM formulation and boundary conditions}

IWM formulation has been widely used for constructing 
quasi-equilibrium initial data for binary compact objects.  
We summarize the basic equations below (for more details see 
e.g.~\cite{Cook:2000vr,ISEN78,WM89}).
The spacetime metric on $\Sigma_t$ is written in 3+1 form as
\beqn
ds^{2} & = & g_{\mu\nu}dx^{\mu}dx^{\nu} \nonumber \\
       & = & -\GA^{2}dt^{2}+\GG_{ij} (dx^{i}+\GB^{i}dt) (dx^{j}+\GB^{j}dt), 
\eeqn
where the spatial three metric $\GG_{ij}$ on the slice $\Sigma_t$ is assumed 
to be conformally flat $\GG_{ij} = \GC^4 f_{ij}$.  
Here, field variables $\psi, \alpha$, and $\beta^i$ are the conformal factor, 
lapse, and shift vector, respectively.  We also assume maximal slicing 
to $\Sigma_t$, so that the trace of the extrinsic curvature 
$K_{ij} := -\frac1{2\alpha}( \Lie_t \GG_{ij}-\Lie_\beta \GG_{ij}) $ vanishes.  
Then, writing its tracefree part $A_{ij}$, 
the conformally rescaled quantity $\tilde A_{ij}$ becomes 
\beq
{\tilde A}_{ij} = 
\frac1{2\alpha} \left(\pd_i{\tilde \GB}_j + \pd_j{\tilde \GB}_i 
- \frac{2}{3}f_{ij}\pd_k{\tilde \GB}^k\right)\ ,
\eeq
where the derivative $\pa_i$ is associated with the flat metric 
$f_{ij}$, and conformally rescaled quantities with tilde are defined by 
$\tilde A_i{}^j= A_i{}^j$ and ${\tilde \beta}^i=\beta^i$, whose indexes 
are lowered (raised) by $f_{ij}$ ($f^{ij}$).  
The system to be solved, which are Hamiltonian and momentum 
constraints and the spatial trace of the Einstein's equation, becomes
\beqn
\Lap \GC &=& -\frac{\GC^5}{8}{\tilde A}_{ij}{\tilde A}^{ij},  
\label{eq:HC}\\
\Lap \GB_i &=& -2\,\alpha\, {\tilde A}_i{}^j \pd_j \ln\frac{\GC^6}{\GA}
-\frac{1}{3}\pd_i\pd_j{\tilde\GB}^j ,
\label{eq:MC}\\
\Lap (\GA\GC) &=& \frac{7}{8}\,\alpha\,\GC^5{\tilde A}_{ij}{\tilde A}^{ij},  
\label{eq:Kdot}
\eeqn
where $\Lap:=\pa_i\pa^i$ is a flat Laplacian.  It is noted that, 
for the shift equation (\ref{eq:MC}), the Cartesian 
components are solved on the spherical coordinates.

As a first set of the boundary conditions 
at the BH excision boundary $S_a$ and 
at the boundary of computational domain $S_b$
for the above system Eqs.~(\ref{eq:HC})-(\ref{eq:Kdot}), 
we choose the following for simplicity, 
\begin{equation}
\begin{aligned}
\left.\GC\right|_{r=r_a} &=n_1  \\
\left.\GB^i\right|_{r=r_a} &= -\,\Omega\,\GP_{\rm cm}^i  \\
\left.\GA\right|_{r=r_a} &=n_0  
\end{aligned}
\qquad\mbox{and}\qquad
\begin{aligned}
\left.\GC\right|_{r=r_b} &=1.0  \\
\left.\GB^i\right|_{r=r_b} &=0.0  \\
\left.\GA\right|_{r=r_b} &=1.0  
\end{aligned}
\label{eq:Initial_bc}
\end{equation}
where $n_1$ and $n_0$ are arbitrary positive constants 
taken as $n_1 \agt 2$ and $n_0\alt 1$, 
$\GP_{\rm cm}^i=(-y_{\rm cm},x_{\rm cm},0)$ is 
the rotational vector with respect to the center of mass of 
the binary associated with coordinates 
$(x_{\rm cm},y_{\rm cm},z_{\rm cm}):=(x-d,y,z)$, and 
$\Omega$ corresponds to the orbital angular 
velocity.  The radius $r_b$ is taken large enough 
as in the test problems in Sec.~\ref{sec:Codetests}.  
Despite the fact that the boundary conditions above may be of 
the simplest type for IWM formulation deduced for acquiring BBH 
data with non-zero orbital angular momentum in an asymptotically 
flat system, they capture the qualitative functional behavior of 
the unknown fields $\{\GC, \GA, \GB^i\}$, as more realistic boundary 
conditions mentioned below.  The solutions calculated from 
these boundary conditions (\ref{eq:Initial_bc}) are compared 
with our previous code \cite{TU2007} that uses 
a different structure for the multiple spherical coordinate patches, 
as well as those solutions of different boundary conditions.

For a second set of the boundary conditions, 
we impose more realistic boundary conditions at the BH boundary $S_a$, 
in particular those that represent apparent horizons in equilibrium 
\cite{Gourgoulhon:2005ng,Booth:2005qc,Cook:2000vr,BBH_QE},   
\begin{eqnarray}
\left.\frac{\pd\GC}{\pd r}+\frac{\GC}{2r}\right|_{r=r_a} 
& = & -\frac{\GC^3}{4}K_{ij}s^i s^j ,  
\label{eq:AHpsiBC} \\
\left.\GB^i\right|_{r=r_a} & = & \frac{n_0}{\GC^2}s^i 
- \Omega\,\GP^i_{\rm cm} - \Omega_{\rm s}\,\GP_{\rm s}^i ,
\label{eq:AHbetaBC} \\ 
\left.\GA\right|_{r=r_a} & = & n_0  ,  
\label{eq:AHalphBC}
\end{eqnarray}
%%% Similar conditions must be applied to the second black hole. 
where $n_0$ is an arbitrary positive constant for which we choose $n_0\alt 0.1$, 
$s^i$ is the unit normal to the sphere $S_a$, 
and $\Omega_{\rm s}$ represents the spin of each 
black hole. The vector $\GP_{\rm s}^i$ is the rotational vectors 
with respect to the coordinate center of BH that generates the BH spin.  
The spin axis is not necessary parallel to the $z$-axis.  
Demanding the sphere $S_a$ to be an apparent horizon (AH) results in 
Eq.~(\ref{eq:AHpsiBC}), while demanding the horizon to be in equilibrium 
results in Eq.~(\ref{eq:AHbetaBC}). 

For the present calculations for BBH initial data, we also assume 
$\pi$-rotation symmetry of the system around the center of mass.  That is, 
two BH have equal masses, and $\pi$-rotation symmetric spins if any.  
In other words, the same boundary conditions are imposed on both BH.  
In those cases, a single patch method discussed in Sec.~\ref{sec:singleBH} 
can be used for simplicity.  As shown in Fig.\ref{fig:bvxy_symmetry}, 
the metric potentials are mapped to the excised sphere $S_e$ from the 
corresponding sphere $S'_e$, taking into account the parity of 
the variables with respect to the $\pi$-rotation.  
As an example, the $\pi$-rotation symmetries 
of the shift components on the $xy$-plane are shown schematically 
(the shift at $A'$ is mapped to point $A$), together with 
the corresponding rules for the derivatives of a function along 
the $x$-axis inside the excised sphere ($B'C'$ mapped to $BC$) 
and along the $y$-axis ($B'D'$ mapped to $BD$).  In terms of 
the center of mass coordinates, the $\pi$-rotation symmetries 
of the components of the shift vector become 
%%%
\begin{eqnarray*}
\GB_x(-x_{\rm cm},-y_{\rm cm},z_{\rm cm}) 
&=& -\GB_x(x_{\rm cm},y_{\rm cm},z_{\rm cm}) \ , \\
\GB_y(-x_{\rm cm},-y_{\rm cm},z_{\rm cm}) 
&=& -\GB_y(x_{\rm cm},y_{\rm cm},z_{\rm cm}) . 
\end{eqnarray*}
%%%
The $z$-component $\GB_z$ is mapped as a scalar quantity. 

Mapped quantities are used in the elliptic equation solver 
when the sources of surface integrals on the excised sphere 
$S_e$, Eq.~(\ref{eq:Green_int}) are evaluated.  
Also, at the end of each iteration step, 
obtained potentials $\{\psi,\alpha,\beta_i\}$ between the spheres 
$S_a$ and $S'_e$ are interpolated inside the sphere $S_e$ 
following the same rules for the parity of each variable.  
At the end we have the solution at every point inside $S_b$ and 
outside of the two black holes of radius $r_a$ positioned
at the origin and at $x=d_s$ on the $x$-axis.

In the case of no spins $\Omega_{\rm s}=0$, or spins are 
parallel to one of coordinate axis, additional 
symmetries with respect to the $xy|_{\rm cm}$ plane occur.  
%%% , $yz|_{\rm cm}$, or $zx|_{\rm cm}$ planes occur.  
%%% 
%%% For example for the case with $\Omega_{\rm s} = 0$, 
%%% \begin{itemize}
%%% \item $\GB_x$: Symmetric with respect to yz-plane, 
%%% xy-plane, and antisymmetric with respect to xz-plane
%%% \begin{eqnarray*}
%%% \GB_x(-x,y,z) &=& \GB_x(x,y,z) \ , \\
%%% \GB_x(x,-y,z) &=& -\GB_x(x,y,z)\ ,   \\
%%% \GB_x(x,y,-z) &=& \GB_x(x,y,z)
%%% \end{eqnarray*}
%%% \item $\GB_y$: Symmetric with respect to xz-plane, xy-plane, 
%%% and antisymmetric with respect to yz-plane
%%% \begin{eqnarray*}
%%% \GB_y(-x,y,z) &=& -\GB_y(x,y,z) \ ,  \\
%%% \GB_y(x,-y,z) &=& \GB_y(x,y,z) \ ,  \\
%%% \GB_y(x,y,-z) &=& \GB_y(x,y,z)
%%% \end{eqnarray*}
%%% \item $\GB_z$: Symmetric with respect to yz-plane, xz-plane, 
%%% and antisymmetric with respect to xy-plane
%%% \begin{eqnarray*}
%%% \GB_z(-x,y,z) &=& \GB_z(x,y,z) \ , \\ 
%%% \GB_z(x,-y,z) &=& \GB_z(x,y,z) \ ,\\
%%% \GB_z(x,y,-z) &=&-\GB_z(x,y,z)
%%% \end{eqnarray*}
%%% \end{itemize}
%%% where $x,y,z$ are coordinates with respect to the center of mass.
%%% 
%%% 
In our previous codes \cite{UryuCF,UryuWL,TU2007}, a part or 
all of these symmetries were encoded in the elliptic solver, 
for a computational domain was reduced by assuming the symmetries 
so that only a part of the whole hypersurface $\Sigma_t$ was solved.  
In {\sc cocal}, we are solving in the whole $\Sigma_t$ 
and therefore such symmetries are satisfied in the solution 
within negligible numerical errors.

%
%%% \begin{figure*}
%%% %  \includegraphics[width=3.1in,clip]{bvxy.ps}
%%%   \includegraphics{bvxy3d_zc.ps}
%%%   \caption{3-dimensional shift vector on the $z=0.1$ plane. Radii of black holes are $r_a=0.2$.
%%%   It is $\GB_z>0$ for $x<1.25$ and $\GB_z<0$ for $x>1.25$. Also the z-component is magnified ten
%%%   times more than the $x$, and $y$ components.}
%%%   \label{fig:Bvxy3d}
%%% \end{figure*}
%
%

\subsection{Solutions for BBH initial data}
\label{sec:AHbc}

Finally, we present the BBH initial data sets computed from 
the above formulation with several parameter sets for boundary 
conditions listed in Table \ref{tab:BBHinitial}.
As we concentrate on testing the {\sc cocal} code, 
we do not discuss much of the physical contents of the 
initial data sets, but display the plots of the fields 
to check their behaviors.

In Fig.~\ref{fig:bvyd_x_TU2007}, we calculated BBH data for 
the same model as shown in our previous paper \cite{TU2007} 
for a comparison.  Parameters in the boundary condition 
(\ref{eq:Initial_bc}) are chosen as $n_1 = 3.0$, $n_0=1.0$, 
and $\Omega=0.3$.  Only in this model, we choose the binary 
separation as $d_s=2.8$ and the BH excision radius $r_a=0.1$.  
We find the solution agrees well with Fig.~11 of \cite{TU2007} 
as expected, although the structures of coordinate patches 
are different in each code.  
In Figs.~\ref{fig:Psi_Alph_x}, and \ref{fig:TU_bvd}, 
We use the same boundary conditions (\ref{eq:Initial_bc})
but with different parameters which may be more common values 
for the BBH data.  In the computation, the grid parameters used 
are D3 in Table \ref{tab:1BHtest_grids}.  When the value of $\Omega$ 
is increased, the magnitude of the lapse and the conformal factor 
remain almost the same, while the magnitude of the components of 
the shift increase with the functional behavior staying the same. 
For example when $\Omega=0.1$, $\GB_x$ on the $y$-axis varies between 
$\pm 0.05$ while $\GB_y$ on the $x$-axis varies between $\pm 0.15$.  
For this boundary conditions, the code blows up when $\Omega$ 
is approximately greater than $0.2$.

Solutions for the BBH initial data with the AH boundary conditions 
(\ref{eq:AHpsiBC})-(\ref{eq:AHalphBC}) are shown 
in Figs.~\ref{fig:Bvxy}--\ref{fig:AH_Bvzd_z}.  
In this calculation, the resolution is D3 in Table \ref{tab:1BHtest_grids} 
with $\Omega=0.08$ and $\Omega_{\rm s}=0$.  
The shift vector plots and the contour plots of 
the conformal factor $\GC$ and the lapse $\GA$ in $xy$-plane 
are for the model with $n_0=0.1$.  The behavior of the field 
$\psi$ and $\alpha$ are analogous to those from the simple boundary 
condition (\ref{eq:Initial_bc}) shown in Fig.~\ref{fig:Psi_Alph_x}.  
Because of the choice $\Omega_{\rm s}=0$, the solution 
satisfies xy-plane symmetry.

From a comparison between the results of parameters $n_0=0.1$ and $0.005$ 
shown in Figs.~\ref{fig:AH_Bvxd}-\ref{fig:AH_Bvzd_z}, it is found 
that the results with $n_0=0.005$ becomes more similar to 
the results of the first boundary conditions (\ref{eq:Initial_bc}) 
shown in Fig.\ref{fig:TU_bvd}.  For example, it is most evident in 
the plot for $\beta_y$ along the $y$-axis, Fig.\ref{fig:AH_Bvyd} middle panel 
and Fig.\ref{fig:TU_bvd} bottom panel.  This seems to be a correct 
behavior because the first term of Eq.~(\ref{eq:AHbetaBC}) contributes 
less as the value of alpha (parameter $n_0$) become smaller
\footnote{
For the first boundary conditions (\ref{eq:Initial_bc}), the solution of the shift 
does not depend much on $n_0$ in the region $n_0 \alt 0.1$.}.
However, it is not expected that there is a solution in the 
limit of $n_0 \rightarrow 0$, in fact, for both types of boundary conditions 
iterations diverge when the $n_0 \alt 0.004$ in the {\sc cocal} code.

All of the solutions above have zero black hole spins.
Setting spins $\Omega_{\rm s}=0.1$ in the same direction
as the orbital motion (rotation on the xy-plane) we get a solution similar to
Figs.~\ref{fig:AH_Bvxd}-\ref{fig:AH_Bvzd_z} except for 
$\GB_x$ along the y-axis and $\GB_y$ along the x-axis.  The results with 
and without spins are compared in Fig.~\ref{fig:spin_Bvxyd} for the case with 
$n_0 = 0.1$.  
For more general black hole spins we obtain Fig.\ref{fig:gspin_Bvxzd} where 
we have taken a spin $\Omega_{\rm s}=0.1$ along the y-axis.
In these plots for the solutions with the spins, we confirm that all 
components of the shift vectors behave correctly along each coordinate axis.

\section{Discussion}

Although the numerical method presented in this paper may seem similar 
to the one presented in our previous paper \cite{TU2007}, 
the robustness of the convergence and the control of the numerical 
errors are largely improved.  In some cases, the previous 
method failed to compute a continuous solution at the interface 
between multiple patches during the iterations, and therefore 
a convergence to a solution couldn't be achieved.  
A major reason for this failure turned out to be a lack of enough 
overlap region between coordinate patches.  The numerical errors 
of the field variables are relatively larger near the boundary of 
computational domain as seen, for example, in Fig.\ref{fig:noneqmBBHtest}.  
Hence if the overlap region is small, those potentials with larger 
numerical errors overlap, which seems to cause the non-convergence 
of the iteration.  
In the {\sc cocal} code, the overlap region is almost as large as 
the whole domain for the two coordinate patch configuration, and 
is large enough even for the three coordinate patch configuration.  
We have never observed so far a discontinuous behavior of a 
field in the solutions of {\sc cocal}.

The {\sc cocal} code currently runs only on a serial processor, 
which is sufficient to maintain the accuracy presented in 
Sec.\ref{sec:Codetests}.  In the computation for BBH initial data 
shown in Sec.IV, the size of main memory and CPU times per 
1 iteration cycle used by {\sc cocal} are about 800MB 50sec 
for D3 grid, and 6GB 8min for D4 grid, and around 50-150 
iterations are needed for a convergence, where the iterations 
start from an initial guess $\psi = \alpha = 1$ and $\beta_i = 0$.  
Because we use second order accurate formulas in {\sc cocal}, 
we can decrease the numerical error by two orders of magnitude with 
$10$ times more grid points in each direction, that is, $10^3$ more 
grids in total.  Considering specs of common parallel computer systems 
it seems to be feasible to achieve this accuracy by parallelizing 
{\sc cocal}.  We have started to develop a prototype of 
such parallelized {\sc cocal} code, whose results would be presented 
elsewhere.

The most advantageous feature of {\sc cocal} would be its 
simplicity in coding.  This helps the users to introduce more complex 
physics on top of the current code.  For example, we have developed 
subroutines for solving spatially conformally flat data (IWM formulation) 
first, and later added subroutines for solving non-conformally flat data 
(waveless formulation) on top.  In the same way, it will be straight 
forward to incorporate subroutines to solve electromagnetic fields, 
which enables us to investigate for example magnetar models.  
We proceed to develop codes for computing various kinds of 
astrophysically realistic equilibriums and quasi-equilibrium data 
and provide the results to applications including initial data for 
numerical relativity simulations.  We plan for making the 
{\sc cocal} code and computed initial data sets available 
for public in the near future.

\acknowledgments
This work was supported by 
JSPS Grant-in-Aid for Scientific Research(C) 23540314 
and 22540287, and MEXT Grant-in-Aid for Scientific Research
on Innovative Area 20105004.  
KU thanks Charalampos Markakis, Noriyuki Sugiyama, and 
members of LUTH at Paris Observatory for discussions.  

\appendix

\section{Computations for homogeneous solutions}
\label{sec:comp_hs}

In this Appendix, we show a concrete derivation 
for the homogeneous solution $\chi(x)$ used in the 
elliptic solver (\ref{eq:solver}) for two cases, 
one with Neumann boundary condition at the inner 
boundary sphere $S_a$ and Dirichlet boundary condition 
at the outer boundary sphere $S_b$, and 
the other with Dirichlet conditions at both $S_a$ and $S_b$.  
We summarize the other cases in the next 
Appendix \ref{sec:SurfInt}.  

As explained in Sec.~\ref{sec:solver}, the solution of 
$\mathcal{L}\Phi=S$ is written 
$ \Phi(x) = \chi(x) + \Phi_{\rm INT}(x)$
to impose certain boundary conditions at the two spheres 
$S_a$ and $S_b$.  
In order to do so, the homogeneous solution 
$\chi$ of the Laplacian is split into two functions 
$\chi_a(x)$ and $\chi_b(x)$ as $\GX(x) = \GX_a(x) + \GX_b(x)$.  
Both are solutions of Laplace equations, and one for the exterior 
of the sphere $S_a$ and the other for the interior of the sphere 
$S_b$ (see Fig.~\ref{fig:SI_ext_int}).  
Since $r^\ell,\ r^{-\ell-1}$ are the solutions of the radial part of 
the Laplacian, the contribution $\GX_a$ is taken to be a series of 
$r^{-\ell-1}$, while the contribution $\GX_b$ to be a series of $r^\ell$.  
Therefore, we write
\beqn
\GX_a(x) & = & \frac{1}{4\pi}\sum_{\ell=0}^{\infty}\sum_{m=0}^\ell \GE_m\frac{(\ell-m)!}{(\ell+m)!}
               P_\ell^m(\cos\GJ)r^{-\ell-1}\times\qquad\nonumber\\
         &   & \times[A_{\ell m}\cos(m\GP)+B_{\ell m}\sin(m\GP)]\qquad\label{eq:chii} 
\label{eq:chi_a}
\\
\GX_b(x) & = & \frac{1}{4\pi}\sum_{\ell=0}^{\infty}\sum_{m=0}^\ell \GE_m\frac{(\ell-m)!}{(\ell+m)!}
               P_\ell^m(\cos\GJ)r^\ell\times\qquad\nonumber\\
         &   & \times[C_{\ell m}\cos(m\GP)+D_{\ell m}\sin(m\GP)]\qquad\label{eq:chio} 
\label{eq:chi_b}
\eeqn
where $A_{\ell m}$, $B_{\ell m}$, $C_{\ell m}$, and $D_{\ell m}$ are constants.  
%%% that depend on the boundary conditions
%%% that will be used on the inner and outer surface. 
One common choice is
\beqn
\left.\frac{\pd\Phi}{\pd r}\right|_{r=r_a} 
& = &
\left.\frac{\pd\Phi_{\rm BC}}{\pd r}\right|_{r=r_a} 
%%% \left.\frac{\pd\Phi_a}{\pd r}\right|_{r=r_a} ,
\label{eq:Nbc}\\
\left.\Phi\right|_{r=r_b} & = & 
%%% \left.\Phi_b\right|_{r=r_b}, \label{eq:Dbc}
\left.\Phi_{\rm BC}\right|_{r=r_b}, \label{eq:Dbc}
\eeqn
where $\frac{\pd\Phi_{\rm BC}}{\pd r}|_{r=r_a}$ and $\Phi_{\rm BC}|_{r=r_b}$ 
are known functions defined at boundaries $S_a$ and $S_b$, respectively; 
%%% where $\Phi_b, \frac{\pd\Phi_a}{\pd r}$ are known functions,
we consider the case with Neumann boundary condition at the inner surface 
$S_a$ and Dirichlet at the outer one $S_b$. 

From boundary condition Eq.~(\ref{eq:Nbc}) with the use of 
Eqs.~(\ref{eq:chii}), (\ref{eq:chio}), (\ref{eq:solver}), (\ref{eq:Green_int}), 
and the orthogonality relations
\begin{eqnarray*}
\int_0^\pi P_\ell^m(\cos\GJ)P_{\ell'}^m(\cos\GJ)\sin\GJ d\GJ = 
\frac{2}{2\ell+1}\frac{(\ell+m)!}{(\ell-m)!}\GD_{\ell\ell'}  \\
\int_0^{2\pi}\sin(m\GP)\cos(m'\GP)d\GP=0 \qquad\qquad\qquad\qquad\\
\int_0^{2\pi}\cos(m\GP)\cos(m'\GP)d\GP=\frac{2\pi}{\GE_m}\GD_{mm'}  \qquad\qquad\quad 
\end{eqnarray*}
we get
\beqn
-\frac{\ell+1}{r_a^{\ell+2}}A_{\ell m}+\ell r_a^{\ell-1}C_{\ell m} = (2\ell+1)\times\qquad\qquad \nonumber\\
\int_0^\pi \int_0^{2\pi} \left(\frac{\pd\Phi_{\rm BC}}{\pd r}-\frac{\pd\Phi_{\rm INT}}{\pd r}\right)_{r=r_a}
P_\ell^m(\cos\GJ)\cos(m\GP) d\Omega  
\nonumber \\ \\
-\frac{\ell+1}{r_a^{\ell+2}}B_{\ell m}+\ell r_a^{\ell-1}D_{\ell m} = \frac{2(2\ell+1)}{\GE_m}
\times\qquad\qquad \nonumber\\
\int_0^\pi \int_0^{2\pi} \left(\frac{\pd\Phi_{\rm BC}}{\pd r}-\frac{\pd\Phi_{\rm INT}}{\pd r}\right)_{r=r_a}
P_\ell^m(\cos\GJ)\sin(m\GP) d\Omega
\nonumber\\
\eeqn
Using now boundary
condition Eq.~(\ref{eq:Dbc}) and again the same equations we get
\beqn
r_b^{-\ell-1}A_{\ell m} + r_b^\ell C_{\ell m} = (2\ell+1)\times\qquad\qquad \nonumber\\
\int_0^\pi \int_0^{2\pi} \left(\Phi_{\rm BC}-\Phi_{\rm INT}\right)_{r=r_b}
P_\ell^m(\cos\GJ)\cos(m\GP) d\Omega  \qquad\\
r_b^{-\ell-1}B_{\ell m} + r_b^\ell D_{\ell m} = \frac{2(2\ell+1)}{\GE_m}\times\qquad\qquad \nonumber\\
\int_0^\pi \int_0^{2\pi} \left(\Phi_{\rm BC}-\Phi_{\rm INT}\right)_{r=r_b}
P_\ell^m(\cos\GJ)\sin(m\GP) d\Omega  \qquad
\eeqn
The system of four equations with respect to $A_{\ell m}$, $B_{\ell m}$, $C_{\ell m}$, $D_{\ell m}$
can be solved to yield
\begin{widetext}
\begin{eqnarray*}
A_{\ell m}\left[1+\frac{\ell}{\ell+1}\left(\frac{r_a}{r_b}\right)^{2\ell+1}\right] &=&
\frac{\ell(2\ell+1)}{\ell+1}r_a^{\ell+1}\left(\frac{r_a}{r_b}\right)^\ell
\int_{S_b}(\Phi_{\rm BC}-\Phi_{\rm INT})P_\ell^m(\cos\GJ)\cos(m\GP) d\Omega  \\
& & -\frac{2\ell+1}{\ell+1}r_a^{\ell+2}
\int_{S_a}\left(\frac{\pd\Phi_{\rm BC}}{\pd r}-\frac{\pd\Phi_{\rm INT}}{\pd r}\right)P_\ell^m(\cos\GJ)\cos(m\GP) d\Omega, \\
C_{\ell m}\left[1+\frac{\ell}{\ell+1}\left(\frac{r_a}{r_b}\right)^{2\ell+1}\right] &=&
(2\ell+1)b^{-\ell}\int_{S_b}(\Phi_{\rm BC}-\Phi_{\rm INT})P_\ell^m(\cos\GJ)\cos(m\GP) d\Omega  \\
& & +\frac{2\ell+1}{\ell+1}\frac{r_a^2}{r_b^{\ell+1}}\left(\frac{r_a}{r_b}\right)^\ell
\int_{S_a}\left(\frac{\pd\Phi_{\rm BC}}{\pd r}-\frac{\pd\Phi_{\rm INT}}{\pd r}\right)P_\ell^m(\cos\GJ)\cos(m\GP) d\Omega, \\
B_{\ell m}\left[1+\frac{\ell}{\ell+1}\left(\frac{r_a}{r_b}\right)^{2\ell+1}\right] &=&
\frac{2\ell(2\ell+1)}{(\ell+1)\GE_m}r_a^{\ell+1}\left(\frac{r_a}{r_b}\right)^\ell
\int_{S_b}(\Phi_{\rm BC}-\Phi_{\rm INT})P_\ell^m(\cos\GJ)\sin(m\GP) d\Omega  \\
& & -\frac{2(2\ell+1)}{(\ell+1)\GE_m}r_a^{\ell+2}
\int_{S_a}\left(\frac{\pd\Phi_{\rm BC}}{\pd r}-\frac{\pd\Phi_{\rm INT}}{\pd r}\right)P_\ell^m(\cos\GJ)\sin(m\GP) d\Omega, \\
D_{\ell m}\left[1+\frac{\ell}{\ell+1}\left(\frac{r_a}{r_b}\right)^{2\ell+1}\right] &=&
\frac{2(2\ell+1)}{\GE_m}r_b^{-\ell}\int_{S_b}(\Phi_{\rm BC}-\Phi_{\rm INT})P_\ell^m(\cos\GJ)\sin(m\GP) d\Omega  \\
& & +\frac{2(2\ell+1)}{(\ell+1)\GE_m}\frac{r_a^2}{r_b^{\ell+1}}\left(\frac{r_a}{r_b}\right)^\ell
\int_{S_a}\left(\frac{\pd\Phi_{\rm BC}}{\pd r}-\frac{\pd\Phi_{\rm INT}}{\pd r}\right)P_\ell^m(\cos\GJ)\sin(m\GP) d\Omega.
\end{eqnarray*}
%%%  \end{widetext}
Substituting the above in Eqs.~(\ref{eq:chii}) and (\ref{eq:chio}), we have
\begin{eqnarray}
\GX_a(x)+\GX_b(x) 
         & = & \frac{1}{4\pi}\sum_{\ell=0}^{\infty}\sum_{m=0}^\ell \GE_m\frac{(\ell-m)!}{(\ell+m)!}
               P_\ell^m(\cos\GJ)\times    \nonumber\\
         &   & \left\{ 
               (2\ell+1)\left(\frac{r_a}{r_b}\right)^\ell
               \frac{\left(\frac{r}{r_a}\right)^\ell+\frac{\ell}{\ell+1}\left(\frac{r_a}{r}\right)^{\ell+1}}
                    {1+\frac{\ell}{\ell+1}\left(\frac{r_a}{r_b}\right)^{2\ell+1}}
               \int_{S_b}[\Phi_{\rm BC}-\Phi_{\rm INT}]P_\ell^m(\cos\GJ')\cos[m(\GP-\GP')] d\Omega' \right. \nonumber \\
         &   & \left. +\frac{-2\ell-1}{\ell+1}\frac{r_a^{\ell+2}}{r_b^{\ell+1}}
               \frac{\left(\frac{r_b}{r}\right)^{\ell+1} - \left(\frac{r}{r_b}\right)^\ell}
                    {1+\frac{\ell}{\ell+1}\left(\frac{r_a}{r_b}\right)^{2\ell+1}}
               \int_{S_a}\left(\frac{\pd\Phi_{\rm BC}}{\pd r}-\frac{\pd\Phi_{\rm INT}}{\pd r}\right)
                         P_\ell^m(\cos\GJ')\cos[m(\GP-\GP')] d\Omega'  
               \right\}. \label{eq:chiioND}
\end{eqnarray}
\end{widetext}
%%% Note that the radial function that multiplies the integral at the outer sphere $S_b$ is identical with 
%%% the radial function of Eq.~(\ref{eq:SIbND}) and the radial function that multiplies the inner sphere $S_a$
%%% is identical with the radial function of Eq.~(\ref{eq:SIaND}).

Similarly, when we have Dirichlet boundary conditions on both the inner and the outer 
spheres $S_a$ and $S_b$, 
\beqn
\left.\Phi\right|_{r=r_a} & = & \left.\Phi_{\rm BC}\right|_{r=r_a}, \label{eq:Dabc}  \\
\left.\Phi\right|_{r=r_b} & = & \left.\Phi_{\rm BC}\right|_{r=r_b}, \label{eq:Dbbc}
\eeqn
the contribution to the potential will be
\begin{widetext}
\begin{eqnarray}
\GX_a(x)+\GX_b(x) 
         & = & \frac{1}{4\pi}\sum_{\ell=0}^{\infty}\sum_{m=0}^\ell \GE_m\frac{(\ell-m)!}{(\ell+m)!}
               P_\ell^m(\cos\GJ)\times    \nonumber\\
         &   & \left\{ 
               (2\ell+1)\left(\frac{r_a}{r_b}\right)^\ell
               \frac{\left(\frac{r}{r_a}\right)^\ell-\left(\frac{r_a}{r}\right)^{\ell+1}}
                    {1-\left(\frac{r_a}{r_b}\right)^{2\ell+1}}
               \int_{S_b}[\Phi_{\rm BC}-\Phi_{\rm INT}]P_\ell^m(\cos\GJ')\cos[m(\GP-\GP')] d\Omega' \right. \nonumber \\
         &   & \left. +(2\ell+1)\left(\frac{r_a}{r_b}\right)^{\ell+1}
               \frac{\left(\frac{r_b}{r}\right)^{\ell+1} - \left(\frac{r}{r_b}\right)^\ell}
                    {1-\left(\frac{r_a}{r_b}\right)^{2\ell+1}}
               \int_{S_a}[\Phi_{\rm BC}-\Phi_{\rm INT}]P_\ell^m(\cos\GJ')\cos[m(\GP-\GP')] d\Omega'  
               \right\} \label{eq:chiioDD}
\end{eqnarray}
\end{widetext}
The final solution will be obtained from the iteration of
\[  \Phi(x)\ =\ \GX(x)\ +\ \Phi_{\rm INT}(x)\ 
=\ \GX_a(x)\ +\ \GX_b(x)\ +\ \Phi_{\rm INT}(x)   \]
where $\GX_a+\GX_b$ are taken from Eq.~(\ref{eq:chiioND}) or (\ref{eq:chiioDD}) 
depending on the boundary condition. For the other boundary value 
problem, $\GX(x)=\GX_a(x)+\GX_b(x)$ is modified accordingly as shown 
in the next Appendix.

\section{Green's functions and surface integrals}
\label{sec:SurfInt}

In this Appendix, we present the explicit forms of 
the kernel functions denoted by $G^{\rm BC}(x,x')$ in 
Eq.(\ref{eq:general_G}), which appear in the surface integrals 
of the homogeneous solution $\chi(x)$ (\ref{eq:Green_chi}).  
Various types of boundary conditions are imposed
on a spherical domain bounded by two concentric spheres 
$S_a$ and $S_b$, and the corresponding kernel functions available 
in {\sc cocal} are tabulated in Table \ref{tab:Greenfn}.  
In \cite{TU2007} we used Green's functions $G^{\rm NB}(x,x')$, 
$G^{\rm DD}(x,x')$, and $G^{\rm ND}(x,x')$. 
Here we construct also $G^{\rm DN}(x,x')$, $G^{\rm NN}(x,x')$, 
$G^{\rm RD}(x,x')$. 

\begin{figure}
  \includegraphics[height=70mm,clip]{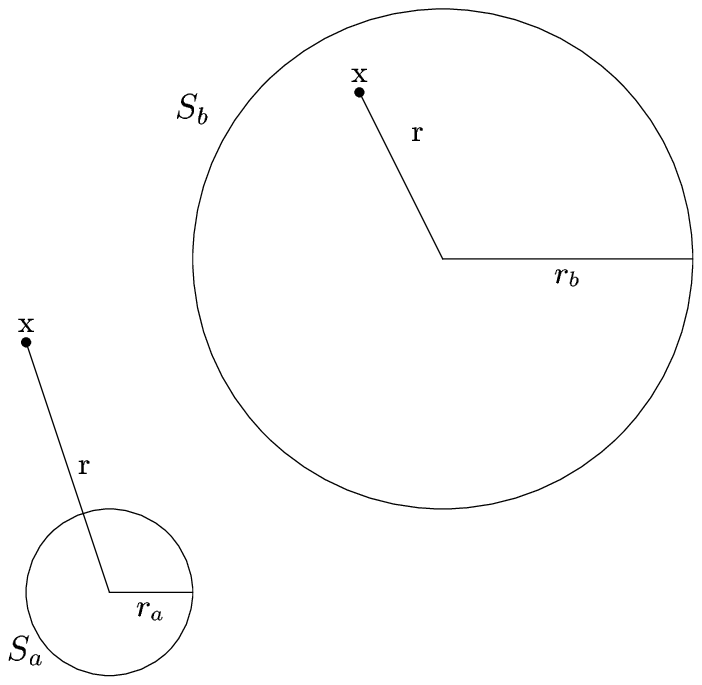}
  \caption{Surface integral for the exterior and the interior of a sphere   }
  \label{fig:SI_ext_int}
\end{figure}

The surface integral 
\beqn
\chi(x)&=&
\frac{1}{4\pi}\int_{S_a \cup S_b}
\left[G^{\rm BC}(x,x')\nabla'^\GA\widehat{\Phi}(x')
\right.
\nonumber\\
&&\qquad \qquad \left.
-\widehat{\Phi}(x')\nabla'^\GA G^{\rm BC}(x,x')\right]dS'_\GA , 
\eeqn
where $\widehat{\Phi}(x'):=\Phi^{\rm BC}(x')-\Phiint(x')$, 
are written below for both the exterior and the interior problem.  
Noticing that $dS'_\GA$ is pointing outward, 
$S_a$ and $S_b$ are the concentric 
spheres, and $\nabla'^\GA f(x')\, dS'_\GA = \pa_{r'} f \,r'^2 d\Omega'$, 
we have 
\beqn
&&
\chi_a = \frac{1}{4\pi}\int_{S_a}
\sum_{\ell=0}^{\infty}
\left[-g_\ell^{\rm BC}(r,r')\pd_{r'}\widehat{\Phi}
\,+\,\pa_{r'} g_\ell^{\rm BC}(r,r')\,\widehat{\Phi}\right]_{r'=r_a}
\!\!\!\!\times
\nonumber\\
&&
\sum_{m=0}^\ell 
\GE_m\frac{(\ell-m)!}{(\ell+m)!}P_\ell^m(\cos\GJ)
P_\ell^m(\cos\GJ')\cos[m(\GP-\GP')]
r_a^2 d\Omega' , 
\nonumber\\
\label{eq:SIaBC}
\eeqn
and
\beqn
&&
\chi_b = \frac{1}{4\pi}\int_{S_b}
\sum_{\ell=0}^{\infty}
\left[g_\ell^{\rm BC}(r,r')\pd_{r'}\widehat{\Phi}
\,-\,\pa_{r'} g_\ell^{\rm BC}(r,r')\,\widehat{\Phi}\right]_{r'=r_b}
\!\!\!\!\times
\nonumber\\
&&
\sum_{m=0}^\ell 
\GE_m\frac{(\ell-m)!}{(\ell+m)!}P_\ell^m(\cos\GJ)
P_\ell^m(\cos\GJ')\cos[m(\GP-\GP')]
r_b^2 d\Omega' . 
\nonumber\\
\label{eq:SIbBC}
\eeqn
Note that these $\chi_a$ and $\chi_b$ in this section are defined 
differently from those in the previous section, Eq.~(\ref{eq:chi_a}) 
and (\ref{eq:chi_b}).  
As shown in Fig.~\ref{fig:SI_ext_int}, 
we denote by $r_a$ the radius of the sphere $S_a$ 
for the exterior problem and by $\chi_a$ the corresponding integral
and by $r_b$ the radius of the sphere $S_b$ for the interior problem and 
$\chi_b$ the corresponding integral.

\subsection{Kernel function $G^{\rm NB}(x,x')$}
\label{sec:SIGNB}

The radial part for the kernel function without boundary 
$G^{\rm NB}(x,x')$, is  
\be
g^{\rm NB}_\ell(r,r')=\frac{r_<^\ell}{r_>^{\ell+1}}, 
\ee
where $r_> := \max\{r,r'\}$, and $r_< := \min\{r,r'\}$.  
The radial part of the kernel function in the surface integral on $S_a$ 
(\ref{eq:SIaBC}) becomes 
\beq
g_\ell^{\rm NB}(r,r_a) = \frac{r_a^\ell}{r^{\ell+1}}, 
\ \ \ \ 
\pa_{r'}g_\ell^{\rm NB}(r,r_a) = \ell\frac{r_a^{\ell-1}}{r^{\ell+1}}, 
\label{eq:SIaNB}
\eeq
while the surface integral on $S_b$, 
\beqn
g_\ell^{\rm NB}(r,r_b) = \frac{r^\ell}{r_b^{\ell+1}}, 
\ \ \ \ 
\pa_{r'}g_\ell^{\rm NB}(r,r_a) = -(\ell+1)\frac{r^{\ell}}{r_b^{\ell+2}}.
\nonumber\\
\label{eq:SIbNB}
\eeqn

\subsection{Kernel function $G^{\rm DD}(x,x')$}
\label{sec:SIGDD}

When the Dirichlet boundary condition is imposed on both $S_a$ and $S_b$, 
the kernel function $G^{\rm DD}(x,x')$ satisfies
\[ \left. G^{\rm DD}(x,x')\right|_{S_a}=0,
\qquad \left. G^{\rm DD}(x,x')\right|_{S_b}=0\ . \]
These conditions lead to the vanishing of the radial part $g^{\rm DD}_\ell(r,r')$ 
on the two spheres $S_a$ and $S_b$,
\[ g^{\rm DD}_\ell(r,r_a)=g^{\rm DD}_\ell(r,r_b)=0,  \]
which result to following formula for $g^{\rm DD}_l(r,r')$:
\beqn
&&g^{\rm DD}_\ell(r,r') \,=\,
\left[1-\left(\frac{r_a}{r_b}\right)^{2\ell+1}\right]^{-1}
\frac{r_a^\ell}{r_b^{\ell+1}}
\nonumber\\
&&\!\!\!\!
\times
\left[
\left(\frac{r_{<}}{r_a}\right)^{\ell} 
-\left(\frac{r_a}{r_{<}}\right)^{\ell+1}
\right]
\left[
\left(\frac{r_b}{r_{>}}\right)^{\ell+1}
-\left(\frac{r_{>}}{r_b}\right)^{\ell}
\right].
\label{eq:GreenDDrad}
\eeqn

The radial kernel function at $S_a$ becomes 
\beqn
g^{\rm DD}_\ell(r,r_a)&=&0,
\nonumber\\
\pa_{r'}g^{\rm DD}_\ell(r,r_a)
&=&
(2\ell+1)\frac{r_a^{\ell-1}}{r_b^{\ell+1}}
\frac{\left(\frac{r_b}{r}\right)^{\ell+1}
-\left(\frac{r}{r_b}\right)^\ell}{1-\left(\frac{r_a}{r_b}\right)^{2\ell+1}}, 
\label{eq:SIaDD}
\eeqn
and at $S_b$, 
\beqn
g^{\rm DD}_\ell(r,r_b)&=&0,
\nonumber\\
\pa_{r'}g^{\rm DD}_\ell(r,r_b)
&=&
-(2\ell+1)\frac{r_a^{\ell}}{r_b^{\ell+2}}
\frac{\left(\frac{r}{r_a}\right)^{\ell}
-\left(\frac{r_a}{r}\right)^{\ell+1}}
{1-\left(\frac{r_a}{r_b}\right)^{2\ell+1}}. \ 
\label{eq:SIbDD}
\eeqn

Special cases are when the surface $S_b$ is absent in the limit 
$r_b\rightarrow\infty$, 
\beq
\pa_{r'}g^{\rm DD}_\ell(r,r_a)
\,=\, (2\ell+1)\frac{r_a^{\ell-1}}{r^{\ell+1}},
\label{eq:SIaD}
\eeq
for the surface integral at $S_a$,
or when the surface $S_a$ is absent in the limit 
$r_a \rightarrow 0$, 
\beq
\pa_{r'}g^{\rm DD}_\ell(r,r_b)
\,=\, 
-(2\ell+1)\frac{r^{\ell}}{r_b^{\ell+2}},  
\label{eq:SIbD}
\eeq
for the surface integral at $S_b$.
The latter will be used for computing neutron stars.  

\subsection{Kernel function $G^{\rm ND}(x,x')$}
\label{sec:SIGND}

When the Neumann and Dirichlet boundary conditions are 
imposed on $S_a$ and $S_b$, respectively,  
the kernel function $G^{\rm ND}(x,x')$ satisfies 
\[ \left.\pa_{r'} G^{\rm ND}(x,x')\right|_{S_a}=0,\qquad 
\left.G^{\rm ND}(x,x')\right|_{S_b}=0 \]
or in terms of $g^{\rm ND}_\ell(r,r')$ 
\[ \pa_{r'} g^{\rm ND}_\ell(r,r_a)=g^{\rm ND}_\ell(r,r_b)=0\ .  \]
Then the radial part of $G^{\rm ND}(x,x')$ is 
\beqn
g^{\rm ND}_\ell(r,r')&=&
\left[
1+\frac{\ell}{\ell+1}
\left(\frac{r_a}{r_b}\right)^{2\ell+1}\right]^{-1}
\frac{r_a^\ell}{r_b^{\ell+1}}\times
\nonumber\\
&&
\times
\left[
\left(\frac{r_<}{r_a}\right)^\ell 
+ \frac{\ell}{\ell+1} \left(\frac{r_a}{r_<}\right)^{\ell+1}
\right]\times
\nonumber\\
&&
\times
\left[
\left(\frac{r_b}{r_>}\right)^{\ell+1}
- \left(\frac{r_>}{r_b}\right)^{\ell}
\right] .
\label{eq:GreenNDrad}
\eeqn

The radial kernel function at $S_a$ becomes 
\beqn
g^{\rm ND}_\ell(r,r_a)&=&
\frac{2\ell+1}{\ell+1}\frac{r_a^{\ell}}{r_b^{\ell+1}}
\frac{\left(\frac{r_b}{r}\right)^{\ell+1}-\left(\frac{r}{r_b}\right)^\ell}
{1+\frac{\ell}{\ell+1}\left(\frac{r_a}{r_b}\right)^{2\ell+1}}, \ 
\label{eq:SIaND}
\\
\pa_{r'}g^{\rm ND}_\ell(r,r_a)&=&0
\nonumber
\eeqn
and at $S_b$, 
\beqn
g^{\rm ND}_\ell(r,r_b)&=&0
\nonumber\\
\pa_{r'}g^{\rm ND}_\ell(r,r_b)&=&
-(2\ell+1)\frac{r_a^{\ell}}{r_b^{\ell+2}}
\frac{\left(\frac{r}{r_a}\right)^{\ell}
+\frac{\ell}{\ell+1}\left(\frac{r_a}{r}\right)^{\ell+1}}
{1+\frac{\ell}{\ell+1}\left(\frac{r_a}{r_b}\right)^{2\ell+1}}. 
\nonumber\\
\label{eq:SIbND}
\eeqn

Special case is when the surface $S_b$ is absent
in the limit $r_b\rightarrow\infty$, 
\beqn
g^{\rm ND}_\ell(r,r_a)&=&
\frac{2\ell+1}{\ell+1}
\frac{r_a^{\ell}}{r^{\ell+1}}
\label{eq:SIaN}
\eeqn
for the surface integral at $S_a$.

\subsection{Kernel function $G^{\rm DN}(x,x')$}
\label{sec:SIGDN}

When the Dirichlet and Neumann boundary conditions are 
imposed on $S_a$ and $S_b$, respectively,  
the kernel function $G^{\rm DN}(x,x')$ satisfies 
\[ \left.G^{\rm DN}(x,x')\right|_{S_a}=0,\qquad \left.\pa_{r'} G^{\rm DN}(x,x')\right|_{S_b}=0 \]
or in terms of $g^{\rm DN}_\ell(r,r')$
\[ g^{\rm DN}_\ell(r,r_a)=\pa_{r'} g^{\rm DN}_\ell(r,r_b)=0\ .  \]
Then the radial part of $G^{\rm DN}(x,x')$ is 
\beqn
g^{\rm DN}_\ell(r,r')&=&
\left[ \frac{\ell}{\ell+1} + 
\left(\frac{r_a}{r_b}\right)^{2\ell+1}\right]^{-1}
\frac{r_a^\ell}{r_b^{\ell+1}}\times
\nonumber\\
&&
\times
\left[\left(\frac{r_<}{r_a}\right)^\ell - \left(\frac{r_a}{r_<}\right)^{\ell+1}\right]\times
\nonumber\\
&&
\times
\left[\left(\frac{r_>}{r_b}\right)^{\ell}+\frac{\ell}{\ell+1}\left(\frac{r_b}{r_>}\right)^{\ell+1} \right] .\qquad
\label{eq:GreenDNrad}
\eeqn

The radial kernel function at $S_a$ becomes 
\beqn
g^{\rm DN}_\ell(r,r_a)&=&0, 
\nonumber \\
\pa_{r'}g^{\rm DN}_\ell(r,r_a)&=&
(2\ell+1)\frac{r_a^{\ell-1}}{r_b^{\ell+1}}
\frac{\left(\frac{r}{r_b}\right)^{\ell}
+\frac{\ell}{\ell+1}\left(\frac{r_b}{r}\right)^{\ell+1}}
{\frac{\ell}{\ell+1}+\left(\frac{r_a}{r_b}\right)^{2\ell+1}}, 
\nonumber\\
\label{eq:SIaDN}
\eeqn
%%% When the surface $S_b$ is absent, we get (\ref{eq:SIaD}). 
and at $S_b$,  
\beqn
g^{\rm DN}_\ell(r,r_b)&=&
\frac{2\ell+1}{\ell+1}\frac{r_a^\ell}{r_b^{\ell+1}}
\frac{\left(\frac{r}{r_a}\right)^{\ell}-\left(\frac{r_a}{r}\right)^{\ell+1}}
{\frac{\ell}{\ell+1}+\left(\frac{r_a}{r_b}\right)^{2\ell+1}}, \ 
\label{eq:SIbDN}
\\
\pa_{r'}g^{\rm DN}_\ell(r,r_b)&=&0.
\nonumber
\eeqn

\subsection{Kernel function $G^{\rm RD}(x,x')$}
\label{sec:SIGRD}

When the Robin and Dirichlet boundary conditions are 
imposed on $S_a$ and $S_b$, respectively,  
the kernel function $G^{\rm RD}(x,x')$ satisfies 
\[ \left[\frac{\pd G^{\rm RD}}{\pd r}+\frac{G^{\rm RD}}{2r}\right]_{S_a}=0,\qquad  
\left.G^{\rm RD}(x,x')\right|_{S_b}=0 \]
or in terms of $g^{\rm RD}_\ell(r,r')$
\[ \left[\frac{\pd g^{\rm RD}_\ell}{\pd r} + \frac{g^{\rm RD}_\ell}{2r}\right]_{r=r_a}=0,\qquad 
  g^{\rm RD}_\ell(r,r_b)=0 .  \]
Then radial part of $G^{\rm RD}(x,x')$ is
\beqn
&&g^{\rm RD}_\ell(r,r') \,=\,
\left[1+\left(\frac{r_a}{r_b}\right)^{2\ell+1}\right]^{-1}
\frac{r_a^\ell}{r_b^{\ell+1}}
\times\nonumber\\
&&\!\!\!\!
\times
\left[
\left(\frac{r_{<}}{r_a}\right)^{\ell} 
+\left(\frac{r_a}{r_{<}}\right)^{\ell+1}
\right]
\left[
\left(\frac{r_b}{r_{>}}\right)^{\ell+1}
-\left(\frac{r_{>}}{r_b}\right)^{\ell}
\right].\quad
\label{eq:GreenPsi}
\eeqn

For the surface integral at $S_a$, it is more convenient to rewrite 
Eq.(\ref{eq:SIaBC}), 
\beqn
&&
\chi_a = \frac{1}{4\pi}\int_{S_a}
\sum_{\ell=0}^{\infty}
\left[-g_\ell^{\rm RD}(r,r')\left(\pd_{r'}\widehat{\Phi}
+\frac{\widehat{\Phi}}{2r'}\right)
\right.
\nonumber\\
&&\qquad \qquad \left.
\,+\,\left(\pa_{r'} g_\ell^{\rm RD}(r,r')+\frac{g_\ell^{\rm RD}(r,r')}{2r'}\right)\,
\widehat{\Phi}\right]_{r'=r_a}
\!\!\!\!\times
\nonumber\\
&&
\sum_{m=0}^\ell 
\GE_m\frac{(\ell-m)!}{(\ell+m)!}P_\ell^m(\cos\GJ)
P_\ell^m(\cos\GJ')\cos[m(\GP-\GP')]
r_a^2 d\Omega' , 
\nonumber\\
\label{eq:SIaRD}
\eeqn
while for $\chi_b$, Eq.~(\ref{eq:SIbBC}) is used.  
Here, the radial kernel function at $S_a$ becomes 
\beqn
&& g^{\rm RD}_\ell(r,r_a)
\,=\,
 2\frac{r_a^{\ell}}{r_b^{\ell+1}}
\frac{\left(\frac{r_b}{r}\right)^{\ell+1}
-\left(\frac{r}{r_b}\right)^\ell}{1+\left(\frac{r_a}{r_b}\right)^{2\ell+1}}
\label{eq:SIaPsi}
\\
&& \pa_{r'} g_\ell^{\rm RD}(r,r_a)+\frac{g_\ell^{\rm RD}(r,r_a)}{2r_a}\,=\, 0.
\nonumber
\eeqn
The radial kernel function at $S_b$ in Eq.(\ref{eq:SIbBC}) becomes 
\beqn
g_\ell^{\rm RD}(r,r_b)&=& 0,
\nonumber\\
\pa_{r'} g_\ell^{\rm RD}(r,r_b) &=&
-(2\ell+1)\frac{r_a^{\ell}}{r_b^{\ell+2}}
\frac{\left(\frac{r}{r_a}\right)^{\ell}
+\left(\frac{r_a}{r}\right)^{\ell+1}}{1+\left(\frac{r_a}{r_b}\right)^{2\ell+1}}.
\ \ \ 
\label{eq:SIbPsi}
\eeqn

Special case is when the surface $S_b$ is absent 
in the limit $r_b\rightarrow\infty$.
In that case, we use in Eq.~(\ref{eq:SIaRD}),
for $\ell=1,2,\ldots$
\beqn
g^{\rm RD}_\ell(r,r_a) \,=\, 
2\frac{r_a^{\ell}}{r^{\ell+1}}. 
\label{eq:SIa0Psi}
\eeqn

\subsection{Kernel function $G^{\rm DR}(x,x')$}
\label{sec:SIGDR}

When the Dirichlet and Robin boundary conditions are 
imposed on $S_a$ and $S_b$, respectively,  
the kernel function $G^{\rm DR}(x,x')$ satisfies 
\[ 
\left.G^{\rm DR}(x,x')\right|_{S_a}=0 ,\qquad  
\left[\frac{\pd G^{\rm DR}}{\pd r}+\frac{G^{\rm DR}}{2r}\right]_{S_b}=0
\]
or in terms of $g^{\rm DR}_\ell(r,r')$
\[ 
g^{\rm DR}_\ell(r,r_a)=0,\qquad   
\left[\frac{\pd g^{\rm DR}_\ell}{\pd r} + \frac{g^{\rm DR}_\ell}{2r}\right]_{r=r_b}=0 .
\]
Then radial part of $G^{\rm DR}(x,x')$ is
\beqn
&&g^{\rm DR}_\ell(r,r') \,=\,
\left[1+\left(\frac{r_a}{r_b}\right)^{2\ell+1}\right]^{-1}
\frac{r_a^\ell}{r_b^{\ell+1}}
\times\nonumber\\
&&\!\!\!\!
\times
\left[
\left(\frac{r_{<}}{r_a}\right)^{\ell} 
-\left(\frac{r_a}{r_{<}}\right)^{\ell+1}
\right]
\left[
\left(\frac{r_b}{r_{>}}\right)^{\ell+1}
+\left(\frac{r_{>}}{r_b}\right)^{\ell}
\right].\quad
\label{eq:GreenDR}
\eeqn

For the surface integral at $S_b$, it is more convenient to rewrite 
Eq.(\ref{eq:SIbBC}), 
\beqn
&&
\chi_b = \frac{1}{4\pi}\int_{S_b}
\sum_{\ell=0}^{\infty}
\left[g_\ell^{\rm DR}(r,r')\left(\pd_{r'}\widehat{\Phi}
+\frac{\widehat{\Phi}}{2r'}\right)
\right.
\nonumber\\
&&\qquad \qquad \left.
\,-\,\left(\pa_{r'} g_\ell^{\rm DR}(r,r')+\frac{g_\ell^{\rm DR}(r,r')}{2r'}\right)\,
\widehat{\Phi}\right]_{r'=r_b}
\!\!\!\!\times
\nonumber\\
&&
\sum_{m=0}^\ell 
\GE_m\frac{(\ell-m)!}{(\ell+m)!}P_\ell^m(\cos\GJ)
P_\ell^m(\cos\GJ')\cos[m(\GP-\GP')]
r_b^2 d\Omega' , 
\nonumber\\
\label{eq:SIbDR}
\eeqn
while for $\chi_a$, Eq.~(\ref{eq:SIaBC}) is used.  
Here, the radial kernel function at $S_a$ becomes 
\beqn
g_\ell^{\rm DR}(r,r_a)&=& 0,
\nonumber\\
\pa_{r'} g_\ell^{\rm DR}(r,r_a) &=&
(2\ell+1)\frac{r_a^{\ell-1}}{r_b^{\ell+1}}
\frac{\left(\frac{r}{r_b}\right)^{\ell}
+\left(\frac{r_b}{r}\right)^{\ell+1}}{1+\left(\frac{r_a}{r_b}\right)^{2\ell+1}}.
\ \ \ 
\label{eq:radkernel_aDR}
\eeqn
The radial kernel function at $S_b$ in Eq.(\ref{eq:SIbBC}) becomes 
\beqn
&& g^{\rm DR}_\ell(r,r_b)
\,=\,
 2\frac{r_a^{\ell}}{r_b^{\ell+1}}
\frac{\left(\frac{r}{r_a}\right)^{\ell}-\left(\frac{r_a}{r}\right)^{\ell+1}}
{1+\left(\frac{r_a}{r_b}\right)^{2\ell+1}}
\label{eq:radkernel_bDR}
\\
&& \pa_{r'} g_\ell^{\rm DR}(r,r_b)+\frac{g_\ell^{\rm DR}(r,r_b)}{2r_b}\,=\, 0.
\nonumber
\eeqn

Special case is when the surface $S_a$ is absent 
in the limit $r_a\rightarrow 0$.
In that case, we use in Eq.~(\ref{eq:radkernel_bDR}),
for $\ell=1,2,\ldots$
\beqn
g^{\rm DR}_\ell(r,r_b) \,=\, 
2\frac{r^{\ell}}{r_b^{\ell+1}}. 
\label{eq:SIa0DR}
\eeqn

Such kernel functions for imposing Robin boundary conditions at 
the outer surface $S_b$ may improve an accuracy of the solution 
especially near the boundary \cite{York_Piran}.

\subsection{Kernel function $G^{\rm NN}(x,x')$}
\label{sec:SIGNN}

When the Neumann boundary condition is imposed on both $S_a$ and $S_b$, 
the kernel function $G^{\rm NN}(x,x')$ satisfies
\[ \left.\pa_{r'} G^{\rm NN}(x,x')\right|_{S_a}=G_a, \qquad 
\left.\pa_{r'} G^{\rm NN}(x,x')\right|_{S_b}=G_b   \]
where $G_a,\ G_b$ cannot be both zero. In that case $G^{\rm NN}(x,x')$ does not exist
since the $\ell=0$ mode cannot be satisfied. Therefore the boundary conditions for the 
radial part will be
\[ \pa_{r'} g^{\rm NN}_\ell(r,r_a)=G_a \GD_{0\ell}, \qquad \pa_{r'} g^{\rm NN}_\ell(r,r_b)=G_b \GD_{0\ell}  \]  
Then for $\ell=1,2,\ldots$ we get 
\beqn
g^{\rm NN}_\ell(r,r')&=&
\left[1-\left(\frac{r_a}{r_b}\right)^{2\ell+1}\right]^{-1} \frac{r_a^\ell}{r_b^{\ell+1}}\frac{\ell+1}{\ell}
\nonumber\\
&&
\times
\left[\left(\frac{r_<}{r_a}\right)^\ell+
\frac{\ell}{\ell+1} \left(\frac{r_a}{r_<}\right)^{\ell+1}\right]
\nonumber\\
&&
\times
\left[\left(\frac{r_>}{r_b}\right)^\ell+
\frac{\ell}{\ell+1} \left(\frac{r_b}{r_>}\right)^{\ell+1}\right] .\qquad
\label{eq:GreenNNrad}
\eeqn
(symmetric in $r,r'$), while for $\ell=0$
\[g^{\rm NN}_0(r,r')=\frac{1}{r_>}-\frac{G_a r_a^2}{r}+h(r')  \]
where $h(r')$ arbitrary function. Symmetry is imposed by choosing
$h(r')=-G_a r_a^2/r'$ therefore
\be
g^{\rm NN}_0 (r,r')=\frac{1}{r_>}-r_a^2G_a\left(\frac{1}{r_>}+\frac{1}{r_<}\right)
\label{eq:GreenNNrad0}
\ee 
Note also that in order to satisfy the $\ell=0$
mode the following condition must hold
\[  1 = G_a r_a^2 - G_b r_b^2  \] 
therefore $G_a$ and $G_b$ cannot be chosen arbitrarily.
The surface integral at $S_a$ is
\beqn
\chi_a= \frac{1}{4\pi}\sum_{\ell=0}^{\infty}\sum_{m=0}^\ell \GE_m\frac{(\ell-m)!}{(\ell+m)!}P_\ell^m(\cos\GJ)
\times\nonumber\\
\times\int_{S_a} \left(-g^{\rm NN}_\ell(r,r_a)\frac{\pd\Phi}{\pd r}+G_a\GD_{0\ell}\Phi\right)
\times\nonumber\\ 
\times P_\ell^m(\cos\GJ')\cos[m(\GP-\GP')]r_a^2 d\Omega'
\ \ \label{eq:SIaNN}
\eeqn
where for $\ell=1,2,\ldots$
\beq
g^{\rm NN}_\ell(r,r_a)= \frac{2\ell+1}{\ell}\frac{r_a^\ell}{r_b^{\ell+1}}
\frac{\left(\frac{r}{r_b}\right)^\ell+
\frac{\ell}{\ell+1}\left(\frac{r_b}{r}\right)^{\ell+1}}{1-\left(\frac{r_a}{r_b}\right)^{2\ell+1}}
\eeq
and $\ell=0$
\beq
g^{\rm NN}_0 (r,r_a)=\frac{1}{r}-r_a^2G_a\left(\frac{1}{r}+\frac{1}{r_a}\right). 
\eeq
The surface integral at $S_b$ is
\beqn
\chi_b= \frac{1}{4\pi}\sum_{\ell=0}^{\infty}\sum_{m=0}^\ell \GE_m\frac{(\ell-m)!}{(\ell+m)!}P_\ell^m(\cos\GJ)
\times\nonumber\\
\times\int_{S_b} \left(g^{\rm NN}_\ell(r,r_b)\frac{\pd\Phi}{\pd r}-G_b\GD_{0\ell}\Phi\right)
\times\nonumber\\ 
\times P_\ell^m(\cos\GJ')\cos[m(\GP-\GP')]r_b^2 d\Omega'
\ \ \label{eq:SIbNN}
\eeqn
where for $\ell=1,2,\ldots$
\beq
g^{\rm NN}_\ell(r,r_b)= \frac{2\ell+1}{\ell}\frac{r_a^\ell}{r_b^{\ell+1}}
\frac{\left(\frac{r}{r_a}\right)^\ell+
\frac{\ell}{\ell+1}\left(\frac{r_a}{r}\right)^{\ell+1}}{1-\left(\frac{r_a}{r_b}\right)^{2\ell+1}},   
\eeq
and $\ell=0$
\beq
g^{\rm NN}_0 (r,r_b)=\frac{1}{r_b}-r_a^2G_a\left(\frac{1}{r_b}+\frac{1}{r}\right). 
\eeq

Special case is when the surface $S_a$ is absent in the limit $r_a\rightarrow 0$. 
In that case, we use in Eq.~(\ref{eq:SIbNN}), 
for $\ell=1,2,\ldots$
\beq
g^{\rm NN}_\ell(r,r_b)= \frac{2\ell+1}{\ell}\frac{r^\ell}{r_b^{\ell+1}},   
\eeq
and $\ell=0$
\beq
g^{\rm NN}_0 (r,r_b)=\frac{1}{r_b} . 
\eeq

\section{Finite difference formulas}
\label{sec:FDformulas}

The second order finite difference formulas used in the elliptic equation solvers 
of {\sc cocal} code are summarized in this section.  In evaluating the 
integrals of the solver, such as Eq.~(\ref{eq:Green_int}), we use the 
mid-point rule.  Hence, we need to evaluate the source terms in the 
integrand at the mid-points $(r_{i-\frac12}, \theta_{j-\frac12}, \phi_{k-\frac12})$ 
of the grid points that may involve values of potentials and their derivatives.  
Those are calculated, respectively, by 
\beqn
&& f(r_{i-\frac12},\theta_{j-\frac12},\phi_{k-\frac12}) 
\nonumber\\
&&\qquad
\,\simeq\, 
\frac18\sum_{I = i-1}^{i} \sum_{J = j-1}^{j} \sum_{K = k-1}^{k} f_(r_I,\theta_J,\phi_K) , 
\eeqn
\beqn
&& \frac{\pa f}{\pa r}(r_{i-\frac12},\theta_{j-\frac12},\phi_{k-\frac12})
\nonumber\\
&&
\,\simeq\, 
\frac14 \sum_{J = j-1}^{j} \sum_{K = k-1}^{k} 
\frac{f(r_i,\theta_J,\phi_K)-f(r_{i-1},\theta_J,\phi_K)}{\Dl r_{i}} , \ \ \ 
\eeqn
\beqn
&& \frac{\pa f}{\pa \theta}(r_{i-\frac12},\theta_{j-\frac12},\phi_{k-\frac12})
\nonumber\\
&&
\,\simeq\, 
\frac14 \sum_{I = i-1}^{i} \sum_{K = k-1}^{k} 
\frac{f(r_I,\theta_j,\phi_K)-f(r_I,\theta_{j-1},\phi_K)}{\Dl \theta_{j}} , \ \ \ 
\eeqn
\beqn
&&\frac {\pa f}{\pa \phi}(r_{i-\frac12},\theta_{j-\frac12},\phi_{k-\frac12})
\nonumber\\
&&
\,\simeq\, 
\frac14 \sum_{I = i-1}^{i} \sum_{J = j-1}^{j}  
\frac{f(r_I,\theta_J,\phi_k)-f(r_I,\theta_J,\phi_{k-1})}{\Dl \phi_{k}}. \ \ \ 
\eeqn
The quadrature formula for the 2nd order mid-point rule at the 
interval $[r_{i-1},r_i]\times[\theta_{j-1},\theta_{j}]\times[\phi_{k-1},\phi_{k}]$
is written 
\beqn
&&\int_{r_{i-1}}^{r_i} dr
\int_{\theta_{j-1}}^{\theta_j} d\theta
\int_{\phi_{k-1}}^{\phi_k} d\phi\, S(r,\theta,\phi)
\nonumber\\
&&
\,\simeq\, 
S(r_{i-\frac12},\theta_{j-\frac12},\phi_{k-\frac12}) \Dl r_i \Dl \theta_j \Dl \phi_k.
\eeqn

%
%

%\cite{Price:2009nv}
%%% \bibitem{Price:2009nv}
%%%   R.~H.~Price, C.~Markakis and J.~L.~Friedman,
%%%   %``Iteration Stability for Simple Newtonian Stellar Systems,''
%%%   arXiv:0903.3074 [astro-ph.SR].
%%%   %%CITATION = ARXIV:0903.3074;%%
%%% 

%%% \bibitem{SG04} 
%%% G. Sch\"afer and A. Gopakumar, Phys. Rev. D {\bf 69}, 021501(R) (2004)
%%% 
%%% \bibitem{BGGN04} 
%%% %\bibitem[Bonazzola et al.(2004)]{2004PhRvD..70j4007B} 
%%% S.~Bonazzola,  E.~Gourgoulhon, P.~Grandcl{\'e}ment, and 
%%% J.~Novak, Phys.\ Rev.\  D {\bf 70}, 104007 (2004). 
%%% 

\end{document}